\documentclass[nofootinbib]{revtex4}
\usepackage{amsmath}
\usepackage{latexsym}
\usepackage{graphicx}
\usepackage{color}
\usepackage{listings}

\begin{document}

\title{Inflation, Cosmic Perturbations and Non-Gaussianities}
\author{Yi Wang}
\affiliation{Kavli Institute for the Physics and Mathematics of the Universe,
Todai Institutes for Advanced Study, University of Tokyo,
5-1-5 Kashiwanoha, Kashiwa, Chiba 277-8583, Japan
}
\affiliation{Centre for Theoretical Cosmology, DAMTP, University of Cambridge, Cambridge CB3 0WA, UK}

\begin{abstract}
  We review the theory of inflationary perturbations. Perturbations at both linear and nonlinear orders are reviewed. We also review a variety of inflation models, emphasizing their signatures on cosmic perturbations.
\end{abstract}

\maketitle
\tableofcontents

\section{An Introduction to Inflation}

Inflation \cite{Guth:1980zm, Linde:1981mu, Albrecht:1982wi} is the leading paradigm of the very early universe cosmology. The quantum fluctuations during inflation are stretched to super-Hubble scales and return to the Hubble horizon after inflation. They provide the seeds of CMB anisotropy and the large scale structure \cite{Mukhanov:1981xt, Hawking:1982cz, Starobinsky:1982ee, Guth:1982ec, Bardeen:1983qw}. 

Here we review the cosmic perturbation theory, which calculates inflationary fluctuations. There have been already many books, reviews and lecture notes in this area \cite{Kodama:1985bj, Mukhanov:1990me, Komatsu:2002db, Riotto:2002yw, Bartolo:2004if, Mukhanov:2005sc, Wands:2007bd, Weinberg:2008zzc, Malik:2008im, Kinney:2009vz, Baumann:2009ds, Liguori:2010hx, Langlois:2010xc, Chen:2010xka}. In the present review we focus on an explicit treatment of the standard theory, including linear and non-linear perturbations. After that, we make a broad survey of models. Before those aspects, we briefly overview the inflationary cosmology in the rest of this section.

\subsection{The hot big bang}

The dynamics of the standard hot big bang cosmology is governed by the Friedmann equation. By assuming the universe is homogeneous, isotropic\footnote{Homogeneity and isotropy are firstly introduced just to simplify the equations. Later it is supported by observations that our observable universe is indeed nearly homogeneous and isotropic above 100 Mpc scales.} and flat\footnote{Unlike homogeneity or isotropy, flatness is not a statement of symmetry. Nevertheless experiments suggest that our universe is very flat. Non-flatness, if exists, should have a curvature radius at least $10H^{-1}$. Thus for algebraic simplicity we also assume flatness unless stated otherwise.}, the metric can be written as the FRW form:
\begin{align}\label{eq:frw-bg}
  ds^2 = -dt^2 + a(t)^2 d\mathbf{x}^2 ~.
\end{align}
Here $a$ is the scale factor of the universe. Note that in a flat universe, the scale factor itself does not have physical meaning. Because we can always rescale $a$ and spatial coordinate $x_i$ together. On the other hand, ratios of scale factors are important. For example the Hubble parameter $H\equiv\dot a/a$ and the redshift $z\equiv a(t_0)/a(t)-1$ are key quantities in cosmology.

With the FRW metric, the 00-component of the Einstein equation is the Friedmann equation:
\begin{align} \label{eq:friedmann}
  3M_p^2 H^2 = \rho~,
\end{align}
where $\rho\equiv\sum\rho_i\equiv\rho_m+\rho_r+\rho_\Lambda+\cdots$ is the energy density. For non-interacting energy components, the time evolution (characterizing how this component dilutes with expansion of the universe) at the homogeneous and isotropic level is determined by the continuity equation
\begin{align}
  \dot\rho + 3H(\rho + p) = 0~,
\end{align}
where $p$ is pressure, which can be determined by $\rho$ (and perhaps other parameters) from the equation of state for this form of matter. For example, for thermal matter, $p/\rho$ can be calculated from the thermal partition function. For the cosmological constant, $p/\rho=-1$ follows from Lorentz invariance.

The early universe was radiation dominated ($p=\rho/3$). However, radiation domination cannot be eternal to the past. This is because from Eq.~(\ref{eq:friedmann}), inserting $\rho=\rho_r\propto a^{-4}$, we get $\dot a\propto1/a$ and $a\propto\sqrt{t-t_\mathrm{i}}$, where $t_\mathrm{i}$ is an integration constant. The moment $t=t_\mathrm{i}$ is a physical singularity, where scale factor sinks to zero and physical energy density diverges. Note that $t$ is the proper time measured by a comoving observer. Thus the radiation dominated universe must start at some finite physical time (13.7 Gyr ago). The start of the radiation dominated universe is known as the ``hot big bang''.

The singularity at $t=t_\mathrm{i}$ requires the hot big bang. But this singularity by itself does not imply what happens before the hot big bang, or even if anything, including spacetime, exists or not before that. Nevertheless, the hot big bang cosmology is sensitive to its initial conditions. There are strong evidences (almost firm proof) showing that new physics exists before the big bang. And inflation is so far the most successful candidate for this very early epoch. In the remainder of the section, we shall focus on those evidences and how inflation explains them.

\subsection{Horizon, flatness, monopole and correlation problems}

It is a surprising fact that we find ourselves living in a huge universe -- about $10^{23}$ kilometers in the observable radius and $10^{10}$ years in age (where kilometers and years are human's natural units). This turns into a greater surprise when we realize that our universe is supported by fundamental physics, where the natural scales are eV, GeV or even Planck scale. In those units the age and size of our universe are about $10^{33}$eV$^{-1}$ ($\sim 10^{42}$GeV$^{-1}\sim 10^{60} M_p^{-1}$).

Those surprises can be formulated in more accurate terms, by the horizon and flatness problems of the standard hot big bang cosmology. To see those problems, let us assume for a time that nothing happened before the hot big bang. In other words assume the universe is radiation dominated all the way back in the early times. Then problems arise:

\begin{itemize}
\item \textit{Horizon problem (I):} Let us start with an example of the cosmic microwave background (CMB).

From observations, we actually see the universe is homogeneous and isotropic in the past, because light (or any other causal signal) needs time to propagate to us. By optical probes, we can observe the homogeneity and isotropy of the universe up to as early as recombination: When the CMB forms, the universe is already very homogeneous and isotropic (up to fluctuations of order $10^{-5}$, which we will discuss soon). 

However, from the hot big bang to the time of recombination, causal signals could only propagate a limited distance in space. This is because those signals could at most propagate at speed of light, with null geodesic $0=ds^2=-dt^2+a^2dr^2$. Thus the physical distance of propagation (measured by today's ruler $a(t_0)$) from the big bang to recombination is at most\footnote{Here we approximated the universe as radiation dominated. Actually the universe has already become matter dominated at $z\simeq 3000$. Nevertheless this doesn't make a big difference for this estimate.}
\begin{align}\label{eq:light-prop-horizon}
  a(t_0)\Delta r = a(t_0)\int_{t_\mathrm{i}}^{t_\mathrm{rec}} \frac{dt}{a(t)} \simeq 2\sqrt{t_\mathrm{rec}t_0}
  \sim \sqrt{\frac{t_\mathrm{rec}}{t_0}} \times \mathrm{(size~of~current~observable~universe)} ~.
\end{align}
To put in numbers, the causal signals, from the hot big bang to recombination, could only propagate as long as about 1/100 of the radius of our observable universe. Thus different patches of the CMB sky were causally disconnected. This situation is illustrated in Fig.~\ref{fig:horizon}.

One could also consider this problem using conformal time. This is because conformal time, together with comoving spatial coordinates, straightforwardly shows the causal structure of spacetime.\footnote{Conformal time is not only important for showing causal structures. There are also a few other important usages of conformal time. For example, the field equations looks simpler and closer to Minkowski field equations in conformal time. Also, it can be used to show that conformal invariant fields do not ``know'' the cosmic expansion. We will discuss those aspects more extensively in later sections} The flat FRW metric can be rewritten as
\begin{align}\label{eq:frw-bg-conformal}
  ds^2 = a(\tau)^2 (-d\tau^2 + d\mathbf{x}^2)~,
\end{align}
where $\tau$ is defined (up to a constant) by $dt = ad\tau$. Using conformal time, the light cones are 45 degree straight lines in the $\tau$-$\mathbf{x}$ plane. Thus the distance luminal signals travel equals to conformal time.

Consider equation of state $p=w\rho$, where $w$ is a constant. One finds $a\propto (t-t_0)^{\frac{2}{3(1+w)}}$. Thus
\begin{align}\label{eq:sol-conformal-time}
  \tau-\tau_0 \propto (t-t_0)^{\frac{1+3w}{3(1+w)} }~.
\end{align}
For a fluid with $w>-1/3$ (including matter and radiation), a very short period of proper time $t-t_0$ corresponds to a very short period of conformal time $\tau-\tau_0$. Thus light cannot propagate far during this period of time.

\begin{figure}[htbp]
\centering
\includegraphics[width=0.6\textwidth]{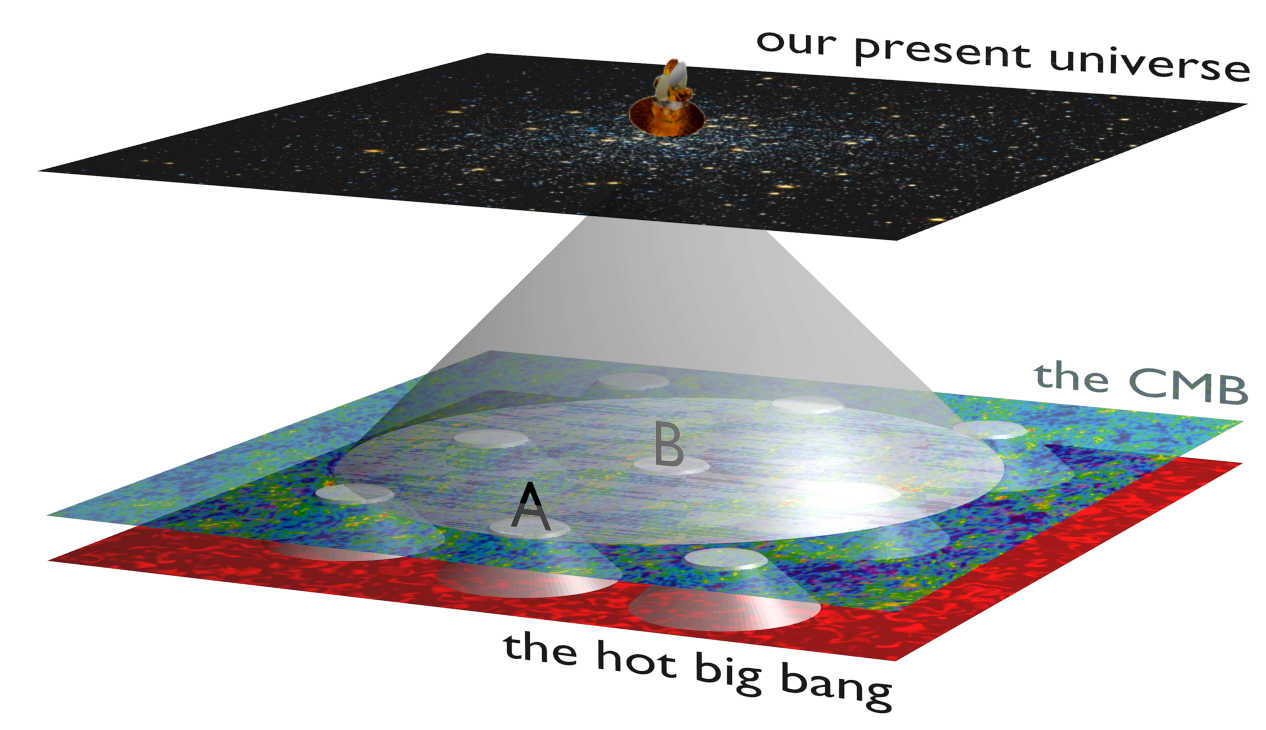}
\caption{\label{fig:horizon}The horizon problem of hot big bang cosmology. The space and time in this diagram should be understood as comoving and conformal respectively. The several smaller light cones are truncated between the hot big bang and recombination (the time of CMB formation). Those light cones have no overlap at the time of hot big bang, thus they are causally unrelated at recombination. But why, they at recombination (for example A and B in the figure), look similar to each other (considering we can observe many of them, as indicated by the greater light cone between recombination and the present time)?}
\end{figure}

Note that in the CMB example of the horizon problem, $t_\mathrm{rec}$ is set by the scale of observation instead of the intrinsic nature of the problem itself. More generally, we could consider different observables (replace $t_\mathrm{rec}$ with other time scales) and generate a family of similar horizon problems. For example, problems of causally disconnect patches of large scale structure. 


\item \textit{Horizon problem (II):} One could alternatively formulate the puzzle of horizon size in another way. Imagine we are observers sitting at $\mathbf{x}=0$, who have existed since the hot big bang, and there exists some structural information $\mathcal{S}$ which are comoving\footnote{The requirement of comoving is not only for simplicity. Motions away from comoving receives Hubble friction (the kinetic energy loss, analog of redshift for a massless particle). Thus on large scales, objects tend to be comoving.} in the universe (including homogeneity, isotropy and flatness, and departures from those ideal conditions). We could observe $\mathcal{S}$ by luminal or sub-luminal signals emitted from the structures on $\mathcal{S}$. Now at time $t_\mathrm{obs}$ (not before or after), how much portion of $\mathcal{S}$ could we observe (in comoving size)?

The comoving distance reached by light since the initial time is $\int_{t_\mathrm{i}}^{t_\mathrm{obs}}dt/a$, which grows with $t_\mathrm{obs}$. A related scale characterizing the evolution of perturbations relative to the expansion is the comoving Hubble radius:

\begin{align}
  \frac{1}{aH}~.
\end{align}

During radiation or matter domination, the comoving Hubble radius grows, so progressively larger comoving scales enter the Hubble radius\footnote{For a constant equation of state $w>-1/3$ and a big-bang origin, the particle horizon is proportional to $1/(aH)$. In general the Hubble radius differs from both the particle horizon and the event horizon.}. This growth reflects decelerating expansion, since $aH=\dot a$.

Now comes the horizon problem: If no one could ever ``see'' (more accurately, interact with) those structures in the early universe, how could they be causally prepared, and prepared in a consistent way? This situation is illustrated in Fig.~\ref{fig:flatness} and Fig.~\ref{fig:horizon-return}.

\begin{figure}[htbp]
\centering
\includegraphics[width=0.6\textwidth]{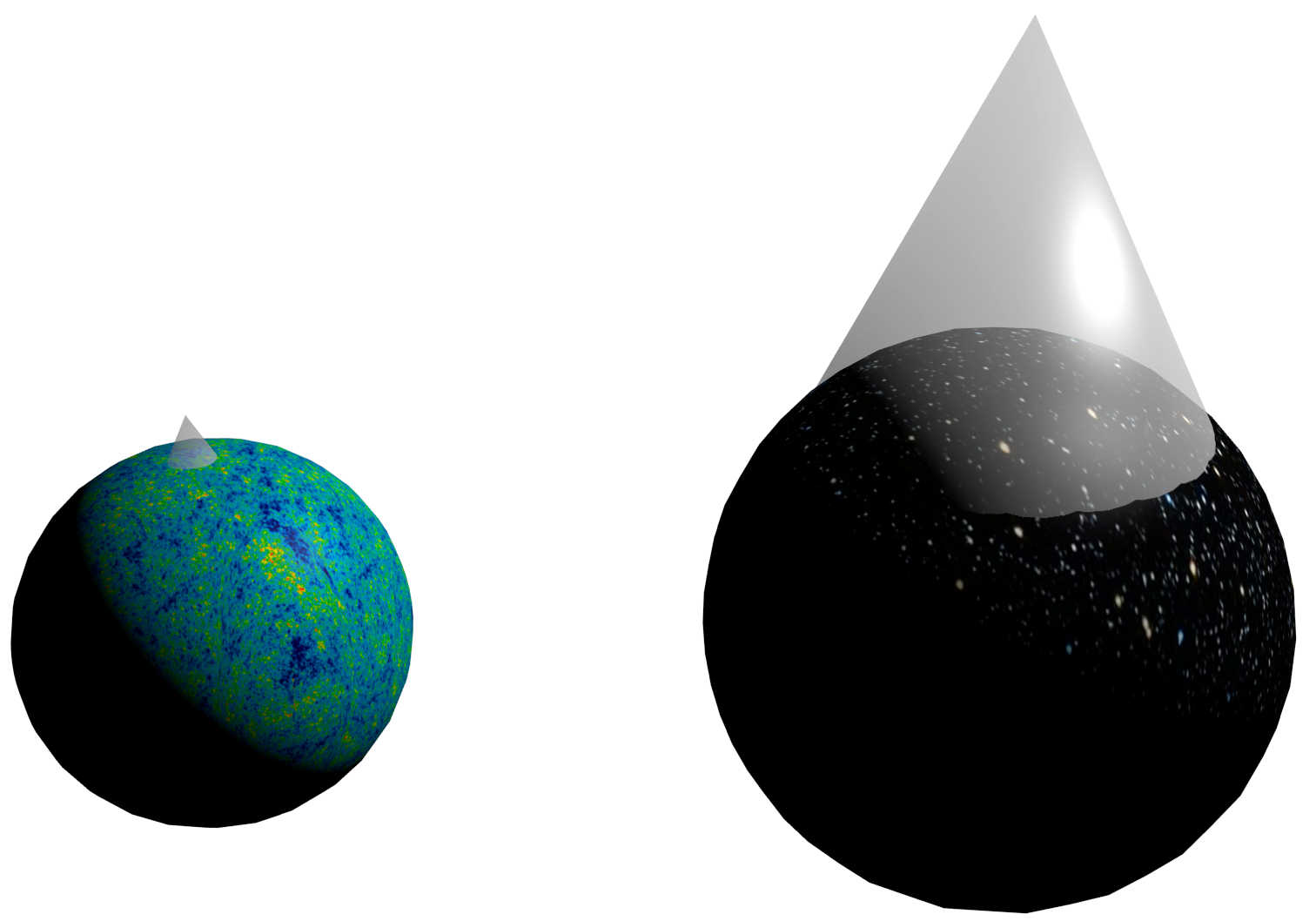}
\caption{\label{fig:flatness}The horizon problem II, or the flatness problem. At early times, the observer could only observe a small portion of the universe thus unaware of larger scale structures (e.g. the spherical topology in the figure). At later times though radiation or matter domination, the universe expands. However, the observer's horizon expands faster. Thus the observer could obtain more large scale structural information (e.g. easier to measure the non-flatness of the universe, if exists). From current observation, the universe is rather flat now. If there were non-flatness in our universe, how could causal process produce them in the early universe, considering no causal process could see this non-flatness in the early universe?}
\end{figure}

\begin{figure}[htbp]
  \centering
  \includegraphics[width=0.95\textwidth]{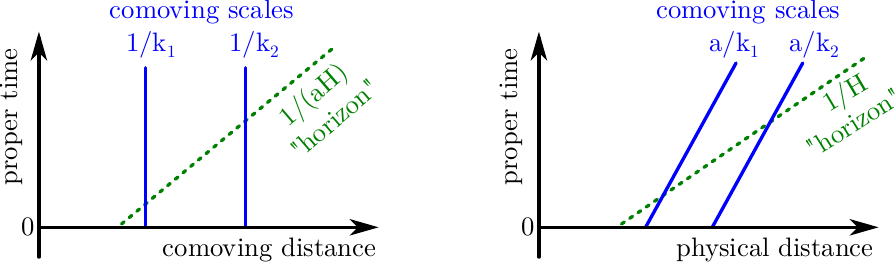}
  \caption{\label{fig:horizon-return} The comoving scales return to the horizon in radiation and matter dominated eras.}
\end{figure}

\item \textit{Flatness problem:} The flatness problem could be considered as an example of the version (II) of the horizon problem. Here the structural information $\mathcal{S}$ is the spatial curvature of the universe. For simplicity, we haven't put spatial curvature into the metric. Nevertheless, we could imagine we live on a sphere instead of a flat plane.

When the universe expands, the curvature radius $\mathcal{R}$ should expand together, keeping $\mathcal{R} = a(t)$ (with proper normalization of $a$). Nevertheless, as horizon problem (II), in matter or radiation dominated universes, the universe does not expand ``fast'' enough thus we still see more and more portion of the sphere as time goes on. 

So far, we haven't seen any curvature ($\mathcal{R}>10H^{-1}$). Thus at a much earlier time, $\mathcal{R}H$ has to be greater than $10$ by a much greater amount. How could such a flat universe be prepared at the beginning, without the help of causal processes?

\end{itemize}

This is not yet the end of the story. Some UV-completion theories of the particle physics standard model predict topological defects. From those theories, another question arises:
\begin{itemize}
\item \textit{Monopole problem}: Phase transitions of the early universe may produce gauge theory monopoles. If produced, they are supposed to dominate the energy density the universe. But we haven't found them. (On the other hand, global monopoles is not necessarily a problem.)
\end{itemize}

So far we have outlined a few problems for the standard picture of the hot big bang cosmology. However, it is worth noticing that sometimes it is hard to judge the seriousness of conceptional problems:

On the one hand, it is because those conceptional problems are paradigm dependent. For example, the monopole problem could be simply solved by starting with a theory without monopole solutions. One could as well argue that the universe could start from some almost deterministic initial conditions where the universe is extremely flat and uniform. Those theories exist, such as the quantum creation of universe \cite{Hartle:1983ai} (see also \cite{Linde:1983cm, Vilenkin:1984wp} for a different approach with different predictions).

On the other hand, solution to some problems may not be complete, or may also raise other (though arguably less serious) problems. We will come back later to the new problems brought by inflation.

However, there does exist physical problem of the hot big bang, without proper extension before the bang. This is the problem\footnote{Historically, the large scale correlation was no longer a problem of the bang, but already the success of inflation. This is because those correlations are detected after inflation was proposed.} of large scale correlations on the CMB (or in the large scale structure). 

\begin{itemize}
\item \textit{Correlation problem:} Again consider Fig.~\ref{fig:horizon}. Actually the CMB map is not absolutely homogeneous and isotropic, but instead has small fluctuations on it, with an amplitude $\sim 10^{-5}$. Now consider the statistical correlation of those fluctuations, in regions A and B. Assuming nothing exists before the hot big bang, A and B shouldn't have correlations. However, those statistical correlations were caught, mode by mode in momentum space, in the CMB experiments \cite{Smoot:1992td}. With nowadays CMB map (WMAP:\cite{Bennett:2012fp, Hinshaw:2012fq} ACT:\cite{Das:2013zf, Dunkley:2013vu, Sievers:2013wk} SPT:\cite{Story:2012wx, Hou:2012xq}), one can actually observe by eyes that the fluctuations are very different from a map generated from white noise.
\end{itemize}

Indeed, historically it is the verified predictions of inflation from the correlation in the cosmic perturbations, instead of the conceptual ``postdictions'', that convinced the majority that inflation should be the theory of the very early universe and produce the dominant seed of CMB and structure formation.

\subsection{Inflation as a solution of problems}

Inflation is a period of time before the hot big bang, during which the universe undergoes accelerating expansion:
\begin{align}
  \ddot a>0~.
\end{align}

For example, consider a constant equation of state $p=w\rho$. In this case acceleration implies $w<-1/3$. And as we shall see, inflation typically require $w\simeq -1$ (approach -1 from the greater side).

Before going into detailed dynamics which could drive the above accelerating expansion, let us first have a look how inflation solves the problems in the previous subsection.

\begin{itemize}
\item \textit{Horizon problem (I):} From equation (\ref{eq:sol-conformal-time}), we find that in case of acceleration ($w<-1/3$), $\tau-\tau_0\propto-(t-t_0)^\mathrm{negative~power}$. Thus small proper time interval $t-t_0\rightarrow0$ means negative and large conformal interval $\tau-\tau_0\rightarrow-\infty$.

Remember that $\tau-\tau_0$ corresponds to the distance that light ray could travel. $\tau-\tau_0\rightarrow-\infty$ means that light ray could travel a very long distance during the $w<-1/3$ epoch (although could be a small interval of proper time $(t-t_0)$) before reaching the time of recombination. Thus Fig.~\ref{fig:horizon} is replaced by Fig.~\ref{fig:horizon-inflation}, where the initial $t_\mathrm{i}$ hypersurface is stretched downwards in conformal distance, and light cones with tips on CMB becomes highly overlapped at $t_\mathrm{i}$.

Alternatively, we can re-calculate the light propagation distance, similarly to Eq.~\eqref{eq:light-prop-horizon}: during inflation
\begin{align}
  a(t_0)\Delta r = a(t_0)\int_{t_\mathrm{ini}}^{t_\mathrm{end}} \frac{dt}{a(t)} \simeq
  \frac{1}{H} \frac{a(t_0)}{a(t_\mathrm{ini})} 
  \simeq \frac{1}{H} \frac{a(t_0)}{a(t_\mathrm{end})} e^{H(t_\mathrm{end} - t_\mathrm{ini})}~.
\end{align}
This corresponds to an exponentially large distance, that light signals could have travelled during inflation. 

\item \textit{Horizon problem (II) and the flatness problem:} Those problems arises because $\dot a=aH$ was a decreasing function all the way since the creation of the universe (i.e. $\ddot a<0$). Now we have a period of inflation $\ddot a>0$ before the hot big bang. Thus the horizon problem (II) and the flatness problem are solved provided that the $\ddot a>0$ period is long enough. This situation is illustrated in Fig.~\ref{fig:horizon-inflation}, and Fig.~\ref{fig:horizon-exit-return}, which show the same physics in a spacetime diagram and in terms of the change of comoving horizon scale, respectively. 

\begin{figure}[htbp]
\centering
\includegraphics[width=0.6\textwidth]
{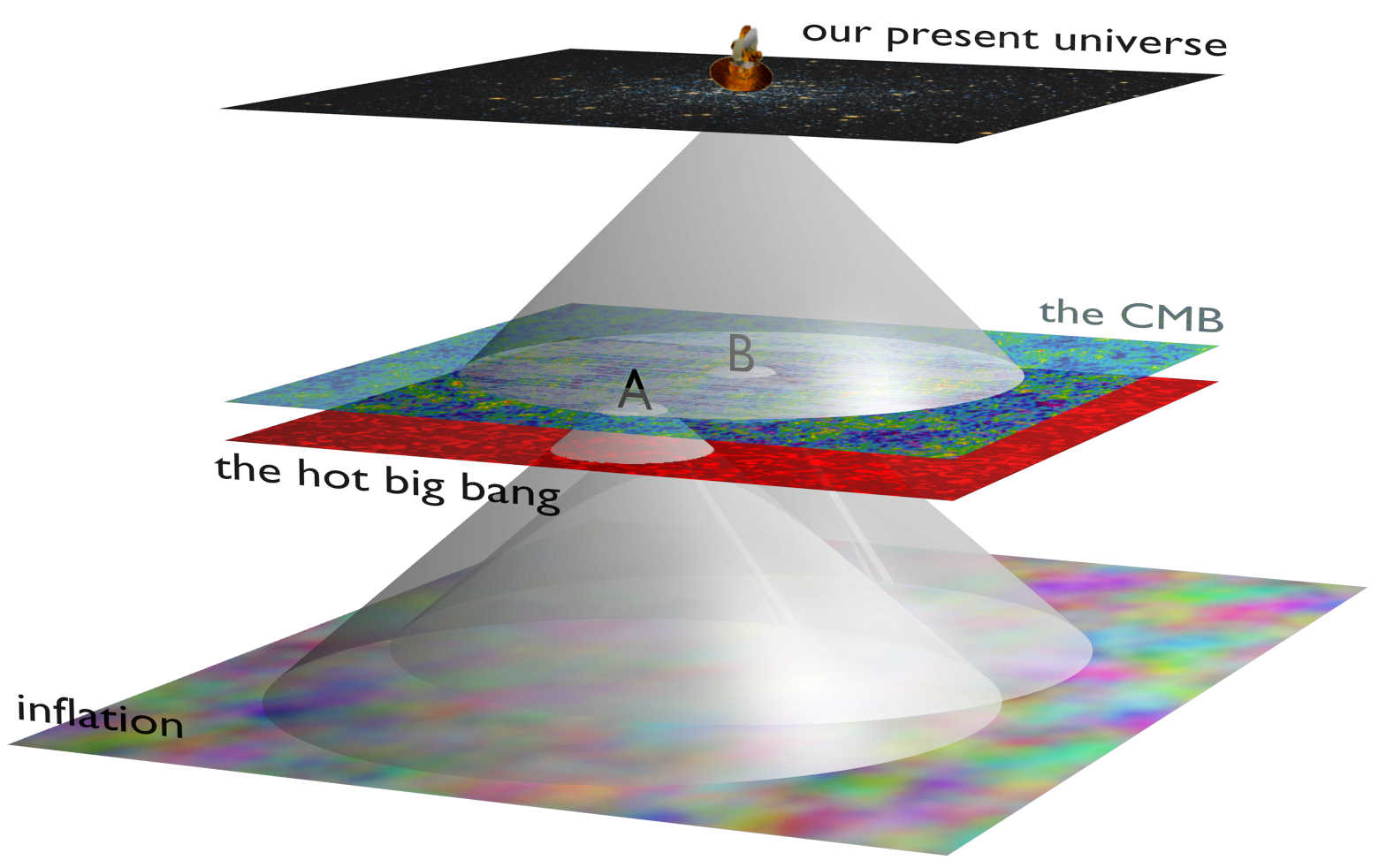}
\caption{\label{fig:horizon-inflation}Inflation solves the horizon problem. Thanks to the accelerated expansion, the conformal time $\tau$ is no longer tightly bounded by the big bang singularity $t=0$. Thus the past light cones on the CMB overlap in the past, which solves the horizon problem.}
\end{figure}

\begin{figure}[htbp]
  \centering
  \includegraphics[width=0.95\textwidth]{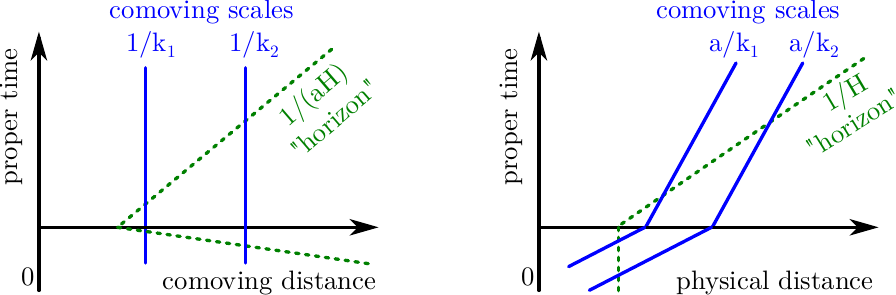}
  \caption{\label{fig:horizon-exit-return} The comoving scales exits the horizon during inflation, and later returns during radiation and matter dominated eras.}
\end{figure}

\item \textit{Monopole problem:} If monopoles were created before inflation, they could easily be inflated away. Thus it is no wonder we can not observe them today. (However, one should still pay attention not to produce gauge theory monopoles, domain walls or very strong cosmic strings after inflation.)

\item \textit{Correlation problem:} How inflation solves this problem is discussed in detail in later sections. The key idea is that, the sub-Hubble quantum fluctuations are stretched by cosmic expansions during inflation, into super-Hubble fluctuations. Those fluctuations return to the horizon later after the end of inflation. Those returned modes cause large scale correlations on the CMB. This is illustrated in Fig. \ref{fig:horizon-inflation}.

\end{itemize}

\subsection{Dynamics of inflation}

Inflation is clearly a very powerful tool for solving problems with standard cosmology, so let us now proceed to see how inflation can be explicitly realized. The requirement $p/\rho=w<-1/3$ is not that easy to obtain. Because our familiar components in the universe have positive pressure. For example, consider a gas of particles in a box. The particles always ``want'' to escape the box, and hence generate positive pressure by hitting the surface of the box.

The same difficulty can be also observed from the continuity equation $\dot\rho + 3H (1 + w) \rho = 0$. Again consider a gas component (but now in the whole universe). Both particle density and momentum of particles get diluted when the universe expands. Thus $\rho$ should dilute at least as quickly as $a^{-3}$. Especially, when $w\simeq -1$, we want a component whose energy density is almost not diluted when the universe expands.

To have such an acceleration, one immediate idea is to use vacuum energy. This is because, the vacuum, after adiabatic expansion, is still vacuum\footnote{Adiabaticity cannot be maintained for the vacuum wave modes as long as $1/H$. We shall come back to this issue later in details.}. Thus we have $w=-1$ and the universe accelerates.\footnote{Indeed our present universe is accelerating, and the vacuum energy is the most popular candidate for our present expansion. For the discussion of late time acceleration, see e.g. \cite{Li:2011sd} and the references therein. }

However, if we use the vacuum energy to accelerate our early universe, problems arise:

\begin{itemize}
\item \textit{Inflation must end.} Our universe is not always in a state of acceleration. Observationally, significant parts of our directly observable history was radiation and matter dominated and the universe decelerates instead of accelerates in expansion. Theoretically, we need the comoving structures to return to the horizon.
\item \textit{The end of inflation must be controlled by a rather accurate clock}. The overlapping of light cones in Fig.~\ref{fig:horizon-inflation} does not guarantee (though makes it possible) that we observe an almost homogeneous and isotropic universe. For example, if one significant part of the universe has much earlier end of inflation, then energy density drops quickly in that part and it will produce a void in our universe, breaking homogeneity and isotropy.
\end{itemize}

The vacuum energy itself does not provide an exit mechanism for inflation. Moreover, even if we add anything, like tunneling to another vacuum, as the end of inflation, it will not solve the problem as long as there is no clock. For example, the tunneling to another vacuum happened spontaneously instead of being controlled by a clock.\footnote{The decay of false vacuum is actually the oldest model of inflation. However, apart from the clock problem, the decay of false vacuum has another more serious problem: Let $\Gamma$ be the decay rate of the false vacuum. Then when $\Gamma>H$, inflation will end within one e-fold, thus do not really solve problems. On the other hand, if $\Gamma<H$, true vacuum bubbles can not percolate because the space between true vacuum bubbles expand too quickly. In this case we found ourselves in an open bubble \cite{Coleman:1977py, Coleman:1980aw}, and the flatness problem returns. Thus inflation will either end too quickly or will never end globally. }

Thus, we have to attach a clock to vacuum energy, and use this clock to end inflation. A clock\footnote{To be more accurate, by ``a clock'', we actually mean a set of clocks, accessible at every spacetime point. Thus the clock is actually a clock field.} is something that takes a number at every physical point in the spacetime manifold, with the numbers increasing along an observer's world line. Thus a clock can be simply realized by a rolling scalar field\footnote{Such a concept of global time appears in conflict with the principles of relativity. Because relativity asserts there is no preferred observer or absolute time. Indeed, Lorentz symmetry is spontaneously broken in cosmology, by the expansion of the universe. Nevertheless, spontaneous breaking of symmetry just means the state breaks the symmetry, which do not reduce the power or (arguably) beauty of the theory.}. In other words we would need an effective vacuum energy, controlled by the motion of a scalar field.

\begin{figure}[htbp]
\centering
\includegraphics[width=0.3\textwidth]
{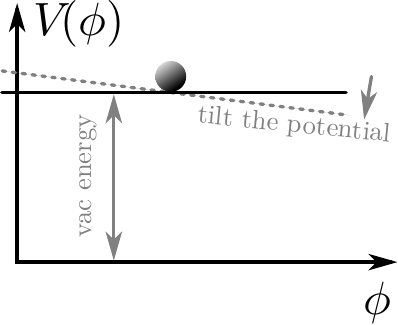}
\hspace{0.2\textwidth}
\includegraphics[width=0.3\textwidth]
{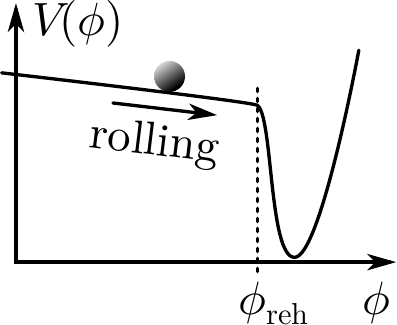}
\caption{\label{fig:inflation-potential} Left panel: a rolling scalar field is a modest deformation of pure vacuum energy. Right panel: the inflationary potential. The shape of the inflationary potential is largely unconstrained, as long as it is flat enough. Inflation ends at where the potential is no longer flat and a process of reheating follows to reheat the universe.}
\end{figure}

The change from a vacuum energy to a rolling scalar field can be thought of as follows: A vacuum energy plus a massless scalar can be viewed in an alternative way as a scalar field lying on a flat potential. By slightly tilting the potential, we have a slowly rolling scalar field acting as an effective vacuum energy. This picture can be visualized by the analog of a rolling ball along a very flat surface, as shown in Fig.~\ref{fig:inflation-potential}. The rolling scalar is called the ``inflaton''.

Further, inflation could be ended by adding some additional mechanisms triggered by some particular value of inflaton. For example, as in Fig.~\ref{fig:inflation-potential}, after $\phi_\mathrm{reh}$, the potential $V(\phi)$ is no longer flat thus inflation ends when $\phi$ rolls to $\phi_\mathrm{reh}$. In this case, the inflaton field is the clock field of the universe.

To be explicit, the Lagrangian for a scalar field is
\begin{align}\label{eq:lag-phi}
  \mathcal{L}_\phi =  \sqrt{-g} \left[ - \frac{1}{2}  g^{\mu\nu} \partial_\mu\phi \partial_\nu\phi - V(\phi) \right]~,
\end{align}
In the homogeneous and isotropic background level, the above equation is reduced to
\begin{align}\label{eq:lag-phi-bg}
  \mathcal{L}_\phi = a^3 \left[ \frac{1}{2} \dot\phi^2 - V(\phi) \right] ~.
\end{align}
The equation of motion for $\phi$ can be obtained from varying this action with respect to $\phi$:
\begin{align}\label{eq:eom-phi-bg}
  \ddot\phi + 3H\dot\phi + V'(\phi) = 0 ~,
\end{align}
where prime denote derivative with respect to $\phi$. 

It is interesting to compare this equation with the flat space equation of motion for a homogeneous and isotropic scalar field: $ \ddot\phi + V'(\phi) = 0$. The new term in Eq.~(\ref{eq:eom-phi-bg}) is a friction term. To see this, consider $V'(\phi)=0$. In this case the solution of $\phi$ is $\phi=$const and $\dot\phi\propto a^{-3}$. The $\phi=$const solution means a ``ball'' which is at rest. The $\dot\phi\propto a^{-3}$ solution means that when there is no additional driving force, the initial velocity decreases. 

To solve Eq.~(\ref{eq:eom-phi-bg}), we need yet to know $H$. Thus we have to couple the $\phi$ field to gravity. There are a few (essentially the same) ways to couple $\phi$ to gravity. One could either vary the gravitational Lagrangian together with $\mathcal{L}_\phi$ with respect to $g_{\mu\nu}$, or vary $\mathcal{L}_\phi$ in Eq.~(\ref{eq:lag-phi}) with respect to $g_{\mu\nu}$ to get the stress tensor\footnote{Using this method, the stress tensor is $T_{\mu\nu}=(\rho+p)u_\mu u_\nu + p g_{\mu\nu}$, where $u_\mu \equiv \partial_\mu\phi / \sqrt{-\partial_\nu\phi\partial^\nu\phi}$, and $\rho$ and $p$ are given in \eqref{eq:ham-phi-bg} and \eqref{eq:pressure-bg}.} and insert this stress tensor into the Einstein equations. Instead, here we take a simpler route:

The physical energy density can be obtained by a Legendre transform:
\begin{align}\label{eq:ham-phi-bg}
  \rho = \mathcal{H}_\phi / a^3 = \frac{1}{2} \dot\phi^2 + V(\phi) ~.
\end{align}
Compare the continuity equation $\dot\rho + 3H(\rho+p)=0$ with the equation of motion (\ref{eq:eom-phi-bg}), one concludes
\begin{align}\label{eq:pressure-bg}
  p = \mathcal{L}_\phi / a^3 = \frac{1}{2} \dot\phi^2 - V(\phi) ~.
\end{align}
Thus we have inflation when potential energy dominates.

Inserting $\rho$ into the Friedmann equation, we get
\begin{align}\label{eq:eom-H-bg}
  3M_p^2H^2 = \rho = \frac{1}{2} \dot\phi^2 + V(\phi) ~.
\end{align}

Now we are in the position to solve this system (with Eqs. (\ref{eq:eom-phi-bg}) and (\ref{eq:eom-H-bg})). It is in general difficult to find solutions for a general $V$. For some special potentials, the system can be solved. For example, one can make an ansatz $w=p/\rho=$const, and this case is called ``power law inflation'' \cite{Lucchin:1984yf}, resulting in an exponential potential of $\phi$.

However, here we expect more general solutions. Thanks to the friction force, there is a slow-roll-approximated solution of Eq.~(\ref{eq:eom-phi-bg}) when the potential is flat enough, regardless the details of the potential. This is the solution that the friction force balances the external force $V'(\phi)$, such that the acceleration of $\phi$ does not change significantly. In addition, to have inflation, we want the potential energy $V(\phi)$ dominates over the kinetic energy $\frac{1}{2} \dot\phi^2$. Thus the approximation is
\begin{align}\label{eq:sr-raw}
  \frac{1}{2} \dot\phi^2 \ll  V(\phi) ~, \qquad |\ddot\phi|\ll3H|\dot\phi|~.
\end{align}
The first condition guarantees inflation because in this case $p\simeq -\rho$. The second condition requires $\dot\phi$ to change slowly such that the first condition can be satisfied for long enough time.

With the above approximations (\ref{eq:sr-raw}), given the value of $\phi$, we can first solve $H$ by
\begin{align}
  H \simeq \sqrt{\frac{V}{3M_p^2}}~,
\end{align}
and then insert the resulting $H$ into $3H\dot\phi + V' \simeq 0$ to obtain 
\begin{align}
  \dot\phi \simeq -\frac{V'}{3H}~.
\end{align}
After obtaining those solutions, we need to step back to verify the assumptions in Eq.~(\ref{eq:sr-raw}). Inserting the solutions of $H$ and $\dot\phi$, we have
\begin{align}
  \epsilon_V = \frac{M_p^2}{2} (\frac{V'}{V})^2 \ll  1 ~, \qquad
  \eta_V = M_p^2 \frac{V''}{V}~, \qquad |\eta_V| \ll  1~.
\end{align}
The subscript $V$ means that those small ``slow roll parameters'' are defined using the inflation potential. This is convenient because by model construction, we typically start with a potential and want to know immediately if the potential is flat enough to have inflation.\footnote{Note the appearance of Planck mass $M_p$ in those \textit{small} parameters. It turns out quite hard to get those slow roll conditions satisfied.}

Especially, when inflation is driven by the mass of the inflaton, we have
\begin{align}
  \eta_V = \frac{m^2}{3H^2} ~.
\end{align}
Thus $\eta_V$ actually denotes the mass square of the inflaton, measured in Hubble units. The condition $|\eta_V| \ll  1$ means, the dynamics of the \textit{inflationary background}, has an energy scale much smaller than Hubble parameter. As we shall see later, the inflationary perturbations actually lives on the energy scale $H$, with initial condition set on an even higher energy scale.

It is also convenient to define another set of slow roll parameters using the dynamics of scale factor\footnote{The convenience will show up practically in calculations like expanding the action with respect to perturbations}:
\begin{align}
  \epsilon \equiv - \frac{\dot H}{H^2} ~, \qquad \eta \equiv \frac{\dot\epsilon}{H\epsilon}~.
\end{align}
When slow roll conditions $\epsilon_V\ll 1$ and $|\eta_V|\ll 1$ are satisfied, those slow roll parameters are related to $\epsilon_V$ and $\eta_V$ by
\begin{align}
  \epsilon = \epsilon_V + \mathcal{O}(\epsilon_V^2, \eta_V^2)~, \qquad \eta  = 4\epsilon_V - 2\eta_V + \mathcal{O}(\epsilon_V^2, \eta_V^2)~.
\end{align}
When slow roll conditions are not satisfied, the relations between ($\epsilon$, $\eta$) and ($\epsilon_V$, $\eta_V$) are non-linear.

\subsection{Attractor}

As we have shown, in the slow roll approximation, given a flat enough potential, there exists a slowly rolling inflationary solution. To take this solution seriously, it is important to consider the naturalness: Could this solution be reached easily without highly fine-tuned initial conditions?

This is in general a hard problem. Because on the one hand, the dynamics before inflation is yet unknown. Thus the a priori naturalness is not well defined. On the other hand, even consider the mathematical naturalness, one need to consider a measure of the phase space. Unfortunately, the phase space measure for inflation diverges and different regularization method leads to different conclusions \cite{Gibbons:1986xk, Gibbons:2006pa, Li:2007rp}. 

Nevertheless the solution of slow roll inflation is an attractor, in the sense that a general point in the phase space of inflation typically flows onto a slow roll trajectory. An attractor solution is natural, at least in a naive way.

There is an intuitive way to see this attractor behavior: the potential energy of inflation acts as an effective vacuum energy, and thus do not significantly redshift during inflation. However, the kinetic energy of inflaton tends to be redshifted away quickly. As discussed in the last section, without continuous injection of energy from the potential, the kinetic energy of $\phi$ decays as $a^{-6}$.

Thus, on the one hand, if the universe starts potential dominated, we are fine to have inflation already. On the other hand, if the universe starts kinetic energy dominated, with a small coherent potential energy\footnote{Actually a small coherent potential energy may be a big claim because the horizon problem is back.}, the kinetic energy will get diluted away and the universe will eventually become potential dominated. Thus inflation is a dynamical attractor solution.

\subsection{Reheating}

As we already emphasized, it is important to end inflation. From local energy conservation, the effective vacuum energy should go somewhere. The simplest case is that this effective vacuum energy decays into matter or radiation, which heats the universe. This process is called reheating\footnote{Reheating, as we can read, seems to mean the universe heat up for a second time. However, we don't know what happens before inflation, thus actually don't know if the universe was hot before inflation.}.

After reheating, the universe is radiation dominated\footnote{It is still possible that the inflaton decays into matter, and the matter later decays into lighter particles which behaves as radiation.}, thus
\begin{align}
  M_p^2 H^2 \sim \rho_r \sim T_\mathrm{reh}^4~, \qquad T_\mathrm{reh} \propto \sqrt{M_p H}~,
\end{align}
where $H$ is the energy scale at reheating. We require the reheating temperature to be compatible with BBN. Thus $T_\mathrm{reh} > 10$MeV. This is considered as a hard lower bound for inflationary energy scale $H > 10^{-22}$GeV. Nevertheless, we expect inflation to happen at much higher energy scales, because it is increasingly harder to build a flat potential at extremely low energy scale.

If reheating is sufficient, most of the vacuum energy goes into radiation. In this case $H$ is close to the energy scale of inflation.

Unlike the dynamics during inflation, the dynamics of reheating is extremely model dependent. Here we shall not come into detailed physics (see \cite{Bassett:2005xm, Allahverdi:2010xz} for a review). Instead we only outline the generic class of models:

\begin{itemize}
\item \textit{Perturbative decay:} In this case, the decay is characterized by a decay rate $\Gamma$ in particle physics \cite{Abbott:1982hn, Dolgov:1982th, Albrecht:1982mp}. A typical scenario of this kind has the inflaton oscillate around its minima of potential after inflation. And when the duration of oscillation is longer than $1/\Gamma$, the inflaton energy decays into other particles. Note that the duration of oscillation is approximately $t\propto 1/H$ at the decay time. Thus the inflaton decay happens on a uniform Hubble (and thus uniform total energy density) hypersurface in spacetime. 

In this case the reheating temperature can be written as $T_\mathrm{reh} \propto \sqrt{M_p \Gamma}$. In case of a long time of oscillation, the Hubble parameter at the inflaton decay can be significantly lower than the inflationary Hubble parameter.

\item \textit{Non-perturbative decay:} This case is model dependent. For example, brane inflation typically end by collision of brane and anti-brane \cite{Sen:2002in, Shiu:2002xp, Cline:2002it}. The decay is very fast due to the tachyon mode between the branes. The decay happens on a uniform inflaton hypersurface in spacetime. This situation is illustrated in Fig.~\ref{fig:brane-inf}.

\begin{figure}[htbp]
  \centering
  \includegraphics[width=0.6\textwidth]{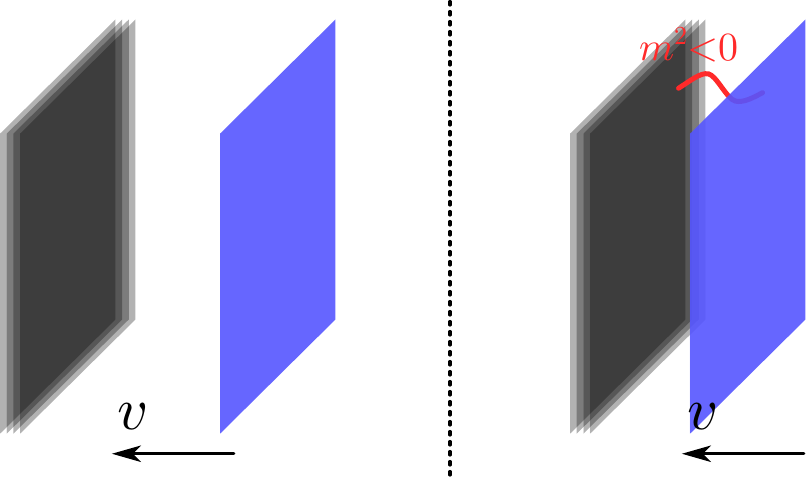}
  \caption{\label{fig:brane-inf} Brane inflation and reheating. When the brane moves close to the stack of anti-branes, the open string mode with end point on both brane and anti-brane become tachyonic, which triggers the decay of the brane. Brane inflation ends on a uniform inflaton hypersurface.}
\end{figure}

\item \textit{Parametric resonance (preheating):} When inflation ends and the inflaton is oscillating around its minima, the oscillation may trigger a resonance for other particles (say, $\chi$) \cite{Traschen:1990sw, Dolgov:1989us, Shtanov:1994ce, Kofman:1994rk}. In the resonance band, the energy density of $\chi$ grows exponentially until $\chi$ has comparable energy density with the inflaton, after which the system becomes non-linear while the decay continues.

\end{itemize}

Note that during reheating, the small scale physics changes. However those scales are so small that the same comoving scale today is not cosmological. Moreover, those sub-Hubble scales shall not go super-Hubble again thus there is no associated conserved quantity. Thus it is harder to probe the physics of reheating than to probe the physics during inflation.

Nevertheless, there do exist a few possibilities that can constrain (and in the future may observe) scenarios of reheating:

\begin{itemize}
\item \textit{Modification of large scale fluctuations:} As we shall show later in Sec.~\ref{sec:non-triv-rehe} and \ref{sec:modulated-reheating}, in some special models of reheating, not only the small scale perturbations but also the large scale ones can get modified. 
\item \textit{Gravitational waves:} The gravitational waves propagate freely after they are produced. Thus the primordial ones carry information of extremely early ages \cite{Khlebnikov:1997di}. Unfortunately, high-scale reheating typically produces present-day gravitational-wave frequencies above the sensitive band of current detectors. Nevertheless, a low scale reheating or a table top detector may uncover the physics of reheating.
\end{itemize}

\subsection{Open questions}
\label{sec:open-questions}

We have discussed the dynamics of inflation, and its success. Before going on to discuss the further success of inflation at the level of perturbations, let us pause and mention a few open questions for inflation.\footnote{Those open questions, from a critical point of view, are problems of inflation.}

\begin{itemize}
\item \textit{The $\eta_V$-problem} \cite{Copeland:1994vg}: As we discussed, for inflation to happen, we need $\epsilon_V\ll 1$. On the other hand, for inflation to happen for enough e-folds to solve any problem, we need $\eta_V\ll 1$. For example, consider a free massive field as the inflaton. In this case, $\eta_V\ll 1$ implies $m\ll H$.

Note that $H$ is a remarkably low energy scale: The inflationary $H$ is high in our today's high energy physics point of view. However, during inflation $1/H$ is the size of the whole ``observable'' universe at that time. Thus it is nontrivial that we need an inflaton, whose Compton wave length is much greater than the size of observable universe, during the era of inflation.

Even worse, the mass of a scalar field is not protected by quantum fluctuations. Like in the case of the Higgs mass hierarchy problem, the quantum correction to the inflaton mass runs to the UV completion scale $\Lambda$ of the theory. Even if the inflaton field only has gravitational self coupling (or coupling to other fields), the mass correction, by power counting \footnote{The gravitational coupling contributes $M_p^{-2}$. Thus there should be the $\Lambda^4$ factor to match the overall mass dimension.}, is
\begin{align}
  \Delta m^2 \sim \frac{\Lambda^4}{M_p^2} ~.
\end{align}
When $\Lambda\sim M_p$, the mass correction runs directly to Planck scale. 

As in the case of Higgs physics, we may expect supersymmetry to save the theory from severe fine tuning. However, the supersymmetry, if exists, has to be broken at the Hubble scale. Thus the cancellation leaves a mass correction $\Delta m\sim H$.

Also, a locally supersymmetric theory typically get mass corrections of order $\Delta m\sim H$. For example, consider a Kahler potential $K\simeq c \phi^2/M_p^2$, where $c$ is a number of order one. The Kahler potential appears in the Lagrangian as
\begin{align}
  e^{K} V(\phi) \simeq V(\phi) + \frac{c\phi^2}{M_p^2} V(\phi) \simeq V(\phi) + 3cH^2\phi^2 ~.
\end{align}
Thus an effective mass $\Delta m^2 = 6cH^2$ is generated.

There are also other sources to generate $m\sim H$ corrections. For example, no symmetry forbids the coupling $\lambda \phi^2R$ in the action, where $R$ is the Ricci scalar. Note that during inflation, $R = 12H^2 + 6 \dot H$. Thus coupling to Ricci scalar generates a mass correction $\Delta m^2 = 24\lambda H^2$ during inflation. 

As another example, if $\phi$ couples to another field $\sigma$ with $g\phi^2\sigma^2$ coupling, the $\sigma$ field gets a vev at the fluctuation level (as we shall show in Sec.~\ref{sec:spectator-field-de}), up to an IR growth factor:
\begin{align}
  \langle\sigma^2\rangle \simeq \frac{H^2}{4\pi^2} \quad\Rightarrow\quad \Delta m^2 = 2g\langle\sigma^2\rangle = \frac{gH^2}{2\pi^2}~.
\end{align}
The above problems are collectively called the $\eta_V$-problem. To avoid this problem, one could introduce a shift symmetry to protect the mass of the inflaton. Nevertheless the shift symmetry has to be broken softly to have inflation \cite{Kawasaki:2000yn, Kawasaki:2000ws}.

\item \textit{The degeneracy problem:} This is a problem at the opposite direction of the $\eta_V$-problem. As we discussed, it is hard to get a flat enough potential, because of the $\eta_V$-problem. However, if we allow some fine tuning to flatten the potential, there are immediately a huge number of inflation models. We shall review those models in later sections. 

As we shall discuss later, the number of observables of inflation is limited. Thus it is considered very hard to break the degeneracy, to select the correct model of inflation from observations.

\item \textit{The trans-Planckian problem} \cite{Martin:2000xs, Brandenberger:2000wr, Brandenberger:2012aj}: Comoving scales expand exponentially during inflation. Thus those scales are exponentially small in the past. Especially those scales can be smaller than Planck scale at the initial time.\footnote{There need to be 10 more e-folds of inflation before the observable stage, for the smallest cosmological scale to reach Planck scale in the past.} How can we trust a theory with such small length scales? Especially how to set the trans-Planckian initial condition?

It is argued that the vacuum state is adiabatic on sub-Hubble scales, and thus does not have this scaling problem. Yet it is useful to show explicitly if there is a problem or not.

One should also note that at the level of free theory, the Planck mass $M_p$ has no physical meaning and can be rescaled away. Only when we consider interactions, $M_p$ becomes physical as the energy scale where gravity is strongly coupled. Thus the trans-Planckian problem is really a problem of interaction. And at the interacting level (which is by definition non-linear), there can be trans-Planckian modes interacting with the super-Hubble modes as well. This situation is illustrated in Fig.~\ref{fig:trans-planckian}.

\begin{figure}[htbp]
  \centering
  \includegraphics[width=0.6\textwidth]{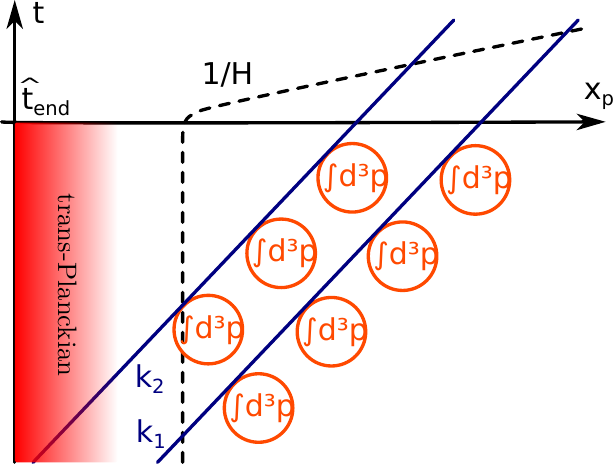}
  \caption{\label{fig:trans-planckian} The trans-Planckian problem, as a problem of non-linear interactions.}
\end{figure}

\item \textit{Initial condition of inflation:} Inflation cannot be the full story of the very early universe. The inflationary FRW metric only covers half of the de Sitter spacetime. As a result, inflation is not geodesic complete (a more general proof can be found in \cite{Borde:2001nh}). We need a start for inflation. This situation is illustrated in Fig.~\ref{fig:deSitter}. It is an open question how to prepare a homogeneous and isotropic Hubble patch, for inflation to happen.

\begin{figure}[htbp]
  \centering
  \includegraphics[width=0.3\textwidth]{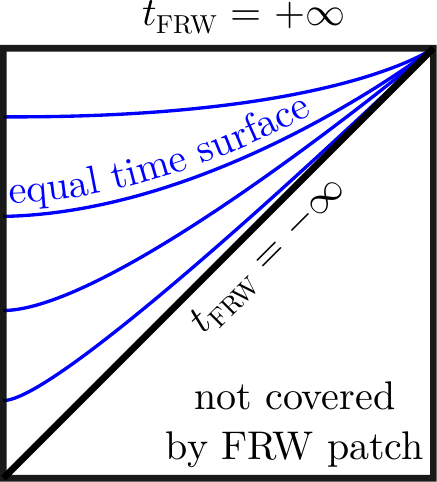}
  \caption{\label{fig:deSitter} The FRW patch is not geodesic complete. It covers only half of the global de Sitter space. }
\end{figure}

There are some candidates for the theories before inflation. For example eternal inflation \cite{Steinhardt:1982kg, Vilenkin:1983xq} (which yet needs an initial condition), or the quantum creation of the universe \cite{Hartle:1983ai, Linde:1983cm, Vilenkin:1984wp}. So far we are still unable to experimentally verify those theories.

\end{itemize}

From now on, we shall move on to the cosmic perturbation theory. This is the field where inflation has already shown its predictive power, and continues to excite the community with forthcoming experiments.

\section{Linear Perturbations}

Now we move on to the perturbations. Perturbations are by definition smaller than background. But for inflation (and many other cases in cosmology) they are by no means less important. Because a perfectly homogeneous and isotropic universe do not tell us a lot (and cannot find anybody to tell).

\subsection{A spectator field in de Sitter space}
\label{sec:spectator-field-de}

To warm up, we first consider a spectator field in de Sitter space ($a=e^{Ht}$) \cite{Parker:1969au}. ``Spectator'' means gravity is not dynamical. Instead, gravity is just an expanding playground and do not receive back-reaction from the spectator field. 

Operationally, one can consider a universe with a cosmological constant $\Lambda$. Then take $M_p\rightarrow\infty$ while keeping $\Lambda/M_p^2$ fixed. Then a finite energy density scalar field will be a spectator field.

For simplicity, we consider the spectator field to be a real scalar, non-interacting and massless. The action of the spectator field is
\begin{align}\label{eq:S-sigma}
  S = \int d^3 x ~ dt ~ a^3  \left[ \frac{1}{2} \dot\sigma^2 - \frac{1}{2a^{2}}(\partial_i\sigma)^2 \right]~.
\end{align}
In Fourier space, the above action is
\begin{align}
  S = \int \frac{d^3 k}{(2\pi)^3} ~ dt ~ a^3  \left[ \frac{1}{2} \dot\sigma_\mathbf{k}\dot\sigma_{-\mathbf{k}} - \frac{k^2}{2a^2} \sigma_\mathbf{k}\sigma_{-\mathbf{k}} \right]~.
\end{align}

The $\sigma$ field can be quantized as
\begin{align}\label{eq:quant-s}
  \sigma_\mathbf{k}(t) = v_k(t) a_\mathbf{k} + v^*_k(t) a^\dagger_\mathbf{-k} ~,
\end{align}
where $a_\mathbf{k}$ and its Hermitian conjugate $a^\dagger_\mathbf{k}$ are time-independent creation and annihilation operators:
\begin{align}\label{eq:quant-a}
  [a_\mathbf{k}, a_\mathbf{k'}] = 0, \qquad [a_\mathbf{k}, a^\dagger_\mathbf{k'}] = (2\pi)^3 \delta^3(\mathbf{k}-\mathbf{k}')~,
\end{align}
and $v_k$ satisfies the classical equation of motion of $\sigma$:
\begin{align}
  \ddot v_k + 3H\dot v_k + \frac{k^2}{a^2} v_k = 0~.
\end{align}
To solve this equation, it is convenient to use conformal time. The same equation of motion now becomes
\begin{align}
  (av_k)'' + \left( k^2 - \frac{a''}{a}\right) (av_k) = 0 ~,
\end{align}
where prime denotes derivative with respect to the conformal time. Note that $a=-1/(H\tau)$ for de Sitter space. Thus $a''/a = 2/\tau^2$. The above equation has simple solution
\begin{align} \label{eq:vk-with-c2}
  v_k = c_1 (1+ik\tau)e^{-ik\tau} + c_2 (1-ik\tau)e^{ik\tau} ~.
\end{align}

Now we have to fix the integration constants $c_1$ and $c_2$. Before a careful treatment, let us first consider a quicker way: Note that $av_k$ is the canonically normalized field. Thus when this $k$-mode is far inside the horizon ($\tau\rightarrow-\infty$), $av_k$ effectively lives in flat space and should have the same solution as in Minkowski space. By comparing with the Minkowski space solution $1/\sqrt{2k}$, we get
\begin{align}
  |c_1| = \frac{H}{\sqrt{2k^3}}~, \qquad c_2 = 0~.
\end{align}
Note that the mode function $v_k$ only coincide with the Minkowski space solution at $\tau\rightarrow-\infty$. When $-k\tau\lesssim 1$, the solution of $v_k$ becomes completely different from the Minkowski one.

Now we shall determine the above integration constants in a more careful way. There are two observations to determine $c_1$ and $c_2$:

\begin{itemize}

\item \textit{Quantization and uncertainty.} The $\sigma$ field is quantized in Eq.~(\ref{eq:quant-a}). Note that in quantum mechanics, quantization and uncertainty principle have close connections. The size of quanta determines how accurate or inaccurate we could measure a system. 

In our situation, the uncertainty principle is
\begin{align}
  \pi_\mathbf{k} \equiv \frac{\delta S}{{\delta\dot\sigma_\mathbf{k}}} = a^3\dot\sigma_{-\mathrm{k}} ~,
  \qquad [\sigma_\mathbf{k}, \pi_{\mathbf{k}'}] = i (2\pi)^3 \delta^3(\mathbf{k}-\mathbf{k}')~.
\end{align}
To compare this relation with (\ref{eq:quant-a}), we get a normalization condition
\begin{align}
  a^2 (v_k v_k^*{}' - v_k' v_k^*) = i ~.
\end{align}
Note that we have rewritten the time derivative in terms of conformal time. Applying the equation of motion for $v_k$, we obtain
\begin{align}\label{eq:vac-cond-from-quanta}
  c_1 c_1^* - c_2 c_2^* = \frac{H^2}{2k^3}~.
\end{align}

\item \textit{Vacuum has lowest energy.} Now we are to fix a ``vacuum'' state. There are two things to do to fix the state:

First, we should relate the vacuum to annihilation operator. A ``vacuum'' is a state satisfying
\begin{align}\label{eq:annihilate-vacuum}
  a_\mathbf{k} |0\rangle = 0~.
\end{align}
However, this condition is actually a precise definition of $a_\mathbf{k}$, among all the possible definitions. More importantly, we still need to know what \textit{is} a vacuum as a physical state. We define the vacuum as the lowest energy state at initial time $\tau\rightarrow-\infty$. To make it more precise, the Hamiltonian (defined with respect to $\tau$) is
\begin{align}
  \mathcal{H} = \frac{1}{2}  \left[ a^4\dot\sigma^2 + a^2 (\partial_i\sigma)^2 \right] 
  = \frac{1}{2}  a^2 (c_1c_1^*+c_2c_2^*) (k^2+2k^4\tau^2)
  + \frac{1}{2} \left[ a^2 c_1c_2^*e^{-2ik\tau}(k^2+2ik^3\tau)+ \mathrm{c.c.} \right]~,
\end{align}
where $\mathrm{c.c.}$ denotes complex conjugate. At $\tau\rightarrow-\infty$, the Hamiltonian is
\begin{align}
  \mathcal{H} \simeq \frac{1}{H^2} (c_1c_1^*+c_2c_2^*) k^4 ~.
\end{align}
We want this Hamiltonian to be minimal with the constraint (\ref{eq:vac-cond-from-quanta}). Thus we get
\begin{align}
  |c_1| = \frac{H}{\sqrt{2k^3}}~, \qquad c_2 = 0~.
\end{align}
The remaining overall phase of the mode function can be absorbed into the definition of the annihilation operator.

To collect the result of quantization, we have
\begin{align}\label{eq:quant-s-2}
  \sigma_\mathbf{k}(\tau) = v_k(\tau) a_\mathbf{k} + v^*_k(\tau) a^\dagger_\mathbf{-k} ~, \qquad v_k(\tau) = \frac{H}{\sqrt{2k^3}} (1+ik\tau)e^{-ik\tau} ~.
\end{align}

So far we have discussed two ways to fix the vacuum state: Instant Minkowski vacuum, as well as the initial lowest energy state. This vacuum state is known as the Bunch-Davies vacuum \cite{Bunch:1978yq}. The same vacuum state can also be obtained in various other ways. For example, the Bunch-Davies vacuum can be obtained by the no boundary wave function of the universe. This is because the Euclidean version of initial time, $i\tau=-\infty$, is a regular point on the sphere, thus shouldn't result in a divergence in the Euclidean action. Alternatively, it is also believed that the Bunch-Davies vacuum is an attractor solution of inflation. Quite general (but not all) initial states are eventually driven to the Bunch-Davies state due to the cosmic expansion.

\end{itemize}

Before ending the discussion of vacuum, some more comments are in order here:
\begin{itemize}

\item \textit{The relation between Eq.~(\ref{eq:annihilate-vacuum}) and minimizing the Hamiltonian:} At first glance, they look like the same condition. This is because, in flat space QFT, $a_k$ annihilate one particle. So Eq.~(\ref{eq:annihilate-vacuum}) means there is no particle, and thus is lowest energy state.\footnote{In flat space QFT, we still need $c_2=0$ to relate $a_k$ to the concept of ``annihilate one particle''.} However, on time dependent background, things become more complicated.

On a time dependent background, matter energy is not globally conserved. A ``empty'' state without particle could evolve into a state with lots of particles in it. Note that we work in the Heisenberg picture. Once we have defined the ``vacuum'' state, the state does not evolve with time. However once we define the vacuum state as the lowest energy state at one time, the same state could be no longer the lowest energy state at another time, due to the time dependence of the matter Hamiltonian. Here, we minimize the energy at $\tau\rightarrow-\infty$, as one essential piece of vacuum selection.

\item \textit{The vacuum is selected in the UV:} Note that $\tau\rightarrow-\infty$ corresponds to the $k$-mode has almost zero wave-length and infinite energy. Thus in this procedure, the vacuum is selected in the UV. The $k/a\rightarrow\infty$ limit can not be reached in QFT because QFT is considered as an effective field theory with a UV completion. The UV completion scale is at most Planck scale. As we have discussed, this is the ``trans-Planckian'' problem of inflation. One might expect the selection of initial state at $k/a\rightarrow\infty$ could be relaxed to a selection of lowest energy state at $k/a\rightarrow\Lambda$, as long as $H\ll\Lambda\ll M_p$. Before $k/a=\Lambda$, the universe can be considered as expanding adiabatically and no particle is produced. 

\item \textit{The vacuum is selected linearly mode by mode:} We first linearized the theory, and consider one $k$-mode without worrying about other modes. In other words, we implicitly set all other $k$-modes to zero. In flat space QFT, we can safely do that because in scattering process we indeed have set other modes to zero, at least locally near the incoming wave packet. However, in a time dependent background, the situation is actually not clear. The perturbations of other wave length, especially longer wave length exist and we cannot just forget about them. The inclusion of those modes will give a contribution at higher order in perturbation theory (fortunately not at the leading order).

\end{itemize}

Now we have determined the vacuum state, using a simple matching together with a more detailed discussion. The next step is to interpret and understand the solution. Suppose we could measure the scalar field $\sigma$ after the end of inflation\footnote{Actually at this time the modes are super-Hubble and not measurable. Nevertheless, we suppose we could measure those super-Hubble modes at the moment. We shall return to this topic when we discuss the conserved quantity.}, what shall we see?

We see random fluctuations and we want to extract statistical information from it. The way to do it is to consider the two point correlation function $\langle\sigma_\mathbf{k}\sigma_{\mathbf{k}'}\rangle$ at $\tau\simeq 0$ (the end of inflation). Inserting the solution of the mode function, we have
\begin{align}
  \langle\sigma_\mathbf{k}\sigma_{\mathbf{k}'}\rangle = (2\pi)^3\delta^3(\mathbf{k}+\mathbf{k}') \frac{2\pi^2}{k^3} P_\sigma(k)~,\qquad P_\sigma(k) = \left(\frac{H}{2\pi}\right)^2 ~,
\end{align}
where $P_\sigma(k)$ is called the power spectrum. $\sqrt{P_\sigma}$ is the fluctuation amplitude of $\sigma$ per Hubble volume per Hubble time. When $P_\sigma(k)$ does not depend on $k$ (as the current case), we call the power spectrum ``scale invariant''.

The scale invariance of a power spectrum means if we zoom the fluctuations into a bigger volume, and compare with the true fluctuations in this bigger volume, statistically we cannot find difference between those fluctuations. The ``zoom'' operation shows up in the above formula as the $k^3$ factor, as a normalization of the volume.

Note that in our present example, the dynamics of the perturbation is linear. Linearity implies that all the odd-point correlation functions (e.g. $\langle\sigma_{\mathbf{k}_1}\sigma_{\mathbf{k}_2}\sigma_{\mathbf{k}_3}\rangle$) vanishes, and all the even-point correlation functions (e.g. $\langle\sigma_{\mathbf{k}_1}\sigma_{\mathbf{k}_2}\sigma_{\mathbf{k}_3}\sigma_{\mathbf{k}_4}\rangle$) can be reduced trivially into their disconnect parts (e.g. $\langle\sigma_{\mathbf{k}_1}\sigma_{\mathbf{k}_2}\rangle\langle\sigma_{\mathbf{k}_3}\sigma_{\mathbf{k}_4}\rangle+$permutations). Thus the power spectrum carries all the statistical information about the perturbations. Considering the unpredictable nature of quantum random phase, $P_\sigma(k)$ is the unique function to connect theory and observations. As we shall show in later sections, in case of a non-linear theory, the above is no longer true and the predictions can become much richer.

The above discussion can be generalized to a massive spectator field. In the massive case, the above discussions can be directly generalized. The equation of motion takes the form
\begin{align}
    \ddot v_k + 3H\dot v_k + \frac{k^2}{a^2} v_k + m^2 v_k = 0~.
\end{align}
The solution, with the same method of vacuum choice, is (up to a phase)
\begin{align}
  v = e^{i\pi\nu/2}\frac{\sqrt{\pi}H}{2}(-\tau)^{3/2} H_{\nu}^{(1)}(-k\tau)~,
\end{align}
where $\nu\equiv\sqrt{9/4-m^2/H^2}$ and $H^{(1)}_\nu(x)\equiv J_\nu(x)+iY_\nu(x)$ is the Hankel function of the first kind. In the $k/a\gg\max(m,H)$ limit the solution returns to the same Minkowski limit as discussed in the massless case. 

Finally, we heuristically apply the above discussion of $\sigma$ to the fluctuation of the inflaton. This ``application'' must be considered with a \textit{big caution} that, considering that inflaton is the field that drives inflation, the inflaton fluctuation couples to gravity dynamically. Thus it is dangerous and non-rigorous to assume the fluctuation of inflaton lives on a non-dynamical gravitational background. We will leave those issues to the next subsections.

Now we postulate that $\sigma$ is the fluctuation of the inflaton field:
\begin{align}
  \phi(\mathbf{x},t) = \phi_0(t) + \delta\phi(\mathbf{x},t)~,
\end{align}
where $\delta\phi(x,t)$ is small and has the same solution as $\sigma$.

Return to the intuitive picture of $\phi$ as a clock. During inflation, on super-Hubble scales different parts of the universe temporarily lost causal contact with each other. It is the clock field within the Hubble patch, instead of the impact from other parts of the universe, that controls the reheat time of the local universe. When $\phi$ has a fluctuation $\delta\phi$, $\delta\phi$ brings an uncertainty to the clock. The time uncertainty, measured in Hubble unit, is $\zeta\equiv -H\delta t = - H\delta\phi/\dot\phi$. This time difference may be conserved, considering different patches of the universe do not talk to each other. If we consider this $\zeta$ as an observable, the power spectrum of $\zeta$ is
\begin{align}
  P_\zeta = \frac{H^2}{\dot\phi^2} P_\phi = \frac{H^4}{4\pi^2\dot\phi^2} = \frac{H^2}{8\pi^2\epsilon M_p^2} ~.
\end{align}
We shall see in the next sections how the above argument could be made rigorous.

\subsection{Decomposition of the metric}

In the last section, we considered a spectator scalar field. In that case we could operationally take $M_p\rightarrow\infty$ keeping $\rho_\sigma/(M_p^2H^2)$ to be small. However, this limit can not be applied to the inflaton. This is because from the Friedmann equation, the inflaton energy density is exactly $\rho_\phi\sim M_p^2H^2$. Thus gravity may not decouple \footnote{One can still dump the major part of $\rho_\phi$ to an effective cosmological constant. It is the time variation of $\rho_\phi$ that invalidates decoupling of gravity. Thus one can expect the gravitational corrections (per e-fold) to the inflaton are slow roll suppressed, compared with the case of a spectator field in de Sitter space. As we shall show it is indeed the case.}. 

To consider the inflationary perturbations, we need to perturb gravity as well. It is not enough to promote the scale factor $a(t)$ in the FRW metric to be spacetime dependent. Instead, we have to consider the full metric, with the full dynamics of gravity. The dynamics of gravity is governed by the Einstein-Hilbert action

\begin{align}
  S_g = \frac{M_p^2}{2} \int d^4x \sqrt{-g} ~ R ~.
\end{align}

To calculate it explicitly in terms of gravitational perturbations, we have to write the metric into an explicit form. Instead of writing $g_{\mu\nu}$ in terms of $g_{00}, g_{01}, g_{02}, \cdots$, it is convenient to write the metric in the ADM form \cite{Arnowitt:1962hi}, without loss of generality: 
\begin{align}
  ds^2 = -N^2dt^2 + h_{ij} (N^idt+dx^i)(N^jdt+dx^j)~.
\end{align}
The convenience that the ADM form of the metric brings include:

\begin{figure}[htbp]
\centering
\includegraphics[width=0.8\textwidth]
{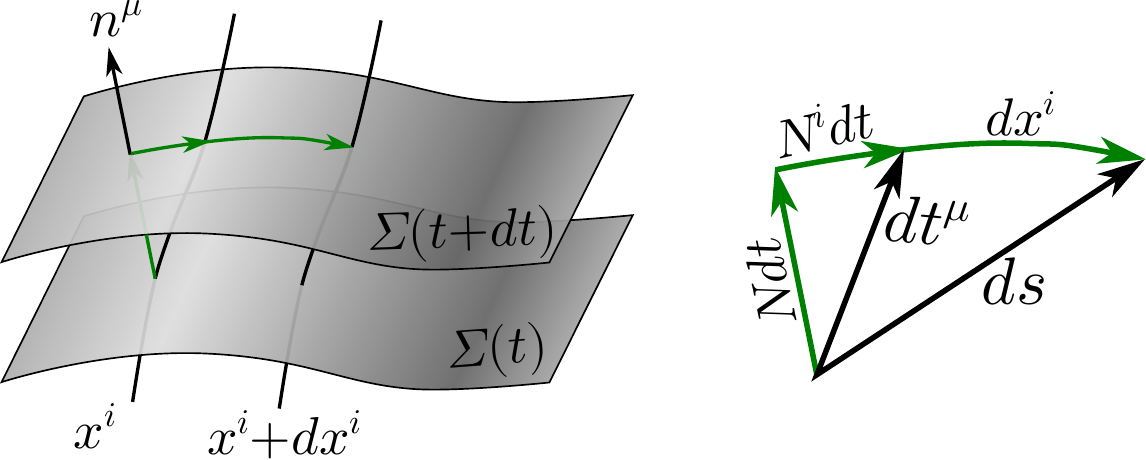}
\caption{\label{fig:adm-metric} The ADM decomposition of the metric. The green arrows are the components of the metric, namely, lapse of time $Ndt$, a shift of hypersurface due to the time flow $N^idt$ and the spatial distance $dx^i$. The distance $N^idt+dx^i$ is first measured by 3-metric $h_{ij}$, and then together with $Ndt$, makes $ds$. If $dx^i=0$, the time part themselves sums up to $dt^\mu$, along the flow direction of time.}
\end{figure}

\begin{itemize}
\item \textit{Clear physical meaning.} The meaning of each component is demonstrated in Fig.~\ref{fig:adm-metric}. 

\item \textit{Convenience of calculation.} Various quantities can be calculated easily using the ADM decomposition. For example, the determinant of the metric is 
  \begin{align}
    \sqrt{-g} = N \sqrt{h}~.
  \end{align}
  And the inverse metric can be written as
  \begin{align}
    g_{\mu\nu}=
    \left( \begin{array}{cc}
        -N^2+N^kN_k & N_i\\
        N_j & h_{ij}
      \end{array} \right)
    \qquad \rightarrow \qquad
    g^{\mu\nu}=
    \left( \begin{array}{cc}
        -1/N^2 & N^i/N^2\\
        N^j/N^2 & h^{ij}-N^iN^j/N^2
      \end{array} \right)~,
  \end{align}  
  where $N_i\equiv h_{ij}N^j$, and $h^{ij}$ is the inverse matrix of $h_{ij}$ (note that in general $h^{ij}\neq g^{ij}$).
\item \textit{Decomposition of the Ricci scalar.} The 4-dimensional Ricci scalar is much more difficult to calculate than the 3-dimensional one. The ADM decomposition provides a connection between the 3- and 4-dimensional Ricci scalars. The derivation can be found in e.g. \cite{Wald:1984rg}. Here we outline the results:

One can construct normal of the hypersurface as
  \begin{align}
    n_\mu =(-N,0,0,0)=-\delta_\mu^0/\sqrt{-g^{00}} ~, \qquad
    n^\mu=(1/N,-N^i/N)~.
  \end{align}
The extrinsic curvature of the hypersurface can be written as
  \begin{align}
    K_{\mu\nu} = (\delta^\sigma_\mu+n^\sigma n_\mu)\nabla_\sigma n_\nu
    = \frac{1}{2} \mathcal{L}_n (g_{\mu\nu}+n_\mu n_\nu)
    ~.
  \end{align}
This $K_{\mu\nu}$ can be calculated explicitly as
\begin{align}
  K^{0\mu} = 0~, \qquad K_{ij} = \frac{1}{2N} \left[ \dot h_{ij} - \mathcal{D}_iN_j - \mathcal{D}_jN_i\right] ~,
\end{align}
where $\mathcal{D}_i$ is the 3-dimensional covariant derivative, defined from $h_{ij}$.

The 3- and 4-dimensional Ricci scalars are related by
  \begin{align} \label{eq:ADM-R}
    R = {}^{(3)}R - K^2 + K^\mu_\nu K^\nu_\mu 
        - 2\nabla_\mu\left(-Kn^\mu+n^\nu\nabla_\nu n^\mu\right)~.
  \end{align}
Note that there is a total divergence term in the RHS. For Einstein gravity, this term is a boundary term. In the case there is no spatial or temporal boundary, this term can be dropped\footnote{Actually, even there is boundary, those terms should also be canceled by explicitly adding boundary terms \cite{York:1972sj, Gibbons:1976ue}. Otherwise the variation principle of the action is not well defined.}.

\item \textit{Constraints easily solvable.} Note that in the above setup, the Einstein-Hilbert action has no derivative acting on $N$, and no time derivative acting on $N^i$. Thus $N$ and $N^i$ are constraints and easily solvable. The disappearance of derivatives on $N$ and $N^i$ are not accidental. The reason is that gravity can be viewed as a gauge theory and there are constraints to be solved. This can be also understood from Fig.~\ref{fig:adm-metric}: $N$ and $N^i$ only tells how to define (instead of dynamically evolve) the spatial hypersurface at the next instant of time. Thus they should be non-dynamical.
\end{itemize}

One can still decompose the ADM metric further. Note that the 3-dimensional rotational symmetry is explicitly preserved during inflation. Thus it is convenient to decompose the metric components according to the irreducible representations of the rotational group. The relevant representations are 3d scalar, 3d vector and 3d tensor representations. Thus it is convenient to decompose the metric into those components. At linear level (i.e. calculation of two point function), those components do not mix with each other, because different representations are orthogonal to each other.

For this purpose, we decompose\footnote{At higher order of perturbations, the $\gamma_{ij}$ part defined here may be not very convenient. See Maldacena's definition.}
\begin{align}
  N_i = \partial_i \beta + b_i ~, \qquad
  h_{ij} = e^{2\zeta} a^2 \left[ \delta_{ij} + \gamma_{ij} + 
      \mathcal{D}_i\mathcal{D}_j 
    E + \mathcal{D}_iF_j + \mathcal{D}_jF_i \right] ~.
\end{align}
Here the vector sector contains $b_i$ and $F_i$, satisfying
\begin{align}
  \mathcal{D}_ib^i = 0~, \qquad \mathcal{D}_iF^i = 0 ~.
\end{align}
The tensor sector contains $\gamma_{ij}$, satisfying
\begin{align}
  \mathcal{D}_i\gamma^{ij} = 0~, \qquad \gamma^i_i = 0~.
\end{align}
The scalar sector contains $N$, $\beta$, $\zeta$ and $E$.

\begin{figure}[htbp]
  \centering
  \includegraphics[width=0.7\textwidth]{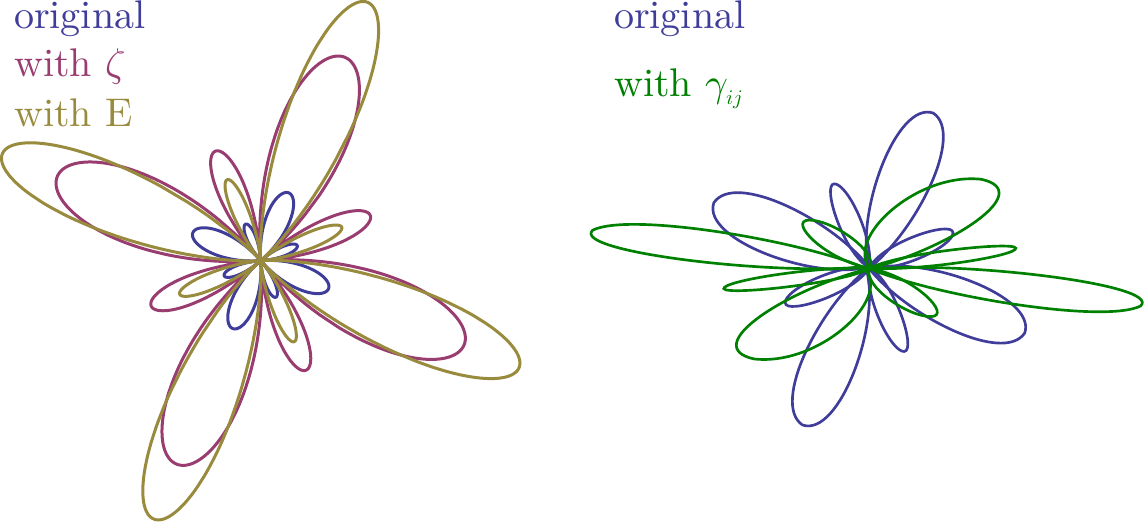}
  \caption{\label{fig:zetaF} The meaning of $\zeta$, $E$ and $h_{ij}$. $\zeta$ is an overall rescaling. $E$ rescale the points with respect to its original radius to the origin. $h_{ij}$ rescale the points with respect to its own direction.}
\end{figure}

From the orthogonal theorem in the theory of group representations, modes in different representations do not propagate into each other. For example\footnote{However, note that if we want to calculate one loop contribution of $\langle\zeta(x)\zeta(y)\rangle$, we do need to consider the loop contribution from $\gamma_{ij}$. Also, more-than-two-point functions can be mixed. For example, $\langle\zeta(x)\zeta(y)\gamma_{ij}(z)\rangle\neq0$.}, $\langle\zeta(x)\gamma_{ij}(y)\rangle=0$.

The above scalar, vector, tensor perturbations need to be considered together with matter perturbations, for example, $\delta\phi$ of the inflaton. Here we shall consider the scalar sector of the perturbations in detail. The tensor sector looks similar to the scalar sector and we shall only discuss it briefly. The vector sector decays in most inflation scenarios. Thus we shall neglect the vector sector here.

\subsection{Gauge invariance and gauge fix}

Einstein's theory of gravity is invariant under general coordinate transformation. The action (and thus physics) do not change under a change of coordinate
\begin{align}
  \widetilde x^\mu = x^\mu + \xi^\mu(x)~.
\end{align}
The metric itself transform as a tensor under this coordinate transformation. Thus components in the metric, $N$, $\beta$, $\zeta$, $\cdots$, indeed change under this transformation. Those transformations can be written collectively as
\begin{align}
  \delta_\xi g_{\alpha\beta}(x) \equiv \widetilde g_{\alpha\beta}(x) - g_{\alpha\beta}(x)
  = \widetilde g_{\alpha\beta}(x) - (\delta^\mu_\alpha+\partial_\alpha\xi^\mu)(\delta^\nu_\beta+\partial_\beta\xi^\nu) \widetilde g_{\mu\nu}(x+\xi)~.
\end{align}
Scalar fields also change under the gauge transformation as
\begin{align}
  \delta_\xi\phi(x) \equiv \widetilde \phi(x) - \phi(x) = \widetilde \phi(x) - \widetilde \phi(x+\xi)~.
\end{align}

Those transformations can be grouped into two parts: the scalar part $\xi^0$ and $\xi$, and the vector part $\hat \xi^i$, where
\begin{align}
  \xi^i = \hat \xi^i + \partial^i\xi~, \qquad \partial_i\hat \xi^i = 0~.
\end{align}
Those two parts transform the scalar sector and vector sector respectively. The tensor sector is gauge invariant by itself.

For example, under a shift of time $\xi^0=\delta t(x)$, and at linear order, $\zeta$ and $\delta\phi$ changes as
\begin{align}
  \widetilde \zeta = \zeta - H \delta t~, \qquad \widetilde {\delta\phi} = \delta\phi - \dot\phi_0 \delta t~.
\end{align}

Consider a featureless spacetime with all those perturbations $N-1$, $\beta$, $\zeta$, $\cdots$ being zero. If we perform a gauge transformation on this spacetime, we find those perturbation quantities no longer zero. Thus there appears to be perturbations. Those apparent perturbations are called gauge artifacts, which need to be eliminated from the calculation. For example, a gauge artifact $\delta t$ brings pure background spacetime into
\begin{align}
  \widetilde \zeta = -H \delta t~, \qquad \widetilde {\delta\phi} = -\dot\phi_0 \delta t~.
\end{align}

There are two ways to eliminate those gauge artifacts. One way is to use gauge-invariant formalism. In this formalism, the gauge artifacts cancels within the gauge-invariant quantities. For example, at linear order, one can choose the gauge invariant quantity as the comoving curvature perturbation
\begin{align}
  \zeta^\mathrm{gi} \equiv \zeta - H \frac{\delta\phi}{\dot\phi} ~.
\end{align}

However, one need to note that free of gauge artifact do not necessarily mean the effect is observable. We still need to identify theoretical observables with what one actually observe in experiments. The situation is particularly dangerous for non-linear perturbations. This is because at the non-linear level, the short wavelength fluctuations, which are accessible by local observers, can be mixed with long wavelength perturbations, which are not accessible by local observers (but still gauge invariant) \cite{Unruh:1998ic, Geshnizjani:2002wp}. 

A related issue with gauge invariant variable is that, at next to leading order, the gauge invariant quantities are non-local and non-unique \cite{Prokopec:2012ug}.

Here we follow a simpler way: to fix the gauge\footnote{The observables, for example the number of photons found in the detectors, are gauge invariant. However, most quantities we talk about, which we have intuition for, are gauge dependent. Those gauge dependent but intuitive quantities include field fluctuations $\delta\phi$, and the fluctuations in the metric. Thus it is still convenient to consider an observer and consider the intuitive picture of the world understood by this observer.}. Once the gauge is fixed, there are no longer available gauge transformations thus pure artifacts cannot appear.

It is convenient to choose a gauge with $E=0$. 
We have yet another gauge condition to choose in the scalar sector. This additional gauge condition fixes spatial slicing. There are a few frequent choices. They give the same result of observables. On the other hand each gauge has pros and cons:

\begin{itemize}
\item \textit{The $\delta\phi$-gauge (spatial flat gauge):} We can choose to let the spatial part of the (scalar sector) metric to be flat: $\zeta=0$. This gauge is convenient to discuss sub-Hubble scale physics. Because on small scales (sub-Hubble, but not as small as Planck scale), gravity is weak. Thus it is intuitive to think about gravity as decoupled. Also, the calculation of the action is simplest in this gauge.

\item \textit{The $\zeta$-gauge (uniform inflaton gauge):} We can choose to let the inflaton field have no fluctuation\footnote{Note that all scalar fields behave similarly under gauge transformations. Thus in the case of multi-field inflation, although we have two gauge conditions in the scalar sector, we could only set one scalar field to zero (even with $E\neq 0$).}: $\delta\phi=0$. This gauge is a comoving gauge, in that the observer moves together with the cosmic fluid and measures no matter flux. 

Note that the three dimensional curvature is
\begin{align}
  {}^{(3)}R = -\frac{4e^{-2\zeta}\partial_i\partial_i\zeta}{a^2}   
  -\frac{2e^{-2\zeta}(\partial_i\zeta)^2}{a^2}~.
\end{align}
Thus $\zeta$ is the perturbation of ${}^{(3)}R$. For this reason, the $\zeta$ variable in the $\zeta$-gauge is known as the comoving curvature perturbation.

This $\zeta$-gauge is convenient to discuss the super-Hubble scale physics. Because on the one hand, gravity accumulates. Thus on large scales it is more intuitive to think about the perturbations as $\zeta$ instead of $\delta\phi$. On the other hand, as we shall show later, $\zeta$ is conserved in the $\zeta$-gauge for single field inflation. This provide a connection between the inflationary physics and current observations.

However, on small scales, $\zeta$-gauge is not intuitive. Let's consider an analogue situation: A table has different height from the ground (if we choose Euclid coordinates). And in our room, gravity is not too strong. However, from general covariance, we can choose a coordinate system where the table and the ground has the same coordinate height. However, in this coordinate system gravity is extremely strongly coupled to cancel the same-height artifact. In the $\zeta$-gauge, on small scales the situation is similar: our intuition gets mixed up by a small physical effect plus a big gauge artifact (this comment also applies for $\delta\phi$-gauge on super-Hubble scales).

Technically, we shall indeed find that in the $\zeta$-gauge, we have to do a lot of integrations by parts to cancel large numbers in the gravitational action.

\item \textit{The Newtonian gauge:} We can choose to let the shift vector vanish: $\beta=0$. In this gauge one can rewrite the scalar sector of the metric as
\begin{align}
  ds^2 = - (1+2\Phi) dt^2 + a^2 (1-2\Psi) d\mathbf{x}^2 ~.
\end{align}
Here $\Phi$ and $\Psi$ are gravitational potentials. The Newtonian gauge allows us to think intuitively about how particle moves in the gravitational field: $\Phi$ is the Newtonian potential that non-relativistic particles move in Newtonian mechanics. Relativistic particles could feel both $\Phi$ and $\Psi$. In Einstein gravity, the traceless spatial Einstein equation yields $\Psi=\Phi$ in the absence of anisotropic stress. A departure can arise from anisotropic stress or modified gravity. For those reasons, the Newtonian gauge is very convenient in analysis of the late time universe. 

\item \textit{Mixed gauges:} One can choose the spatially flat gauge near Hubble crossing and the uniform inflaton gauge near reheating, using a smooth time shift to interpolate between the slices. For example, at linear order, starting from the flat gauge, take $\xi^0=s(t)\delta\phi/\dot\phi_0$, where $s(t_\mathrm{crossing})=0$ and $s(t_\mathrm{reheating})=1$. Then $\widetilde{\delta\phi}=(1-s)\delta\phi$ and $\widetilde\zeta=-sH\delta\phi/\dot\phi_0$. We shall use such a choice of initial and final slices when reviewing the $\delta N$ formalism.

\end{itemize}

It is useful to note that, sometimes those intuitive gauge choices do not completely fix the gauge. For example, in the $\zeta$-gauge, there is still residue global symmetry, known as the large gauge transformation \cite{Hinterbichler:2012nm}. For single field inflation,
\begin{align}\label{eq:global-shift}
  \mathbf{x} \rightarrow \widetilde \mathbf{x} = \mathbf{x} e^{\lambda}~, \quad 
  \zeta \rightarrow \widetilde \zeta(\widetilde \mathbf{x}) = \zeta(\mathbf{x}) - \lambda~,
\end{align}
\begin{align}
  \mathbf{x} \rightarrow \widetilde \mathbf{x} = \mathbf{x} + 2(\mathbf{b}\cdot\mathbf{x})\mathbf{x} - x^2\mathbf{b}~, \quad 
  \zeta \rightarrow \widetilde \zeta(\widetilde \mathbf{x}) = \zeta(\mathbf{x}) 
  - 2\mathbf{b}\cdot\mathbf{x}~,
\end{align}
For multiple field inflation, the change of other fields should be considered together. One could further fix this residue gauge (see, e.g. \cite{Pimentel:2012tw}) but we shall not need those further fixings at the moment.

\subsection{The 2nd order gravitational action, $\delta\phi$-gauge}

Now we are in the position to calculate the perturbations in the gravitational theory. There are typically two ways to do this. One way is to start from the classical Einstein equation, and perturb the Einstein equation to the linear order. However, the cosmic perturbations are quantum in nature. As we observed in the spectator field case, we need to know how to normalize the field canonically (how large one quanta is). Thus it is better to start from the action. In the action approach, the first order action is just a variation principle of the background and thus vanishes. To get the linear equation of motion, we need the second order action, and the variation principle on top of that. 

We start from the $\delta\phi$-gauge. In the $\delta\phi$-gauge, we set $\zeta=0$ and $E=0$. The perturbation variables in the scalar sector are $\delta\phi$, $\alpha$, $\beta$ defined as:
\begin{align}
  \delta\phi\equiv \phi(x)-\phi_0(t) ~,\qquad
  N \equiv 1+\alpha~, \qquad N_i \equiv \partial_i\beta ~.
\end{align}
The gravitational plus inflaton action is
\begin{align}\label{eq:single-field-action}
  S = \int d^4x \sqrt{-g} \left[ \frac{M_p^2R}{2} - \frac{1}{2}g^{\mu\nu}\partial_\mu\phi \partial_\nu\phi - V(\phi) \right] ~.
\end{align}
To calculate the Ricci scalar $R$ by hand, one had better go to the ADM formalism Eq.~(\ref{eq:ADM-R}) and drop the total derivative. Or if a computer algebra system is used, it is technically easier to start directly from the 4-dimensional Ricci scalar. 

In the ADM case the 2nd order action in momentum space takes the form
\begin{align} \label{eq:S2-ADM}
  S_2 = \int dt \frac{d^3k}{(2\pi)^3}
  \Big[ &
    a^3(\epsilon-3)M_p^2H^2\alpha^2 
    +2 k^2M_p^2 a H \alpha \beta 
    -a^3V'\alpha\delta\phi
    -a^3\dot\phi_0\alpha\dot{\delta\phi}
    - k^2 a \dot\phi_0 \beta\delta\phi 
\nonumber\\ &
    + \frac{1}{2} a^3 \dot{\delta\phi}^2 
    - \frac{1}{2} a (k^2+a^2V'')\delta\phi^2
  \Big] ~.
\end{align}

In 4-dimensional Ricci scalar case, the 2nd order action in momentum space takes the form
\begin{align}
  S_2 = \int dt \frac{d^3k}{(2\pi)^3}
  \Big[ &
    2a^3(3-\epsilon)M_p^2H^2\alpha^2 
    + 6M_p^2 a^3H\alpha \dot\alpha
\nonumber\\ &
    + k^2M_p^2 a H \alpha \beta 
    -k^2M_p^2 a \dot\alpha \beta 
    - k^2M_p^2a\alpha\dot\beta
    -a^3V'\alpha\delta\phi
    -a^3\dot\phi_0\alpha\dot{\delta\phi}
    - k^2 a \dot\phi_0 \beta\delta\phi 
\nonumber\\ &
    + \frac{1}{2} a^3 \dot{\delta\phi}^2 
    - \frac{1}{2} a (k^2+a^2V'')\delta\phi^2
  \Big] ~.
\end{align}
The above two actions can be related by some integrations by parts, as desired from the boundary term of (\ref{eq:ADM-R}). Note that in the action we have not written the k-dependence of the perturbation variables explicitly. Those k-dependence should be understood as, e.g. $2a^3(3-\epsilon)M_p^2H^2\alpha^2 ~\Leftrightarrow~ 
2a^3(3-\epsilon)M_p^2H^2\alpha_\mathbf{k}\alpha_\mathbf{-k}$. The $\alpha_\mathbf{k}\alpha_\mathbf{-k}$ part follows from momentum conservation.

From Eq.~(\ref{eq:S2-ADM}), we observe that there is no time derivative acting on $\alpha$ or $\beta$. This is as expected from the general principles we discussed when we introduce the ADM formalism. Thus $\alpha$ and $\beta$ can be solved:
\begin{align}
  \alpha = \frac{\dot\phi_0 \delta\phi}{2HM_p^2}~, \qquad 
  \beta = \frac{a^2}{2HM_p^2k^2}(\dot\phi_0\dot{\delta\phi}-H\epsilon\dot\phi_0\delta\phi-\ddot\phi_0\delta\phi)~.
\end{align}
Actually we shall not need the solution of $\beta$ when only the linear perturbations are considered. This is because Eq.~(\ref{eq:S2-ADM}) is linear in $\beta$. Thus inserting the solution of $\alpha$ into Eq.~(\ref{eq:S2-ADM}) eliminates $\alpha$ and $\beta$ together. Then after an integration by part to eliminate the $\delta\phi\dot{\delta\phi}$ term, we get the second order action
\begin{align}\label{eq:S2-fg}
  S = \int dt \frac{d^3k}{(2\pi)^3}
  \left\{ \frac{1}{2} a^3\dot{\delta\phi}^2 - \frac{1}{2} k^2a\delta\phi^2 
    - \frac{1}{2} a^3\left[ 
      V'' -2H^2 (3\epsilon-\epsilon^2+\epsilon\eta)
    \right] \delta\phi^2 \right\} 
  ~.
\end{align}
A few comments are in order:
\begin{itemize}
\item The 3rd term in Eq.~(\ref{eq:S2-fg}) is a mass term for the perturbations. The mass is $m_\mathrm{eff}^2 = V'' -2H^2 (3\epsilon-\epsilon^2+\epsilon\eta)$. Due to gravitational corrections, this mass is different from the naively expected $m^2 = V''$. Nevertheless they are of the same order.

\item Note that this mass term is rather small. To see this, we recall that the energy scale for inflationary perturbations is $H$. And $m_\mathrm{eff}^2 = \mathcal{O}(\eta) H^2$.

\item Could this small mass be neglected? The answer is actually subtle. Although $m_\mathrm{eff}\ll H$, it may take effect on super-Hubble scales. For the mode corresponding to the current cosmological scales, the mass term takes effect for about 60 Hubble times (the observable e-folding number of inflation). The same subtlety can be also observed in that, the $\dot{\delta\phi}^2$ term in the action would vanish exponentially on super-Hubble scales, and eventually become comparable with the ``small'' mass term. We for the time being neglect this mass term, and leave the subtleties to the next subsection. 
\end{itemize}

(Doubtfully) neglecting the mass term, the discussion for the $\delta\phi$ perturbations become completely parallel to the discussion of a massless spectator scalar field on de Sitter background. The power spectrum of the $\delta\phi$ perturbations can be calculated as
\begin{align}
  P_{\delta\phi} = \left(\frac{H}{2\pi}\right)^2~.
\end{align}

\subsection{The 2nd order gravitational action, $\zeta$-gauge}
\label{sec:2nd-order-grav}
As we have discussed in the previous subsection, in the $\delta\phi$-gauge, perturbations could evolve on super-Hubble scales. This invalidates the naive slow roll approximation. Actually the problem is more serious: Inflation takes place at an extremely high energy scale. The physics from after inflation until TeV scale is unknown. How could we evolve the inflationary perturbations (or any other observables produced by inflation) through this unknown epoch and make predictions for inflation? 

Fortunately, for single field inflation on an attractor background, the $\zeta$-perturbation in the $\zeta$-gauge is conserved on super-Hubble scales. We shall show the conservation at linear level in this subsection, and at non-linear level in the later sections. It is the conserved perturbation on super-Hubble scales, that brings connection between the very early time physics and very late time physics, after Hubble-reentering. 

For this consideration, it is important to consider the perturbations in the $\zeta$-gauge. Here we set $\delta\phi=0$ and $E=0$. Inserting the perturbations into the gravitational plus inflaton action (\ref{eq:single-field-action}), we get using the ADM Ricci scalar:
\begin{align} \label{eq:S2-ADM-zeta}
  S_2 = M_p^2\int dt \frac{d^3k}{(2\pi)^3}
  \Big[ &
  - a^3(3-\epsilon)H^2\alpha^2 
  + (2aHk^2\beta+2ak^2\zeta+6a^3H\dot\zeta)\alpha
  - 2k^2a \beta\dot\zeta
\nonumber\\ &
  -3a^3\dot\zeta^2
  - 18a^3H\zeta\dot\zeta
  + a(-9(3-\epsilon)a^2 H^2+k^2)\zeta^2
  \Big] ~.
\end{align}
Or from the 4-dimensional Ricci Scalar:
\begin{align}
  S_2 = M_p^2\int dt \frac{d^3k}{(2\pi)^3}
  \Big[ &
  2a^3(3-\epsilon)H^2\alpha^2 
  + (-k^2a\beta-9a^3H\zeta-3a^3\dot\zeta)\dot\alpha
\nonumber\\&
  + (6a^3H\dot\alpha+k^2aH\beta-k^2a\dot\beta+a(2k^2-9(3-\epsilon)a^2H^2)\zeta
    -12a^3H\dot\zeta-3a^3\ddot\zeta)\alpha
\nonumber\\&
  - 2k^2a \beta\dot\zeta
  + (36a^3H\dot\zeta+9a^3\ddot\zeta)\zeta
  + 6a^3\dot\zeta^2
  + \frac{1}{2} a(9(3-\epsilon)a^2H^2+2k^2)\zeta^2
  \Big] ~.
\end{align}
Again they are related by integration by parts. The constraints can be solved as
\begin{align}\label{eq:cons-zeta-1}
  \alpha = \frac{\dot\zeta}{H} ~, \qquad \beta = -\frac{\zeta}{H} - \frac{a^2\epsilon\dot\zeta}{k^2}
  ~.
\end{align}
Insert the constraints into the 2nd order action, and do some integration by parts to eliminate cross terms. Eventually one finds the action to be in a very simple form
\begin{align} \label{eq:action-zeta}
  S = M_p^2\int dt \frac{d^3k}{(2\pi)^3} \epsilon (a^3\dot\zeta^2 - k^2 a\zeta^2)~.
\end{align}

Note that unlike the case of $\delta\phi$-gauge, now there is no mass term for the $\zeta$-gauge perturbations. Thus the awkward super-Hubble evolution problem no longer exists in the $\zeta$-gauge. On an attractor background, there is a decaying mode and a constant mode for $\zeta$. After a few e-folds after Hubble exit, the decaying mode goes away and $\zeta$ becomes conserved. We shall verify this conservation argument explicitly at the end of this section.

It is useful to note two facts:
\begin{itemize}
\item No slow roll approximation is assumed in the above derivation. Thus the above action applies even $\epsilon\sim 1$.
\item In general, even if the cosmic fluid is not a scalar field, as long as $p=p(\rho)$, one can choose the $\zeta$-gauge.\footnote{For an adiabatic fluid with sound speed $c_s^2=dp/d\rho$, the quadratic action is \cite{Garriga:1999vw}
\begin{align}
S_2=M_p^2\int dt\,d^3x\,\frac{a^3\epsilon}{c_s^2}
\left[\dot\zeta^2-\frac{c_s^2}{a^2}(\partial_i\zeta)^2\right].
\end{align}
The canonical scalar result corresponds to $c_s=1$.}
\end{itemize}
Thus the conservation of $\zeta$ should hold after inflation (with the assumption of $p=p(\rho)$). This answers the question at the beginning of this subsection: The conservation of $\zeta$ makes inflation predictive.

Now we are in the place to calculate the quantum perturbations of $\zeta$. The equation of motion of $\zeta$ is
\begin{align} \label{eq:zeta-eom}
  \ddot\zeta + (3+\eta) H \dot\zeta + \frac{k^2}{a^2} \zeta = 0~.
\end{align}
To the lowest order of slow roll approximation, the $\eta$ factor in the above equation can be ignored and approximately $a\propto e^{Ht}$. The solution of the above equation, as a function of conformal time, is
\begin{align}
  \zeta = \frac{H}{M_p\sqrt{4\epsilon k^3}} (1+ik\tau)e^{-ik\tau}~.
\end{align}
In terms of the quantized fields, the above solution is equivalent to
\begin{align} \label{eq:zeta-mode-function}
  \zeta_\mathbf{k} = u_k a_\mathbf{k} + u^*_k a^\dagger_{-\mathbf{k}}~,
  \qquad u_k = \frac{H}{M_p\sqrt{4\epsilon k^3}} (1+ik\tau)e^{-ik\tau}~.
\end{align}

As in the case of a scalar field in de Sitter space, the ``negative frequency'' mode is disregarded and the constant factor is determined from the consistency of quantization together with the assumption that the initial state of the k-mode has lowest energy.

Before closing this subsection, we would like to verify the conservation of $\zeta$ on super-Hubble scales, without the reference of slow roll or inflationary background. Note that \eqref{eq:zeta-eom} is an exact equation without making slow roll approximation. 

On super-Hubble scales, $k\ll aH$ thus the $(k^2/a^2) \zeta$ term can be neglected in \eqref{eq:zeta-eom}. Thus the equation of motion can be approximated as
\begin{align}
  \ddot\zeta + (3+\eta) H \dot\zeta \simeq 0~.
\end{align}
Considering the exponential growth of $a$, this equation is exponentially accurate on super-Hubble scales. The solution is
\begin{align}
  \zeta(t) = c_1 + c_2 \int^t dt_1~ \exp\left\{ -\int^{t_1}[3+\eta(t_2)]H(t_2) dt_2 \right\} ~.
\end{align}
Thus there exists a constant mode $c_1$. In a quasi-de Sitter background, the other mode is decaying for approximately constant $\eta>-3$. More generally, its behavior is determined by $\int^t dt_1/[a(t_1)^3\epsilon(t_1)]$.\footnote{However, it is interesting to note that in case $\eta<-3$, $\zeta$ finds itself in an effectively contracting universe, although the real metric is expanding.}

\subsection{Power spectrum, scale invariance and tilt}

Similarly to the case of massless scalar field, the power spectrum of comoving curvature perturbation can be defined as
\begin{align} \label{eq:scalar-power}
  \langle\zeta_\mathbf{k}\zeta_{\mathbf{k}'}\rangle = (2\pi)^3 \delta^3(\mathbf{k}+\mathbf{k}') 
  \frac{2\pi^2}{k^3} P_\zeta(k)~, \qquad 
  P_\zeta(k) = \frac{H^4}{4\pi^2\dot\phi^2} = \frac{H^2}{8\pi^2\epsilon M_p^2} ~.
\end{align}
The observational value of the power spectrum is $P_\zeta(k)\simeq 2\times 10^{-9}$. This value\footnote{The value $P_\zeta(k)\simeq 2\times 10^{-9}$ is the power spectrum on large scales, spanning between the scale of the whole observable universe, and scale about $e^{-10}$ smaller.  On the other hand, there are constraints that the power spectrum should be perturbative on small scales all the way down to the comoving Hubble scale at the end of inflation, by not observing the primordial black holes (see \cite{Khlopov:2008qy} for a review, and the references therein) formed by those small scale fluctuations. } is known as the COBE normalization \cite{Smoot:1992td}. Sometimes one also use $\Delta_\mathcal{R}$ defined as $\Delta_\mathcal{R} \equiv \sqrt{P_\zeta}$.

Again, the power spectrum does not depend on $k$ at the lowest order. Thus the inflationary perturbations are nearly scale invariant. 

Although the power spectrum is nearly scale invariant, it is still important to consider the small non-scale-invariant correction to the power spectrum. This correction is called the tilt of the spectrum, or the spectral index. According to the WMAP result, the exact scale invariant power spectrum is almost ruled out and the deviation from scale invariance should be a few percents.

To describe the scale dependence, the spectral index $n_s$ is defined as
\begin{align}
  n_s - 1 \equiv \frac{d\ln P_\zeta(k)}{d\ln k} ~.
\end{align}

To calculate this quantity, it is helpful to have a closer look at the equation of motion (\ref{eq:zeta-eom}). There are two ways to take the scale dependence into account in (\ref{eq:zeta-eom}). The most straightforward way is to consider the small deviation from $a\propto e^{Ht}$ together with no longer neglecting $\eta$. The equation of motion can be solved in this case and a deviation from scale invariance is observed \cite{Stewart:1993bc, Lidsey:1995np}. 

On the other hand, there is an easier way to calculate the scale dependence. Weak scale dependence means that the parameters in (\ref{eq:zeta-eom}) would vary by a small amount per Hubble time. Within a few e-folds near the Hubble crossing, Eq.~(\ref{eq:zeta-eom}) should still be applicable, with $H$ considered as a constant and $\eta$ neglected. A few e-folds before the Hubble crossing the solution is linked with the instant Minkowski solution, and a few e-folds after the Hubble crossing the solution is linked with the conserved quantity. Both the sub-Hubble instant Minkowski solution and the super-Hubble conserved quantity are not slow roll approximated but instead exact result.

In other words, the prescription should be to take the Hubble parameter in the equation of motion as a function of momentum $k$: $H=H(k)$, and calculate the value of $H$ at Hubble crossing $k=aH$. Thus up to higher order slow roll corrections, we have $d\ln k \simeq d\ln a = H dt$. The spectral index can be calculated as
\begin{align}
  n_s - 1 = \frac{d\ln P_\zeta(k)}{H dt} = - 2\epsilon - \eta 
  = -6\epsilon_V + 2\eta_V ~.
\end{align}
The observational constraint for $n_s-1$ is \cite{Ade:2013rta}
\begin{align}
  n_s = 0.9603\pm 0.0073 \quad \mbox{(68\% CL)}~.
\end{align}
Thus although inflation is nearly scale invariant, an exact scale invariance with $n_s=1$ is ruled out at over 5$\sigma$.

One can also define the running of spectral index, as
\begin{align}
  \alpha_s \equiv \frac{d n_s}{d\ln k} ~.
\end{align}
For the above single field inflation model we are considering, we have
\begin{align}
  \alpha_s = -2\eta\epsilon - \frac{\dot\eta}{H} \equiv -2\eta\epsilon - \eta_2 \eta ~, \qquad \eta_2\equiv\frac{\dot\eta}{H\eta} ~.
\end{align}
Thus $\alpha_s$ is one order smaller than $n_s-1$ in terms of slow roll parameters in the above simple model. The observational constraint for $\alpha_s$ is \cite{Ade:2013rta}
\begin{align}
  \alpha_s = -0.0134\pm 0.0090\quad \mbox{(68\% CL)}~.
\end{align}

\subsection{Comoving curvature perturbation as the e-folding number difference}
\label{sec:comov-curv-pert}

As we discussed at the end of Sec.~\ref{sec:2nd-order-grav}, on super-Hubble scales the equation of motion simplify. Actually, the $k/(aH)$ parameter itself can be used as an expansion parameter. One can set up perturbation theory with respect to  $k/(aH)$, which could be exact to all orders of $\zeta$ or the slow roll parameter. This method is called the gradient expansion.

There is an intuitive way to view the gradient expansion, at least at the leading order $k/(aH)\rightarrow0$. During inflation, the universe broken up into exponentially many ``local universes'', which temporarily cannot talk to each other. Each local universe is quite homogeneous and isotropic and thus can be locally approximated as a FRW universe.

\begin{figure}[htbp]
  \centering
  \includegraphics[width=0.7\textwidth]{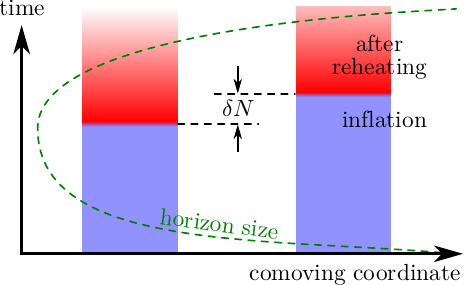}
  \caption{\label{fig:deltaN} Time difference between local FRW universes. After reheating, the universe become radiation dominated and the energy density dilutes quickly with the expansion of the universe. The change of color illustrate the drop of energy density.}
\end{figure}

The local properties of those local FRW universes have no difference between one another. However, when we compare those local universes in spacetime, there could be a difference of expansion between them at a fixed time, or a difference of time at a fixed rate of expansion. This difference can be produced by the quantum fluctuations. For example, in one local universe, if there is a quantum fluctuation to drive the inflaton a little bit up on the potential, this local universe is then driven farther away from reheating.
As we shall show, this difference is actually the comoving curvature perturbation. This situation is illustrated in Fig.~\ref{fig:deltaN}.

It is useful to make this separate universe idea explicit. There are a few ways to do this and here we follow the method of \cite{Lyth:2004gb}. Consider a local universe with scale factor
\begin{align}
  \widetilde a(t,\mathbf{x}) = a(t) e^{\zeta(t,\mathbf{x})}~,
\end{align}
where $\zeta(t,\mathbf{x})$ is almost a constant on scales comparable with the Hubble scale, and only change significantly on super-Hubble scales. The local Hubble expansion can be defined as\footnote{Note that the time derivative here can denote a general time instead of restricted into physical time. The lapse function cancels in this equation and later related equations.} $\widetilde H \equiv \dot{\widetilde a}/ \widetilde a$. The continuity equation is
\begin{align}
  \dot\rho + 3\widetilde H (\rho+p) = 0~.
\end{align}
In terms of $a$ and $\zeta$, this equation can be written as
\begin{align} \label{eq:local-energy-conservation}
  \frac{\dot a}{a}  + \dot\zeta = -\frac{1}{3} \frac{\dot\rho}{\rho+p} ~.
\end{align}
So far we haven't specified any gauge yet. Now we shall look at this equation in the uniform energy density gauge. (For regular adiabatic attractor modes, this gauge agrees with the comoving gauge at leading order in spatial gradients.) If there is no entropy perturbation, i.e. $p=p(\rho)$, then in Eq.~\eqref{eq:local-energy-conservation} only $\dot\zeta$ is a perturbation variable. The precise definition and meaning of entropy perturbation will be discussed in Sec.~\ref{sec:curv-isoc-pert}. All other variables are background variables, independent of $\mathbf{x}$. Considering a perturbation variable should not have zero mode, we have
\begin{align}
  \dot\zeta = 0~.
\end{align}
We have thus confirmed the conservation of uniform-density curvature perturbation on super-Hubble scales. And this time the confirmation is exact for all orders of perturbations in $\zeta$ (but assuming that $\zeta$ is classical, both in external lines and in loops. For a quantum treatment, see \cite{Assassi:2012et}).

Now let us make connection between $\zeta$ and the perturbation of e-folding number. Note that this part is independent to the above discussion, in the sense that we do not assume $p=p(\rho)$. Thus the result also applies for a time dependent $\zeta$ on super-Hubble scales.

The connection is intuitive because $\zeta$ is the perturbation of scale factor, and on the exponent. To see this connection more precisely, we note that the e-folding number of a local FRW universe can be written as
\begin{align}
  \widetilde N(t_2,t_1,\mathbf{x}) = \int_{t_1}^{t_2} dt \widetilde H
  = \int_{t_1}^{t_2} dt \left( \frac{\dot a}{a} + \dot\zeta  \right)
  = \ln \left[ \frac{a(t_2)}{a(t_1)} \right] + \zeta(t_2,\mathbf{x}) - \zeta(t_1,\mathbf{x})~.
\end{align}
Again this equation does not depend on gauge choice. Now we choose a mixed gauge: At an initial time $t_1$, $\zeta(t_1)=0$ (flat slice), and at a final time $t_2$ and later times, $\rho=\rho(t)$ (uniform energy density slice). In this mixed gauge, we have
\begin{align} \label{eq:zeta-delta-N}
  \zeta(t_2,\mathbf{x})
  = \widetilde N(t_2,t_1,\mathbf{x}) - N(t_2,t_1) \equiv \delta N ~,
\end{align}
where $\delta N$ is the e-folding number difference between the initial flat slice and the final uniform energy density slice. This relation is known as the $\delta N$-formalism \cite{Starobinsky:1986fxa, Sasaki:1995aw, Lyth:2005fi}.

Finally, if needed, the gauge-invariance can be restored by adding back the gauged counter-part of \eqref{eq:zeta-delta-N}:
\begin{align} \label{eq:zdn-gi}
  \zeta^\mathrm{(gi)}(t) = \delta N(t) + \frac{1}{3} \int_{\bar \rho(t)}^{\rho(t,\mathbf{x})} \frac{d\rho}{\rho+p} ~,
\end{align}
Here $\delta N$ is measured from the initial flat slice to an arbitrary final slice with density $\rho(t,\mathbf{x})$. The pressure $p$ in the integral is evaluated along the same local expansion history, with the target density $\bar\rho(t)$ held fixed.

As a side remark, the above discussions could actually be made stronger. Consider there are $n$ components of fluids (say, $\rho_i$, $i=1,\ldots,n$) in the universe. Assume that those components do not exchange energy with each other and that each is barotropic, $p_i=p_i(\rho_i)$. Then the energy of those components are conserved separately. In this case we have $n$ conserved quantities: one can define $\zeta_i$ as the $\delta N$ on uniform $\rho_i$ slices, respectively. The above discussion holds component-wise.

The $\delta N$-formalism is convenient in several aspects:
\begin{itemize}
\item It is straightforward to calculate the inflaton fluctuation on the initial flat slice. As we have shown, the inflaton fluctuation at Hubble crossing is $\delta\phi=H/(2\pi)$ plus calculable slow roll corrections. 
\item The curvature perturbation is conserved after the final uniform energy density slice if the subsequent evolution is adiabatic.
\item For single field inflation, the uniform energy density slice is the reheating surface. Thus the $\delta N$ defined in the above way does coincide with the e-folding number difference during inflation. 
\item The $\delta N$-formalism is non-linear in orders of $\zeta$. Thus the formalism is still convenient to use for non-linear perturbations.
\end{itemize}

The convenience can be observed by a sample calculation. During inflation, the e-folding number can be calculated as $N(t)= \int_t^{t_\mathrm{reh}} Hdt$. A variation of this equation is
\begin{align}
  \zeta = \delta N = - H\delta t = -\frac{H\delta\phi}{\dot\phi} ~.
\end{align}
Thus
\begin{align}
  P_\zeta = \frac{H^2}{\dot\phi^2} \left( \frac{H}{2\pi}  \right)^2~.
\end{align}
This agrees with the power spectrum we have calculated in \eqref{eq:scalar-power}.

However, there is price for the simplicity. One has to pay attention to the assumptions of the above derivation, and be careful at the situations where those assumptions break down. For example,
\begin{itemize}
\item \textit{Non-conservation of $\zeta$}. When there are entropy perturbations, $p$ is no longer a function of $\rho$ thus $\zeta$ may not be conserved on super-Hubble scales.
\item \textit{Non-trivial reheating surface}. When there are entropy perturbations, the reheating surface may not be the uniform total energy density slice. In this case $\delta N$ no longer has the physical meaning of e-folding number difference of inflation. Instead, a correction term coming from the difference between uniform energy density slice and the reheating surface must be considered.
\item \textit{Nontrivial initial correlation functions}. If the power spectrum or higher order perturbations of $\delta\phi$ is modified at Hubble-crossing, the modification needs to be considered as an input of the $\delta N$-formalism.
\item \textit{Ignorance of the decaying mode}. The requirement $p=p(\rho)$ is not the same as the requirement of single field. This is because a scalar field has second order equation of motion and thus two modes. On super-Hubble scales one mode decays away thus $p=p(\rho)$. However, in case both modes needs to be considered for some reason, one need to examine if $p=p(\rho)$ still hold.
\item \textit{Validity of local FRW metric}. This is typically not a problem for inflation. Because locally, non-FRW energy components are diluted. However, for alternative models to inflation, this is not necessarily true.
\item \textit{Break down of classicality}. At the non-linear level, quantum fluctuations on sub-Hubble scales may interact with the super-Hubble fluctuation. $\delta N$-formalism is classical and do not carry this quantum nature.
\end{itemize}

Finally, let us make one more clarification of the gradient expansion method and the $\delta N$-formalism: There exist sub-Hubble fluctuations. When considering local FRW universes, why we neglect those sub-Hubble fluctuations? Actually, those sub-Hubble fluctuations, as high energy modes in a general QFT consideration, can be integrated out. As a result, the coarse-grained inflaton obeys a Langevin equation in the slow roll approximation \cite{Starobinsky:stochastic}:
\begin{align}
  \dot\phi = -\frac{V'}{3H} + \frac{H^{3/2}}{2\pi} \eta(t)~,
\end{align}
where $\eta(t)$ is a Gaussian noise obeying
\begin{align}
  \langle \eta(t_1) \eta(t_2) \rangle = \delta(t_1-t_2)~.
\end{align}
This is exactly the origin of the $\delta\phi=H/(2\pi)$ fluctuation at Hubble crossing.

\subsection{Observable inflation,  de Sitter space and eternal inflation}

So far we have been talking about the observable stage of inflation. When calculating the dynamics of inflationary perturbations, we have been using the lowest order slow roll approximation. Especially, when calculating the equation of motion for the perturbations, the mode functions are the same as that obtained in (the FRW patch of) de Sitter space (see our toy model of spectator field). 

However, despite of the approximations, we should be careful in thinking about the observable stage of inflation as de Sitter space. 

For example, one note that the power spectrum of comoving curvature perturbation is $P_\zeta \propto H^2 / (\epsilon M_p^2)$. When $\epsilon\rightarrow 0$ with $H$ fixed finite, the power spectrum diverges. This is very different from the observed power spectrum  $P_\zeta \simeq 2\times 10^{-9}$.

The problem not only arises as the divergence in the number. But also, $\zeta$ (or other observables stands for density perturbation of inflation) manifests itself only after the end of inflation and after the perturbation modes returns to the horizon. The exponential expansion of de Sitter space never end thus one can not observe the density perturbations. 

Even worse, in de Sitter space, the ``super-Hubble'' density perturbations are not only non-observable. The whole semi-classical picture of cosmic perturbation theory may break down in the de Sitter limit. It is argued that, like the black hole complementarity \cite{Susskind:1993if}, de Sitter space may have complementarity as well \cite{Goheer:2002vf}. 

Consider two comoving observers A and B, who start near each other with distance nearer than Hubble. Because of cosmic expansion, at a later time, $B$ will exit the horizon of $A$. The horizon exit process of $B$, as seen by $A$, is a disaster of $B$. The observer $A$ measures the Gibbons-Hawking radiation with a temperature $T=H/(2\pi)$. Blue shift this temperature to the near horizon regime, $A$ actually finds $B$ destroyed by the Gibbons-Hawking radiation when approximating the de Sitter horizon. This picture is different from that experienced by $B$, who finds nothing special because of the equivalence principle. The different semi-classical pictures observed by $A$ and $B$ contradicts. But this is not a problem as long as the observers no longer return to each other's horizon to compare notes. 

Applied to cosmic perturbations: for a local observer, the physical picture of perturbations in de Sitter space should be that, the perturbations approaches the de Sitter horizon, get burnt by the blue shifted Gibbons-Hawking radiation, and get emitted back by the horizon. There is no other copy of information which actually ``exits'' the horizon.

Thus the physical picture is different between the perturbations in a stage of observable inflation, and that in de Sitter space. The change of physical picture actually not only happens at the exact value $\epsilon=0$, but takes place when $\epsilon$ becomes so small that $P_\zeta \geq 1$. Note that the power spectrum can be also written as $P_\zeta = H^4 / (2\pi\dot\phi)^2$. Thus when $P_\zeta \geq 1$ we have
\begin{align}
  \frac{|\dot\phi|}{H} \lesssim \frac{H}{2\pi} ~.
\end{align}
The LHS is the classical motion of inflaton per Hubble time, and the RHS is the quantum fluctuation per Hubble time. Thus the above equation can be interpreted as that, the quantum fluctuation of the inflaton dominates over the classical motion.

The classical motion of the inflaton always roll down the potential in the slow roll regime. However, the quantum fluctuations has equal probability to fluctuate up or down. After one Hubble time, a Hubble volume self-reproduces into $e^3\simeq 20$ copies. If in one or more of those $20$ new Hubble volumes, the inflaton field has greater value due to the large quantum fluctuation, the dynamics in those patches are driven away from the reheating surface. The self-reproduction continues and inflation will never globally end. This situation is known as eternal inflation \cite{Steinhardt:1982kg, Vilenkin:1983xq}. 

During eternal inflation, when following an observer's comoving world line, the observer finds that inflation eventually ends locally. However, due to the lack of global end, the global structure of eternal inflation is analogue to that of de Sitter space.

As in the case of de Sitter space, the semi-classical picture of cosmic perturbations break down during eternal inflation. Again this is not only because of $P_\zeta\geq 1$, and not only because of lack of global end of inflation, but also that the whole picture cannot be trusted.

To see this, it is convenient to consider (the slow roll analog of) the static patch of de Sitter space, instead of the FRW patch. In the static patch, the metric can be written as
\begin{align}\label{eq:static-metric}
  ds^2 = -(1-H^2r^2)d\tau^2 +\frac{dr^2}{1-H^2r^2}+r^2d\Omega^2~.
\end{align}

The stretched horizon is defined as the surface with an area smaller than the classical horizon by a Planck-scale amount:
\begin{align}\label{eq:stretched}
  r_\mathrm{SH} = \sqrt{\frac{1}{H^2}-\frac{1}{4\pi^2 M_p^2}}~.
\end{align}This definition is reasonable because stretched horizon $r_\mathrm{SH}$, the Gibbons-Hawking temperature is the Planck temperature. The temperature has a redshift
\begin{align}
  \frac{T}{H/(2\pi)} = \frac{d\tau}{\sqrt{-ds^2}} = \frac{1}{\sqrt{1-H^2r^2}}~.
\end{align}Setting $T=M_p$, one get the $r_\mathrm{SH}$ in equation (\ref{eq:stretched}). Complementarity becomes important as soon as objects (e.g. perturbation modes) reach the stretched horizon.

Now we want to estimate when complementarity becomes important for slow roll inflation. Roughly, we could use the result for pure de Sitter space, and let $H$ vary with respect to time. Now the field excitation propagate towards the stretched horizon and the stretched horizon is also expanding itself. One should compare which of the two goes faster. 

The field excitation near the stretched horizon falls towards the stretched horizon at speed of light:
\begin{align}
  v = \frac{dr}{d\tau} = 1-H^2r_\mathrm{SH}^2 = \frac{H^2}{4\pi^2 M_p^2}~.
\end{align}
Measured by an observer at the center, the stretched horizon itself moves at a velocity (in terms of $r$ and $\tau$) as
\begin{align}
  v_\mathrm{SH} = \frac{dr_\mathrm{SH}}{d\tau} = \frac{dr_\mathrm{SH}}{dt}
  \simeq \frac{d}{dt}\left(\frac{1}{H}\right) \equiv \epsilon~.
\end{align}Here one should note that $v$ is local process near the stretched horizon thus there is a huge redshift factor. however, $v_\mathrm{SH}$ is measured at the center of static patch, thus $d\tau = dt$ when calculating $v_\mathrm{SH}$.

To compare those velocities, one note
\begin{align}
  \frac{v}{v_\mathrm{SH}} \simeq \frac{H^2}{4\pi^2\epsilon M_p^2} = 2P_\zeta~,
\end{align}where $P_\zeta$ is the power spectrum for curvature perturbation during inflation. For the observable stage of inflation, $P_\zeta \simeq 2.4\times 10^{-9}$. Thus during the observable stage of inflation the field excitations would never reach the stretched horizon. Thus the apparent difference of the static patch and the FRW patch is merely a coordinate transformation. However, for eternal inflation, $P_\zeta \geq 1$. Thus during eternal inflation the Hubble radius is expanding so slowly that the field excitations could reach the stretched horizon.

On the other hand, even one believed the semi-classical picture of eternal inflation, the density perturbation generated during eternal inflation could never be observed even in principle. This is because when the $\zeta\sim 1$ modes returns to the horizon, they form black holes. The global structure of spacetime is not accessible by an observer in a black hole \cite{ArkaniHamed:2007ky}.

To conclude this subsection, when taking the de Sitter space limit, one hits eternal inflation and a shift of semi-classical physical picture. The calculations of power spectrum and non-Gaussianities are not affected (although the de Sitter approximation is used for the mode function), because the perturbations never reaches the stretched horizon.

\subsection{Gravitational waves}

The inflaton is not the unique light field during inflation. Instead, there is at least another light field -- the gravitational waves \cite{Rubakov:1982df, Fabbri:1983us, Abbott:1984fp} (for modern conventions, see, e.g. \cite{Boyle:2005se}). The calculation of gravitational waves goes similarly to the scalar sector. One could either do a brute force calculation of the tensor part, or first calculate the Ricci scalar on Minkowski background and then do a conformal transformation into an inflationary background. The result is
\begin{align}
  S = \frac{M_p^2}{8} \int \frac{d^3k}{(2\pi)^3}d\tau~a^2
  \left( \gamma^i_j{}' \gamma^j_i{}' - k^2 \gamma^i_j \gamma^j_i\right) ~.
\end{align}
Note that the traceless and transverse conditions eliminates 4 components from the 6 components of a symmetric 3-tensor. Thus there are only 2 physical modes in $\gamma^i_j$. Those two modes can be decomposed by introducing the polarization tensors
\begin{align} \label{eq:tensor-dec}
  \gamma_{ij}(\mathbf{k}) = \frac{\sqrt2}{M_p} \left[ \gamma_+(\mathbf{k}) e^+_{ij}(\mathbf{k}) 
    + \gamma_\times (\mathbf{k}) e^\times_{ij}(\mathbf{k}) \right]~,
\end{align}
where the symmetric, traceless and transverse properties are carried over by $e^+_{ij}$ and $e^\times_{ij}$ as
\begin{align}
  e^{+,\times}_{ij} = e^{+,\times}_{ji}~, 
  \qquad k^i e^{+,\times}_{ij} = 0, 
  \qquad e^{+,\times}_{ii}= 0~.
\end{align}
Further from the reality conditions we have
\begin{align}
  e^{+,\times}_{ij}(-\mathbf{k}) = (e^{+,\times}_{ij}(\mathbf{k}))^*~.
\end{align}
And we have the freedom to normalize $e^{+,\times}_{ij}$. The normalization we take is
\begin{align}
  e^+_{ij} (e_+^{ij})^* = e^\times_{ij} (e_\times^{ij})^* = 2~.
\end{align}

With the above decomposition, the action of the tensor mode becomes two decoupled components. Theoretically this is because they are different irreducible representations. Technically this is because $e^+_{ij}(e^\times_{ij})^*=0$. The tensor mode action takes the form
\begin{align}
  S = \frac{1}{2} \int \frac{d^3k}{(2\pi)^3}d\tau~a^2
  \left[ \left( \gamma_+' \gamma_+{}' - k^2 \gamma_+ \gamma_+\right) 
    + \left( \gamma_\times' \gamma_\times{}' - k^2 \gamma_\times \gamma_\times\right) \right]~.
\end{align}
For each polarization mode, the action is the same as that of a massless spectator scalar field. Thus
\begin{align} \label{eq:tensor-power}
  P^{+,\times}_\gamma = \left( \frac{H}{2\pi} \right)^2 ~.
\end{align}
The tensor power spectrum is defined as
\begin{align}
  \delta^{ik} \delta^{jl} \langle \gamma_{ij}(\mathbf{k})\gamma_{kl}(\mathbf{k}') \rangle = 
  (2\pi)^3 \delta^3(\mathbf{k}+\mathbf{k}') \frac{2\pi^2}{k^3} P_\gamma~,
\end{align}

Compared with Eq.~(\ref{eq:tensor-dec}), we have
\begin{align}
  P_\gamma = \frac{2H^2}{\pi^2 M_p^2}~. 
\end{align}

To compare with the scalar power spectrum, a tensor to scalar ratio is defined and calculated as
\begin{align} \label{eq:tensor-to-scalar-ratio}
  r \equiv \frac{P_\gamma}{P_\zeta} = 16\epsilon = 16 \epsilon_V~.
\end{align}
The observational constraint on $r$ is \cite{Ade:2013rta}
\begin{align}
  r < 0.11 \quad\mbox{(95\% CL)}~.
\end{align}

It is important to note that, unlike the scalar power spectrum (\ref{eq:scalar-power}), the tensor power spectrum (\ref{eq:tensor-power}) does not depend on $\epsilon$. In other words, $P^{+,\times}_\gamma$ is solely determined by $H$, the energy scale of the inflationary perturbations. Thus if the gravitational wave power spectrum is detected, we immediately know the inflationary Hubble scale.

Also, for the minimal single field inflation, the slow roll parameters $\epsilon_V$ and $\eta_V$ can be calculated as functions of observables $n_s$ and $r$:
\begin{align}
  \epsilon_V = \frac{r}{16} ~, \qquad \eta_V = \frac{1}{2} (n_s -1) + \frac{3}{16} r~.
\end{align}
Thus measurements on $n_s$ and $r$ can be considered as measurements of the first two derivatives of the inflationary potential. Especially, the sign of  $\eta_V$ determines if the inflationary potential is convex or concave. The latter is favored by current data \cite{Ade:2013rta}. One can plot the single field inflation models (as a function of e-folding number) as line intervals on a two dimensional $n_s$-$r$ plot.

The spectral index of gravitational waves is
\begin{align}
  n_T \equiv \frac{d\ln P_\gamma}{d\ln k} = -2\epsilon = -r/8~.
\end{align}
Note that both the very LHS and very RHS of the above equation are expressed in terms of observables. Thus $n_T = -r/8$ is a consistency relation for single field inflation.

It is also interesting to note that, the tensor to scalar ratio is related to the distance that the inflaton rolled in field space during inflation. The inflaton's rolling distance per e-fold is
\begin{align}
  \left|\frac{d\phi}{dN}\right| = \frac{|\dot\phi|}{H} = \frac{M_p}{4} \sqrt{2r}~.
\end{align}
For a monotonic trajectory, the field excursion is $\Delta\phi=M_p\int dN\sqrt{r(N)/8}$. If $r$ is approximately constant over 60 e-folds, this gives $\Delta\phi\simeq15M_p\sqrt{2r}$. A detectable tensor signal therefore suggests a large field excursion, known as the Lyth bound \cite{Lyth:1996im}. One can correspondingly classify inflation models into large field and small field models, with $\Delta\phi\geq M_p$ and $\Delta\phi<M_p$ respectively.

\subsection{Other fields}

As we have seen, the construction of inflation requires two types of light fields: the inflaton field and the gravitational waves. However, we know from QFT that there are many other fields. What do other fields behave during inflation? Here we discuss several possibilities:

\begin{itemize}

\item \textit{Gauge fields:} Non-interacting gauge fields in 3+1 dimension have a conformal symmetry. The action of a non-interacting gauge field can be written as (take U(1) as an example)
\begin{align}
  S = - \frac{1}{4} \int d^4x \sqrt{-g} g^{\mu\alpha}g^{\nu\beta} F_{\mu\nu}F_{\alpha\beta}~.
\end{align}
Note that $F_{\mu\nu} = \partial_\mu A_\nu - \partial_\nu A_\mu$ does not really contain metric or covariant derivative. Moreover, the scale factor dependence (using conformal time) in $\sqrt{-g}$ and $g^{\mu\alpha}g^{\nu\beta}$ cancels. Thus the gauge field does not ``know'' it lives on an expanding background instead of on a Minkowski background. This is due to the conformally flat nature of the FRW metric. As a result, the large scale fluctuations of the gauge field is suppressed by $k^2/a^2$ and thus not important or observable.

On the other hand, if for some reasons we want the gauge field to have large scale fluctuations, then we have to break the conformal invariance of the field (see, e.g. \cite{Fujita:2012rb}).

\item \textit{Accidental, or other protected massless fields:} In this case, our previous discussion of a spectator field in de Sitter space applies. During each e-fold, the field has a fluctuation $\sigma = H/(2\pi)$. Thus on long time scales, the field has a random walk behavior $\sigma(x) = H \sqrt{Ht}/(2\pi)$.

However, one should note that, it is not easy to have a massless field without a symmetry. This is because, without the protection from symmetry, the quantum corrections generates a mass for the field. For the same reason as the Higgs mass hierarchy problem, the quantum correction for a unprotected mass parameter runs to the UV completion scale of the effective field theory. This generates a mass correction $\Delta m^2\sim \Lambda^4/M_p^2$. If there is supersymmetry (which has to be broken at Hubble scale), the mass correction is brought down to $\Delta m \sim H$.

\item \textit{Field with a small mass $m\ll m_\phi$:} The dynamics of such a light field depends on the field value. If the field is located at the origin of the potential, there is no much difference between the current case and the case of $m=0$. However, as we discussed, because of the random walk behavior, the field cannot naturally stay at the minimum of the potential if there were a large amount of inflation before the observable stage. To estimate the field value, we note that there is a competition between the random walk and a classical roll-back of the field. The classical roll-back can be estimated using the field equation
\begin{align}
  3H\dot\sigma \simeq -V' = -m^2\sigma~.
\end{align}
We may approximate $H$ to be constant (considering that $\sigma$ is not the field that drives inflation, which modifies $H$). In this case, $\sigma$ can be solved as
\begin{align}
  \sigma \propto \exp\left( -\frac{m^2\Delta t}{3H}  \right)~.
\end{align}
Thus at large field value, the field rolls back exponentially fast and cannot random walk. On the other hand, when the exponential factor is small, random walk dominates. Given enough e-folds of inflation (including inflation happened at the non-observable stage), the balance is reached at
\begin{align}
  \Delta t \sim \frac{H}{m^2}~.
\end{align}
During the above time (corresponds to $H^2/m^2$ e-folds), the field randomly walks by a distance
\begin{align}
  \Delta\sigma \sim \frac{H}{2\pi} \times \sqrt{H\Delta t} \sim \frac{H^2}{2\pi m} ~.
\end{align}
This is the typical distance for $\sigma$, away from the origin. On large scales, the very light field forms a curvaton web \cite{Linde:2005yw}. The dynamics of the curvaton web can be analyzed in detail using the Langevin equation and the Fokker-Planck equation. If there were no enough e-folds of inflation, a cutoff of e-folding number needs also to be imposed.

At perturbation level, the present case share similar equations with the massless case. However, due to the non-trivial background motion, the conversion mechanism from entropy perturbation to the comoving curvature perturbation could be very model dependent. The definition of entropy perturbation together with the illustrations of mechanisms are discussed in Sec.~\ref{sec:curv-isoc-pert}.

\item \textit{Field with a small mass $m\sim m_\phi$:} 

This case is again different from the $m\ll m_\phi$ case. This is because the $\sigma$ field now rolls fast enough that it does affect the variation of the Hubble parameter. Thus both fields contribute to the background evolution. In this case we have multi-field inflation.

\item \textit{Field with a mass $m_\phi \ll  m\lesssim H$:} In this case, the $\sigma$ field is heavy compared with the background energy scale -- the inflaton mass. However, $\sigma$ is not heavier than the energy scale of perturbations -- the Hubble scale. Thus the field stays at the origin of the potential at the level of homogeneous and isotropic background, but the perturbations could couple to the inflaton and thus observable. In this case, we have quasi-single field inflation.

Note that if there is minimally broken supersymmetry during inflation, then the standard model fields should get a mass correction of order $H$. This is known as the ``$\eta_V$-problem'' for the inflaton \cite{Copeland:1994vg}, and opens up a variety of possibilities for other fields.

\item \textit{Massive fields with $m\gg H$:} Those fields decouple from the light fields at the tree level. However, considering that inflation is UV-sensitive, it is still an open question if loop corrections would make the imprint from those very massive fields observable in the sky \cite{Chen:2012ye}.

\end{itemize}

\section{Nonlinear Perturbations}

As we have shown in the previous section, linear perturbations already carry interesting information about the physics during inflation. However, the information that linear perturbation carries is quite limited. This is because, for each field, the power spectrum is a function of $k$ only. Even worse, nearly scale invariance brings this function down to a few numbers with increasing difficulties to measure. 

For example, consider the comoving curvature perturbation $\zeta$. Currently we have measured two numbers: the amplitude $P_\zeta$ itself at a given $k$, and the tilt $n_s-1$. If there were no features in the power spectrum, even a third number (the running of spectral index $\alpha_s$) would be very difficult to measure. The gravitational wave sector is even worse. Even the amplitude $P_h$ is not detected yet, leaving only an upper bound.

If the above two numbers are the only things that one could get from inflation (which actually comes directly from the nearly scale invariance of the background, not the detail dynamics), then it is very hard to distinguish between inflation and alternatives, or between inflation models, let alone to use inflation as an ultra high energy laboratory of physics. 

One expects quite a few possibilities to overcome the above lack of prediction problem. For example, a detection of tensor mode will be of much help. We shall also discuss other possibilities such as features and entropy modes in later sections. Here, we consider another class of possibilities -- interactions, which produce non-Gaussian fluctuations in the sky.

\subsection{Interaction picture and in-in formalism}

To calculate the interactions, it is convenient to use the interaction picture\footnote{Historically the advantage of interaction picture perturbation theory versus the old fashioned perturbation theory (Lippmann-Schwinger equation) is that the former explicitly preserves Lorentz invariance. In the cosmological background, there is a preferred time direction anyway. However, it is nontrivial to find the eigenstates which diagonalize the time dependent Hamiltonian for perturbations. This is the major difficulty for Lippmann-Schwinger equation in cosmic perturbation theory.}. Let us start by reviewing how the Hamiltonian formalism and the interaction picture are defined on a time dependent background \cite{Weinberg:2005vy}.

\subsubsection{The Hamiltonian for the perturbations}

Given a theory, we have a Hamiltonian $H[\phi(\mathbf{x},\tau),\pi(\mathbf{x},\tau)]$\footnote{The $\mathbf{x}$ dependence in the Hamiltonian is a functional dependence, because the Hamiltonian has an integration over $\mathbf{x}$. On the other hand, the Hamiltonian inherits implicit time dependence from $\phi(\mathbf{x},\tau)$ and $\pi(\mathbf{x},\tau)$.}\footnote{Here we haven't specify the meaning of $\tau$ yet. Nevertheless, we shall mainly have the conformal time in mind thus we denote it as $\tau$ as we did in previous sections.}, with equal time commutator
\begin{align}
  \left[ \phi_a(\mathbf{x},\tau),\pi_b(\mathbf{y},\tau) \right] = i \delta_{ab}\delta^3(\mathbf{x}-\mathbf{y})~,
  \quad \left[ \phi_a(\mathbf{x},\tau),\phi_b(\mathbf{y},\tau) \right] = 0~,
  \quad \left[ \pi_a(\mathbf{x},\tau),\pi_b(\mathbf{y},\tau) \right] = 0~,
\end{align}
where $a,b$ indices denote different fields or field components. The equation of motion can be written as
\begin{align}\label{eq:h-eom-bg}
  \phi_a' = i \left[ H,\phi_a \right]~, \qquad \pi_a' = i \left[ H,\pi_a \right]~,
\end{align}
where as usual, prime denotes time derivative with respect to the time $\tau$. To do perturbation theory, we need to split $\phi_a$ and $\pi_a$ into background and perturbations. The background is assumed to be c-number valued functions which satisfy the classical equation of motion\footnote{Note that there may be no unique classical background for some inflation models (e.g. \cite{Li:2009sp, Duplessis:2012nb}). In this case, the definition of in-in formalism becomes tricky.}. For our purpose, typically it is convenient to define the background as the homogeneous and isotropic part of the fields, which are the FRW metric and the zero mode of the matter fields. The perturbations $\delta\phi_a(\mathbf{x},\tau)$ and $\delta\pi_a(\mathbf{x},\tau)$ are define by 
\begin{align}
  \phi_a(\mathbf{x},\tau) = \bar \phi_a(\tau) + \delta\phi_a(\mathbf{x},\tau)~, 
  \qquad 
  \pi_a(\mathbf{x},\tau) = \bar \pi_a(\tau) + \delta\pi_a(\mathbf{x},\tau)~.
\end{align}
The commutators for perturbations are not affected by the c-number valued background. Thus
\begin{align}
  \left[ \delta\phi_a(\mathbf{x},\tau),\delta\pi_b(\mathbf{y},\tau) \right] = i \delta_{ab}\delta^3(\mathbf{x}-\mathbf{y})~,
  \quad \left[ \delta\phi_a(\mathbf{x},\tau),\delta\phi_b(\mathbf{y},\tau) \right] = 0~,
  \quad \left[ \delta\pi_a(\mathbf{x},\tau),\delta\pi_b(\mathbf{y},\tau) \right] = 0~,
\end{align}
The Hamiltonian can be expanded with respect to the perturbations as
\begin{align}
  H[\phi,\pi] = H[\bar \phi, \bar \pi] 
  + \sum_a \int d^3x \left.\frac{\delta H[\phi,\pi]}{\delta\phi(\mathbf{x},\tau)}\right|_{\bar \phi, \bar \pi} \delta\phi(\mathbf{x},\tau) 
  + \sum_a \int d^3 x \left.\frac{\delta H[\phi,\pi]}{\delta\pi(\mathbf{x},\tau)}\right|_{\bar \phi, \bar \pi} \delta\pi(\mathbf{x},\tau) 
  + \widetilde H[\delta\phi,\delta\pi; \tau] ~,
\end{align}
where the terms which are linear in $\delta\phi(\mathbf{x},\tau)$ and $\delta\pi(\mathbf{x},\tau)$ do not contribute to perturbation equation of motion, since in the equation of motion level, they are cancelled by the background equation of motion. The term $\widetilde H[\delta\phi,\delta\pi; \tau]$ is the Hamiltonian for perturbations, which is of second and higher orders of $\delta\phi$ and $\delta\pi$. Note that $\widetilde H[\delta\phi,\delta\pi; \tau]$ is explicitly time dependent because now we write it as a functional of $\delta\phi$ and $\delta\pi$, and the time dependence coming from $\bar \phi(t)$ and $\bar \pi(\tau)$ appears as an explicit time dependence.

One can verify the consistency of taking $\widetilde H[\delta\phi,\delta\pi; \tau]$ as the Hamiltonian for perturbations, by calculating from Eq.~(\ref{eq:h-eom-bg}):
\begin{align} \label{eq:time-evolution-from-H}
  {\delta\phi}_a' = i \left[ \widetilde H,\delta\phi_a \right]
  \quad \rightarrow \quad &
  {\delta\phi}_a(\tau) = U^{-1}(\tau,\tau_0) {\delta\phi}_a(\tau_0) U(\tau,\tau_0)~,
\nonumber \\
  {\delta\pi}_a' = i \left[ \widetilde H,\delta\pi_a \right]
  \quad \rightarrow \quad &
  {\delta\pi}_a(\tau) = U^{-1}(\tau,\tau_0) {\delta\pi}_a(\tau_0) U(\tau,\tau_0)~,
\end{align}
where $U(\tau,\tau_0)$ is the time evolution operator, satisfying\footnote{Let $\widetilde H_H(\tau)=\widetilde H[\delta\phi(\tau),\delta\pi(\tau);\tau]$ and $\widetilde H_S(\tau)=\widetilde H[\delta\phi(\tau_0),\delta\pi(\tau_0);\tau]$. The two operators obey $\widetilde H_H=U^{-1}\widetilde H_SU$. Therefore $\partial_\tau U=-i\widetilde H_SU=-iU\widetilde H_H$. The same relation applies to the free evolution operator in \eqref{eq:time-evolution-operator-odep}.
}
\begin{align} \label{eq:time-evolution-operator-ode}
  \partial_\tau U(\tau,\tau_0) = -i \widetilde H[\delta\phi(\tau_0),\delta\pi(\tau_0); \tau] U(\tau,\tau_0)~, \qquad 
  U(\tau_0,\tau_0)=1~.
\end{align}

\subsubsection{The interaction Hamiltonian}
\label{sec:inter-hamilt}

The Hamiltonian $\widetilde H[\delta\phi,\delta\pi; \tau]$ contains quadratic terms and higher order terms in $\delta\phi$ and $\delta\pi$. The quadratic terms result in linear equations of motion, which typically lead to simple solutions. However, the non-quadratic terms in $\widetilde H[\delta\phi,\delta\pi; \tau]$ results in non-linear equations of motion, and in most cases cannot be solved exactly. Perturbation theory is in position to solve this problem.

For this purpose, we split $\widetilde H[\delta\phi,\delta\pi; \tau]$ into
\begin{align}
  \widetilde H[\delta\phi,\delta\pi; \tau] = H_0[\delta\phi,\delta\pi; \tau] + H_I[\delta\phi,\delta\pi; \tau]~,
\end{align}
where $H_I[\delta\phi,\delta\pi; \tau]$ is the interaction Hamiltonian.\footnote{Note that this split is just for the convenience of calculation and thus contains some ambiguities in it, which shouldn't affect physics. For example, in the case of field with modified sound speed (which we shall discuss in detail in later sections), we have
\begin{align}
  \widetilde H = \int \frac{d^3k}{(2\pi)^3}~\frac{a^2}{2}
  \left[\sigma'_{\mathbf{k}}\sigma'_{-\mathbf{k}}+c_s^2k^2\sigma_{\mathbf{k}}\sigma_{-\mathbf{k}}\right]~,
  \quad \sigma'_{\mathbf{k}}\equiv\pi_{\sigma,\mathbf{k}}/a^2
\end{align}
We could solve $\widetilde H[\delta\phi,\delta\pi; \tau]$ directly, which is a straightforward generalization of $c_s=1$ case. Or instead, when $|c_s-1|\ll 1$ we could split $\widetilde H[\delta\phi,\delta\pi; \tau]$ into
\begin{align}
  H_0 = \int \frac{d^3k}{(2\pi)^3}~\frac{a^2}{2}
  \left[\sigma'_{\mathbf{k}}\sigma'_{-\mathbf{k}}+k^2\sigma_{\mathbf{k}}\sigma_{-\mathbf{k}}\right]~,
  \\ H_I = \int \frac{d^3k}{(2\pi)^3}~\frac{a^2}{2}
  k^2(c_s^2-1)\sigma_{\mathbf{k}}\sigma_{-\mathbf{k}}~.
\end{align}
The two methods of calculation agree as they should at zeroth and 1st order of $|c_s-1|$, with the usage of first order perturbation theory.
} 
According to the split of $H_0$ and $H_I$, we define the time evolution of the interaction picture fields ${\delta\phi}^I_a$ as
\begin{align}
  {\delta\phi}^I_a{}' \equiv  i \left[  H_0,\delta\phi^I_a \right]
  \quad \rightarrow \quad &
  {\delta\phi}^I_a(\tau) = U_0^{-1}(\tau,\tau_0) {\delta\phi}^I_a(\tau_0) U_0(\tau,\tau_0)~,
\nonumber \\
  {\delta\pi}^I_a{}' \equiv  i \left[  H_0,\delta\pi^I_a \right]
  \quad \rightarrow \quad &
  {\delta\pi}^I_a(\tau) = U_0^{-1}(\tau,\tau_0) {\delta\pi}^I_a(\tau_0) U_0(\tau,\tau_0)~,
\end{align}
where $U_0(\tau,\tau_0)$ is the time evolution operator of the free theory, which is supposed to know explicitly from a given $H_0$ via
\begin{align} \label{eq:time-evolution-operator-odep}
  \partial_\tau U_0(\tau,\tau_0) = -i H_0[\delta\phi^I(\tau_0),\delta\pi^I(\tau_0); \tau] U_0(\tau,\tau_0)~, \qquad U_0(\tau_0,\tau_0)= 1~.
\end{align}
To fix ${\delta\phi}^I_a$ completely, we have to choose the initial condition as well. Formally, this can be done as in the flat space QFT, by requiring that at some initial time $\tau_0$,
\begin{align}\label{eq:init-int-pict}
  {\delta\phi}^I_a(\tau_0) = {\delta\phi}_a(\tau_0)~, \qquad 
  {\delta\pi}^I_a(\tau_0) = {\delta\pi}_a(\tau_0)~.
\end{align}
Combining Eqs. \eqref{eq:time-evolution-operator-ode} and \eqref{eq:time-evolution-operator-odep} (which implies $\partial_\tau (U_0^{-1}) = i U_0^{-1}H_0$), and making use of \eqref{eq:init-int-pict}, we have
\begin{align}
  \partial_\tau [U_0^{-1}(\tau,\tau_0)U(\tau,\tau_0)] 
  = -i U_0^{-1}(\tau,\tau_0) H_I[\delta\phi(\tau_0), \delta\pi(\tau_0); \tau] U(\tau,\tau_0)
  = -i H_I[\delta\phi^I(\tau), \delta\pi^I(\tau); \tau] U_0^{-1}(\tau,\tau_0) U(\tau,\tau_0)~.
\end{align}
As one can directly check, the solution of this equation is
\begin{align} \label{eq:solution-F}
  F(\tau,\tau_0) \equiv  & U_0^{-1}(\tau,\tau_0)U(\tau,\tau_0) 
  \nonumber \\
  = & 1 + (-i) \int_{\tau_0}^\tau d\tau_1 ~ H_I(\tau_1)
  + (-i)^2 \int_{\tau_0}^\tau d\tau_1 \int_{\tau_0}^{\tau_1} d\tau_2 ~ H_I(\tau_1) H_I(\tau_2)    
  \nonumber \\ &
  + (-i)^3 \int_{\tau_0}^\tau d\tau_1 \int_{\tau_0}^{\tau_1} d\tau_2 \int_{\tau_0}^{\tau_2} d\tau_3 ~ 
    H_I(\tau_1) H_I(\tau_2) H_I(\tau_3) + \cdots
  \nonumber \\
  = & T e^{-i \int_{\tau_0}^\tau H_I(\tau')d\tau'}~.
\end{align}
In the last line, the time ordered integrals have been compactly written into a time ordered exponential. However the exponential still needs to be calculated order by order in the form of the second and third line.

Now an operator in Heisenberg picture can be expressed via the interaction picture field as
\begin{align} \label{eq:rel-hei-int}
  Q(\tau) = F^{-1}(\tau,\tau_0)Q^I(\tau) F(\tau,\tau_0) = 
  \left[  \bar T e^{i \int_{\tau_0}^\tau H_I(\tau')d\tau'}\right]  Q^I(\tau) 
  \left[  T e^{-i \int_{\tau_0}^\tau H_I(\tau')d\tau'}\right]  ~,
\end{align}
where $\bar T$ is anti-time ordering and $H_I(\tau)$ denotes $H_I[\delta\phi^I(\tau),\delta\pi^I(\tau);\tau]$. 

\subsubsection{The time evolution of the vacuum}

As in flat space QFT, $F(\tau,\tau_0)$ is the time evolution operator of the quantum states in the interaction picture. In the aspect of quantum states, there is one state of particular importance: the ``vacuum'' state. 

We have discussed the vacuum state in section \ref{sec:spectator-field-de}. This discussion is reused for a few times for the inflationary scalar and tensor perturbations. However, here new ingredients come in, because of the interactions. In flat space QFT, there is already a difference between the interaction vacuum and free vacuum. The most part of this section is parallel to flat space QFT. Comments are made when the expansion of the background brings a difference.

Let us use $|0\rangle$ to denote the vacuum state of the free theory, which is defined by annihilation operators. We use $|\Omega\rangle$ to denote the interaction vacuum, which is to be constructed using $|0\rangle$. As discussed in section \ref{sec:spectator-field-de}, it is important to note that the vacuum is the lowest energy state. 

We first expand $|0\rangle$ using a complete set of energy eigenstates of the full theory. On expanding background matter energy is not conserved and we have yet to define gravitational energy. Nevertheless, we could consider the energy eigenstates at an initial time for the perturbation modes.\footnote{Strictly speaking this requires all modes to be sub-Hubble. However, during inflation there are always Hubble-crossing and super-Hubble modes. Also, the hierarchy between Hubble scale $H$ and Planck scale $M_p$ may not be large enough to contain all those modes safely.} Near $\tau_0\rightarrow-\infty$, we have
\begin{align}
  e^{-i H (\tau-\tau_0)}|0\rangle = \sum_n e^{-i H (\tau-\tau_0)} |n\rangle \langle n|0\rangle = \sum_n e^{-i E_n (\tau-\tau_0)} |n\rangle \langle n|0\rangle
  = e^{-i E_\Omega (\tau-\tau_0)} |\Omega\rangle \langle\Omega|0\rangle + \sum_{n'} e^{-i E_{n'} (\tau-\tau_0)} |n'\rangle \langle n'|0\rangle ~,
\end{align}
where in the last line $n'$ denotes the states excluding the ground state. Now we add a small imaginary part to time:
\begin{align}
  \tau \rightarrow \widetilde \tau = \tau (1 - i \epsilon)~.
\end{align}
Near $\tau_0\rightarrow-\infty$, the $i\epsilon$ factor introduces a term $e^{-\infty\times\epsilon E_n}$. Thus only the $|\Omega\rangle$ state survives from this $i\epsilon$ factor.\footnote{There are a few assumptions here:
\begin{itemize}
\item $\langle\Omega|0\rangle\neq 0$. 
  This may be fine considering $H_I$ is a small perturbation. 
  However things may get worse during a long time of interaction.
\item If $\langle\Omega|0\rangle$ grows with time, the growth near $\tau\rightarrow-\infty$ is slower than exponential.
\item It is good to interpret $\tau$ as conformal time. Because if we interpreted the $\tau$ here as proper time $t$, there is no limit $t\rightarrow-\infty$. However, during inflation the energy scale $H$ drops slowly. This indicates that $Ht\simeq \mathrm{const}\times t$ may be a better measure of time than $\tau\propto -e^{-Ht}$.
\end{itemize}
Unfortunately, we know very little about the interaction vacuum. Thus it is hard to explicitly verify those assumptions.
}
In the vacuum-projection limit, the normalized interaction-picture state is, up to an overall phase,
\begin{align} \label{eq:solve-vacuum}
  |\Omega(\tau)\rangle_I =
  \frac{F(\widetilde\tau,\widetilde\tau_0)|0\rangle}{\sqrt{\mathcal N_\epsilon}}~,
  \qquad \mathcal N_\epsilon\equiv
  \langle0|F^\dagger(\widetilde\tau,\widetilde\tau_0)F(\widetilde\tau,\widetilde\tau_0)|0\rangle~.
\end{align}
From now on, we shall neglect the tilde on $\tau$. In other words the $\tau$ in later sections should be understood with a small imaginary part.

\subsubsection{The in-in expectation value}

\begin{figure}[htbp]
  \centering
  \includegraphics[width=0.8\textwidth]{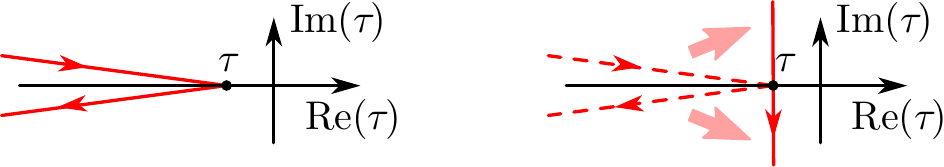}
  \caption{\label{fig:in-in-contour} The left panel is the contour of integration in the in-in formalism. In the right panel, the contour is deformed to a vertical line running top-down. The deformed contour is used numerically in \cite{Chen:2009zp, Chen:2012ge} and more generally in \cite{Behbahani:2012be, Green:2013rd}.}
\end{figure}

Now combine \eqref{eq:rel-hei-int} and \eqref{eq:solve-vacuum}, the time dependent expectation value of an operator can be written as
\begin{align}
  \langle\Omega|Q(\tau)|\Omega\rangle = \frac
  {\langle 0|
    \left[  \bar T e^{i \int_{\tau_0}^\tau H_I(\tau')d\tau'}\right]  Q^I(\tau) 
    \left[  T e^{-i \int_{\tau_0}^\tau H_I(\tau')d\tau'}\right]
  |0\rangle}
  {\mathcal N_\epsilon} ~.
\end{align}
Note that in the anti-time ordering $[\bar T \cdots]$, the $i\epsilon$ prescription is actually with a different sign (i.e. $\widetilde \tau \rightarrow \tau(1+i\epsilon)$) from the time ordering part $[T \cdots]$. This is because the $\bar T$ part is the Hermitian conjugate of $T$ part, which flips the sign of $i\epsilon$ term. This contour of integration, specified by the above $i\epsilon$ prescription, is known as the in-in contour (Fig.~\ref{fig:in-in-contour}).

Taking $Q(t)$ to be the identity, the normalization is
\begin{align}
  \mathcal N_\epsilon = \langle 0|
    \left[  \bar T e^{i \int_{\tau_0}^\tau H_I(\tau')d\tau'}\right]   
    \left[  T e^{-i \int_{\tau_0}^\tau H_I(\tau')d\tau'}\right]
  |0\rangle~.
\end{align}
This denominator is the norm of the vacuum projected along the complex-time contour. The projection is not unitary, so this norm need not be one. The normalized master formula, known as the factorized form of in-in formalism, is
\begin{align}\label{eq:in-in}
  \langle\Omega|Q(\tau)|\Omega\rangle =
  \frac{\langle 0|
    \left[\bar T e^{i\int_{\tau_0}^{\tau} H_I(\tau')d\tau'}\right] Q^I(\tau)
    \left[T e^{-i\int_{\tau_0}^{\tau} H_I(\tau')d\tau'}\right]
  |0\rangle}
  {\langle 0|
    \left[\bar T e^{i\int_{\tau_0}^{\tau} H_I(\tau')d\tau'}\right]
    \left[T e^{-i\int_{\tau_0}^{\tau} H_I(\tau')d\tau'}\right]
  |0\rangle}~.
\end{align}
The denominator cancels vacuum-disconnected diagrams. In the perturbative expressions and contour variants below, these diagrams are understood to be removed, implementing this normalization.
Practically one expands the above formula order by order in perturbation theory. For example, the first order term in the expansion is
\begin{align}
  \langle\Omega|Q(\tau)|\Omega\rangle_1 = 2\mathrm{Im} \int_{\tau_0}^{\tau} d\tau_1 \langle0| Q^I(\tau) H_I(\tau_1) |0\rangle~,
\end{align}
and the second order terms are
\begin{align}
  \langle\Omega|Q(\tau)|\Omega\rangle_2 = \int_{\tau_0}^\tau d\tau_1 \int_{\tau_0}^\tau d\tau_2 \langle0| H_I(\tau_1) Q^I(\tau) H_I(\tau_2) |0\rangle
  - 2\mathrm{Re} \int_{\tau_0}^\tau d\tau_1 \int_{\tau_0}^{\tau_1}d\tau_2 \langle0| Q^I(\tau) H_I(\tau_1) H_I(\tau_2) |0\rangle ~.
\end{align}

However, one should note that the order in $H_I$ does not necessarily agree with the $\mathcal{O}(\zeta^n)$ expansion. For example, when calculating the four-point function of general single field inflation, both the first order expansion of in-in formalism (with $H_I$ being the $\mathcal{O}(\zeta^4)$ vertex)  and the second order expansion of in-in contributes, with comparable amount \cite{Chen:2009bc}.

Before moving on, let us note that there are a few variants of the in-in formula \eqref{eq:in-in}:

\begin{itemize}

\item \textit{The Wick-rotated factorized form} \cite{Behbahani:2012be, Green:2013rd}: In the right panel of Fig.~\ref{fig:in-in-contour}, if there are no poles or cuts between the dashed contour and the solid contour, both contours give the same result. Using the solid contour, the in-in formalism can be written into a more compact form:
\begin{align}\label{eq:in-in-wick}
  \langle\Omega|Q(\tau)|\Omega\rangle & = 
  \langle 0| \bar{T} \left[  Q^I(\tau) 
    \exp \left( i \int_{-i\infty+ \tau}^{i\infty+\tau}H_I(\tau_E)d\tau_E \right) \right]   
  |0\rangle
  \nonumber\\
  & =
  \langle 0| \bar{T} \left[  Q^I(\tau) 
    \exp \left( - \int_{-\infty}^{\infty}H_I(\tau+i\tau_E)d\tau_E \right) \right]   
  |0\rangle ~,
\end{align}
where in the second line a further change of variable is applied to simplify the integral. The Wick-rotated factorized form makes the (linear) UV convergence of in-in integral more explicitly.

\item \textit{The commutator form} \cite{Weinberg:2005vy}: Using a common real-time contour and Hermitian evolution with the same prepared initial state, Eq.~\eqref{eq:in-in} can be written as
\begin{align}\label{eq:in-in-comm}
  \langle\Omega|Q(\tau)|\Omega\rangle = \sum_{n=0}^\infty  i^n 
  \int_{\tau_0}^{\tau} d\tau_1 \int_{\tau_0}^{\tau_1} d\tau_2 \cdots \int_{\tau_0}^{\tau_{n-1}} d\tau_n
  \left\langle [H_I(\tau_n), [H_I(\tau_{n-1}), \cdots,
  [H_I(\tau_1),Q^I(\tau)]\cdots]]\right\rangle~.
\end{align}
To see this, one first observe the $n=0$ order term agrees with \eqref{eq:in-in}. Then assuming the $(n-1)$th order term also agrees, the time derivative of the $(n-1)$th order terms generates the agreement of the $n$th order terms.

The advantage of \eqref{eq:in-in-comm} is that, there are some cancellations in the IR (i.e. $\tau\rightarrow0$) from the commutators, which are not obvious in the form of \eqref{eq:in-in}. For example, the factorized terms in the second-order expansion of \eqref{eq:in-in} could have cancellations between each other in the IR.

UV convergence of \eqref{eq:in-in-comm} requires an adiabatic regulator compatible with this Hermitian evolution and the chosen initial state.

Similarly, the commutator form can also be written as
\begin{align}
  \langle\Omega|Q(\tau)|\Omega\rangle = \sum_{n=0}^\infty  i^n 
  \int_{\tau_0}^{\tau} d\tau_1 \int_{\tau_1}^{\tau} d\tau_2 \cdots \int_{\tau_{n-1}}^{\tau} d\tau_n
  \left\langle [H_I(\tau_1), [H_I(\tau_{2}), \cdots,
  [H_I(\tau_n),Q^I(\tau)]\cdots]]\right\rangle~.  
\end{align}

\item \textit{The mixed form} \cite{Chen:2009zp}: To keep both the advantage of \eqref{eq:in-in} in the UV and \eqref{eq:in-in-comm} in the IR, one can choose an intermediate time $\tau_c$ (which is typically $|k \tau_c|\sim 1$) and re-group the terms before $\tau_c$ using the factorized form and the terms after $\tau_c$ using the commutator form as
\begin{align}
  \sum_{i=0}^n \int_{\tau_c}^\tau d\tau_1 \cdots \int_{\tau_c}^{\tau_{i-1}} d\tau_i ~(\text{commutator form})
  \times \int_{-\infty}^{\tau_c} d\tau_{i+1} \cdots \int_{-\infty}^{\tau_{n-1}} d\tau_n ~(\text{factorized form})~.
\end{align}
The result is independent on the intermediate time $\tau_c$, and well behaved both in the UV and in the IR.

\end{itemize}

\subsection{3rd order gravitational action}

We now use the in-in formalism to calculate the three point function for inflation. For the minimal model of inflation, the three point function is small. On the other hand it is dominated by pure gravitational actions, instead of $V'''(\phi)$ from the explicit non-linearity in the inflationary potential.

To calculate this, we need to derive the third order gravitational action. As an example, here we only do the calculation in $\zeta$-gauge and using the ADM form of Ricci scalar.

For simplicity, here we take the Planck unit: $M_p=1$.

\subsubsection{Constraints}
For calculating the three point function, we have to solve the constraint equations. Naively, we would expect that we have to solve the constraint equations up to third order in $\zeta$. However, fortunately, we still only need to solve the constraints up to first order. 

In general, to derive the second order action, one needs to solve the first order constraint equations. To derive the $n$th order action ($n\geq 3$), it is sufficient to solve the constraint equations through order $n-2$ \cite{Maldacena:2002vr}.

To see this, recall that there are no time derivatives acting on constraint variables.\footnote{if one used the 4-dimensional Ricci scalar for calculation, there exist time derivatives on constraint variables. But those time derivatives can be eliminated by integration by parts.} Also the spatial derivatives on constraint variables can be converted to momentum $k$ by going into momentum space (to save writing, we shall not do it explicitly). Taking the lapse function for example, in general we have
\begin{align}
  S = \int d^4x ~ \mathcal{L}(N)~, \qquad 
  \delta S = \int d^4x ~ \frac{\partial\mathcal{L}}{\partial N} \delta N = 0~.
\end{align}
Let us expand $N = N_0 + N_1 + \cdots + N_n + \cdots$, where $N_n$ denotes the $\mathcal {O}(\zeta^n)$ order terms. The constraint equation $\partial\mathcal{L}/\partial N=0$ can be expanded (up to $\mathcal{O}(\zeta)$) as
\begin{align}\label{eq:cons1}
  \left. \frac{\partial \mathcal{L}}{\partial N} \right|_{0,\zeta^0} = 0 ~,
\end{align}
\begin{align}\label{eq:cons2}
  \left. \frac{\partial \mathcal{L}}{\partial N} \right|_{0,\zeta^1}
  + {N_1} \left. \frac{\partial^2 \mathcal{L}}{\partial N^2} \right|_{0,\zeta^0} = 0~,
\end{align}
where the subscript ${0,\zeta^a}$ denotes the term does not contain $N$, and is $a$th order in $\zeta$ (or the shift vector).

Now consider the $n$th order action ($n\geq 3$). All the terms that contains $N_n$ are
\begin{align}
  S_n \supset  \int  d^4x ~ 
  \left. N_n \frac{\partial \mathcal{L}}{\partial N} \right|_{0,\zeta^0} =0 ~.
\end{align}
This term vanishes because of \eqref{eq:cons1}. All the terms that contains $N_{n-1}$ are
\begin{align}
  S_n \supset  \int  d^4x ~ 
   N_{n-1} \left(
     \left.\frac{\partial \mathcal{L}}{\partial N} \right|_{0,\zeta^1} 
     + N_1 \left. \frac{\partial^2 \mathcal{L}}{\partial N^2} \right|_{0,\zeta^0} 
   \right) =0 ~.
\end{align}
Again this vanishes because of \eqref{eq:cons2}. However, from $N_{n-2}$ on, the coefficients in the constraint equation no longer coincides with the coefficients in $S_n$. The above proof also goes through for the shift vector $N_i$. Thus solving the constraints through order $\mathcal{O}(\zeta^{n-2})$ is sufficient. Now we want to calculate $S_3$. Thus Eq.~\eqref{eq:cons-zeta-1} is enough for solving the constraints (and $b_i=0$ at this order).

\subsubsection{Action}

The second order action is already given in \eqref{eq:S2-ADM-zeta}. The third order action takes the form \cite{Maldacena:2002vr}:
\begin{align} \label{eq:s3-zetab}
  S_3 = & \int  dt d^3x ~ a^3 
  \Big\{
    3H^2\alpha^3 - \epsilon H^2\alpha^3 - 9H^2\alpha^2\zeta + 3\epsilon H^2\alpha^2\zeta + \frac{27}{2} H^2\alpha\zeta^2
    - \frac{9}{2} \epsilon H^2\alpha\zeta^2 - \frac{27}{2} H^2\zeta^3 + \frac{9}{2} \epsilon H^2\zeta^3
    - 6H\alpha^2\dot\zeta 
\nonumber\\ &
    + 18H\alpha\zeta\dot\zeta - 27H\zeta^2\dot\zeta + 3\alpha(\dot\zeta)^2 - 9\zeta (\dot\zeta)^2 
    - \frac{2}{a^2} H\alpha b_i\partial_i\zeta + \frac{2}{a^2} H\zeta b_i\partial_i\zeta
    - \frac{2}{a^2} H\alpha\partial_i\beta\partial_i\zeta + \frac{2}{a^2} H\zeta\partial_i\beta\partial_i\zeta
    + \frac{2}{a^2} b_i \dot\zeta \partial_i\zeta 
\nonumber\\ &
    + \frac{2}{a^2} \partial_i\beta \dot\zeta \partial_i\zeta
    - \frac{1}{a^2} \alpha (\partial_i\zeta)^2 - \frac{1}{a^2} \zeta (\partial_i\zeta)^2
    - \frac{1}{a^4} b_j \partial_i\zeta\partial_jb_i - \frac{1}{a^4} \partial_j\beta \partial_i\zeta\partial_jb_i
    - \frac{1}{a^4} b_i\partial_j\zeta\partial_jb_i
    - \frac{1}{a^4} \partial_i\beta\partial_j\zeta\partial_jb_i 
\nonumber\\ &
    - \frac{1}{4a^4} \alpha (\partial_jb_i)^2
    - \frac{1}{4a^4} \zeta (\partial_jb_i)^2 
    - \frac{1}{4a^4} \alpha \partial_jb_i \partial_ib_j
    - \frac{1}{4a^4} \zeta \partial_jb_i \partial_ib_j
    + \frac{2}{a^2} H\alpha^2\partial^2\beta - \frac{2}{a^2} H\alpha\zeta\partial^2\beta 
    + \frac{1}{a^2} H\zeta^2\partial^2\beta 
\nonumber\\ &
    - \frac{2}{a^2} \alpha\dot\zeta\partial^2\beta
    + \frac{2}{a^2} \zeta\dot\zeta\partial^2\beta - \frac{2}{a^4} b_j \partial_i\zeta\partial_i\partial_j\beta
    - \frac{2}{a^4} \partial_j\beta \partial_i\zeta\partial_i\partial_j\beta - \frac{1}{a^4} \alpha\partial_jb_i\partial_i\partial_j\beta
    - \frac{1}{a^4} \zeta \partial_jb_i\partial_i\partial_j\beta - \frac{1}{2a^4} \alpha (\partial_i\partial_j\beta)^2
\nonumber\\ &
    - \frac{1}{2a^4} \zeta (\partial_i\partial_j\beta)^2 + \frac{1}{2a^4} \alpha (\partial^2\beta)^2
    + \frac{1}{2a^4} \zeta (\partial^2\beta)^2 - \frac{2}{a^2} \alpha\zeta\partial^2\zeta 
    - \frac{1}{a^2} \zeta^2\partial^2\zeta - \frac{9}{2} V \alpha\zeta^2 - \frac{9}{2} V \zeta^3
  \Big\}~.
\end{align}
Note that we can set $b_i=0$. This action is simplified into
\begin{align} \label{eq:s3-zeta}
  S_3 = & \int  dt d^3x ~ a^3 
  \Big\{
    3H^2\alpha^3 - \epsilon H^2\alpha^3 - 9H^2\alpha^2\zeta + 3\epsilon H^2\alpha^2\zeta + \frac{27}{2} H^2\alpha\zeta^2
    - \frac{9}{2} \epsilon H^2\alpha\zeta^2 - \frac{27}{2} H^2\zeta^3 + \frac{9}{2} \epsilon H^2\zeta^3
    - 6H\alpha^2\dot\zeta 
\nonumber\\ &
    + 18H\alpha\zeta\dot\zeta - 27H\zeta^2\dot\zeta + 3\alpha(\dot\zeta)^2 - 9\zeta (\dot\zeta)^2 
    - \frac{2}{a^2} H\alpha\partial_i\beta\partial_i\zeta + \frac{2}{a^2} H\zeta\partial_i\beta\partial_i\zeta
\nonumber\\ &
    + \frac{2}{a^2} \partial_i\beta \dot\zeta \partial_i\zeta
    - \frac{1}{a^2} \alpha (\partial_i\zeta)^2 - \frac{1}{a^2} \zeta (\partial_i\zeta)^2
    + \frac{2}{a^2} H\alpha^2\partial^2\beta - \frac{2}{a^2} H\alpha\zeta\partial^2\beta 
    + \frac{1}{a^2} H\zeta^2\partial^2\beta 
\nonumber\\ &
    - \frac{2}{a^2} \alpha\dot\zeta\partial^2\beta
    + \frac{2}{a^2} \zeta\dot\zeta\partial^2\beta
    - \frac{2}{a^4} \partial_j\beta \partial_i\zeta\partial_i\partial_j\beta
    - \frac{1}{2a^4} \alpha (\partial_i\partial_j\beta)^2
\nonumber\\ &
    - \frac{1}{2a^4} \zeta (\partial_i\partial_j\beta)^2 + \frac{1}{2a^4} \alpha (\partial^2\beta)^2
    + \frac{1}{2a^4} \zeta (\partial^2\beta)^2 - \frac{2}{a^2} \alpha\zeta\partial^2\zeta 
    - \frac{1}{a^2} \zeta^2\partial^2\zeta - \frac{9}{2} V \alpha\zeta^2 - \frac{9}{2} V \zeta^3
  \Big\}~.
\end{align}

\subsubsection{Integration by parts}

Inserting the constraint Eq.~\eqref{eq:cons-zeta-1} into the third order action \eqref{eq:s3-zeta}, and do a lot of integration by parts\footnote{Doing the integration by parts is quite difficult. The readers can save some time by expanding the boundary terms in Eq.~\eqref{eq:s3-bdry}, to keep track of what integrations by parts we have performed.}, we get
\begin{align} \label{eq:s3-solved}
  S_3 = \int dt d^3x \Big\{
   & 
   a^3 \epsilon^2 \zeta\dot\zeta^2 + a\epsilon^2 \zeta (\partial\zeta)^2 - 2a\epsilon \dot\zeta \partial\zeta\partial\chi
   \nonumber\\ &
   + \frac{a^3\epsilon}{2}\dot\eta\zeta^2\dot\zeta + \frac{\epsilon}{2a} \partial\zeta\partial\chi\partial^2\chi
   + \frac{\epsilon}{4a} \partial^2\zeta (\partial\chi)^2
   \nonumber\\ &
   + 2f(\zeta) \left. \frac{\delta\mathcal{L}}{\delta\zeta} \right|_{1} + \mathcal{L}_b
  \Big\}~,
\end{align}
where $\chi$ is defined by
\begin{align}
  \partial^2\chi \equiv a^2 \epsilon \dot\zeta~,
\end{align}
thus the first line of \eqref{eq:s3-solved} is $\mathcal{O}(\epsilon^2)$ in slow roll and the second line is $\mathcal{O}(\epsilon^3)$. In the third line, $\left.\frac{\delta\mathcal{L}}{\delta\zeta}\right|_{1}\equiv\partial_t(a^3\epsilon\dot\zeta)-a\epsilon\partial^2\zeta=-\frac12\delta S_2/\delta\zeta$ vanishes on the free solutions used for the tree-level three point function\footnote{Note that we are doing perturbation theory, thus it is allowed to use the linear equation of motion, when calculating the third or higher order actions. On the other hand, one could insist not to use the equation, but instead do field redefinition $\zeta_n \equiv \zeta + \mathcal{O}(\zeta^2)$ to eliminate $\left.\frac{\delta\mathcal{L}}{\delta\zeta} \right|_{1}$. In this case, all the $\ddot\zeta$ terms can be eliminated without using equation of motion, and there is no boundary term contributing to the action, in terms of $\zeta_n$.}. Just for keeping track, $f(\zeta)$ is
\begin{align}
  f(\zeta) = \frac{\eta}{4}\zeta^2 + \frac{1}{H} \zeta\dot\zeta 
  + \frac{-\partial\zeta\partial\zeta+\partial^{-2}[\partial_i\partial_j(\partial_i\zeta\partial_j\zeta)]}{4a^2H^2} 
  + \frac{\partial\zeta\partial\chi-\partial^{-2}[\partial_i\partial_j(\partial_i\zeta\partial_j\chi)]}{2a^2H} ~.
\end{align}
And $\mathcal{L}_b$ is a collection of boundary terms\footnote{In calculation, one does not need to write down all of them explicitly. In this way it is clear to see which integrations by parts we have performed. On the other hand, for the purpose of calculation, only the boundary terms that both (a) contain $\dot\zeta$ on the boundary (i.e. $\ddot\zeta$ in the bulk), and (b) non-vanishing on the boundary, need to be considered. }:
\begin{align}\label{eq:s3-bdry}
 \mathcal L_b=\partial_tB+\partial_iJ_i~.
\end{align}
The time boundary is
\begin{align}\label{eq:s3-bdry-time}
  B = &
  \Big[
    -9a^3H\zeta^3 + \frac{a}{H} \zeta(\partial\zeta)^2 - \frac{1}{4aH^3} (\partial\zeta)^2\partial^2\zeta
    - \frac{a\epsilon}{H} \zeta(\partial\zeta)^2 - \frac{\epsilon a^3}{H} \zeta\dot\zeta^2
    + \frac{1}{2aH^2} \zeta(\partial_i\partial_j\zeta\partial_i\partial_j\chi - \partial^2\zeta\partial^2\chi)
    \nonumber \\ & \qquad
    - \frac{\eta a}{2} \zeta^2\partial^2\chi 
    - \frac{1}{2aH} \zeta(\partial_i\partial_j\chi\partial_i\partial_j\chi-\partial^2\chi\partial^2\chi)
  \Big]
\end{align}

The spatial current is
\begin{align}\label{eq:s3-bdry-space}
J_i={}&aH\zeta^2\partial_i\chi+\frac a2(\epsilon\eta-4)\zeta^2\partial_i\zeta
-\frac{2a}{H}\zeta\dot\zeta\partial_i\zeta
\nonumber\\ &
+\frac{a\epsilon}{2H^2}(\zeta\dot\zeta\partial_i\dot\zeta-\dot\zeta^2\partial_i\zeta)
+\frac{\epsilon-3}{4aH^2}(\partial\zeta)^2\partial_i\zeta
-\frac{\epsilon}{4a}(\partial\chi)^2\partial_i\zeta
\nonumber\\ &
+\frac{(\partial\zeta)^2}{aH}\partial_i\chi+\frac{(\partial\zeta)^2}{4aH^3}\partial_i\dot\zeta
+\frac{\zeta+\dot\zeta/H}{2aH^2}(\partial_i\zeta\partial^2\zeta-\partial_j\zeta\partial_i\partial_j\zeta)
\nonumber\\ &
+\frac{\epsilon\zeta}{aH}(\partial_i\chi\partial^2\zeta-\partial_j\chi\partial_i\partial_j\zeta)
+\frac{3\zeta+\dot\zeta/H}{2aH}\partial_j\zeta\partial_i\partial_j\chi
\nonumber\\ &
-\frac{\zeta}{2aH^2}\partial_j\dot\zeta\partial_i\partial_j\chi
+\frac{\zeta}{2aH}\partial_i\zeta\partial^2\chi
+\frac{2-\epsilon}{2a}\zeta(\partial_i\chi\partial^2\chi-\partial_j\chi\partial_i\partial_j\chi)
\nonumber\\ &
+\frac{\zeta}{2aH^2}\left[(\partial_i\zeta-2H\partial_i\chi)\partial^2\dot\chi
-(\partial_j\zeta-2H\partial_j\chi)\partial_i\partial_j\dot\chi\right]
\nonumber\\ &
-\frac{1}{2aH^2}\left[\mathcal P\partial_i\mathcal D-\mathcal D\partial_i\mathcal P
+\mathcal D\partial_j\mathcal K_{ij}-(\partial_j\mathcal D)\mathcal K_{ij}\right]~,
\end{align}
where
\begin{align}
 \mathcal D&\equiv\dot\chi+H\chi-\epsilon\zeta~,\qquad
 \mathcal P\equiv\partial^{-2}\partial_i\partial_j\mathcal K_{ij}~,
 \nonumber\\
 \mathcal K_{ij}&\equiv\partial_i\zeta\partial_j\zeta
 -H(\partial_i\zeta\partial_j\chi+\partial_i\chi\partial_j\zeta)~.
\end{align}
The spatial part $\partial_i[\cdots]$ do not contribute to the three point function. Because there is no spatial boundary in our inflationary universe. However, we have to be more careful about the time total derivative. This is because for in-in formalism, we do have a time boundary, typically chosen at a time where $\zeta$ becomes conserved after that. In the $a\rightarrow\infty$ limit, only one of these time-boundary terms contributes to the tree-level three point function at leading order in slow roll\footnote{It appears that this boundary term contains $\zeta''$ when expanded in the bulk, and thus the variation principle is not well defined. However, actually there is no problem because we are allowed to use linear equations of motion in the bulk. If we choose not to use the linear equations of motion, but instead perform field redefinitions to get rid of high derivative terms in the 3rd order action, then this boundary term does not appear \cite{Maldacena:2002vr} .} \cite{Arroja:2011yj, RenauxPetel:2011sb} :
\begin{align}
  \mathcal{L}_b \supset \partial_t\left( -\frac{\epsilon\eta}{2} a^3\zeta^2\dot\zeta \right)~.
\end{align}
To conclude, the third order Lagrangian, which contributes to the three point function is
\begin{align}\label{eq:l3}
  \mathcal{L}_3 = &
   a^3 \epsilon^2 \zeta\dot\zeta^2 + a\epsilon^2 \zeta (\partial\zeta)^2 - 2a\epsilon \dot\zeta \partial\zeta\partial\chi 
    + \partial_t\left( -\frac{\epsilon\eta}{2} a^3\zeta^2\dot\zeta \right) 
   \nonumber\\ &
   + \frac{a^3\epsilon}{2}\dot\eta\zeta^2\dot\zeta + \frac{\epsilon}{2a} \partial\zeta\partial\chi\partial^2\chi
   + \frac{\epsilon}{4a} \partial^2\zeta (\partial\chi)^2
\end{align}
where the first line dominates over the second line in slow roll approximation (note $\chi$ is proportional to $\epsilon$).

Before proceeding, it is important to note that, the dominant contribution to $\mathcal{L}_3$ come from gravitational interactions, instead of the inflaton's direct coupling. To see this, consider the inflaton's direct coupling $V''' \delta\phi^3$. The third derivative $V'''$ can be estimated as
\begin{align}
  V'\simeq -3H\dot\phi~, \quad \partial_t(V') = V''\dot\phi \sim H^2\dot\phi\times \mathcal{O}(\epsilon)~,\quad
  \partial_t(V'') = V'''\dot\phi \sim H^3 \times \mathcal{O}(\epsilon^2)~,
\end{align}
where $\mathcal{O}(\epsilon)$ collectively denote order of slow roll parameters. Thus
\begin{align}\label{eq:Vppp-sr}
  V''' \sim \frac{H^3}{\dot\phi} \times \mathcal{O}(\epsilon^2) \sim H \mathcal{O}(\epsilon^2)\sqrt{P_\zeta}~.
\end{align}
Gauge transform $V''' \delta\phi^3$ into $\zeta$-gauge, using $\delta\phi = -\dot\phi \zeta / H$, we have the Lagrangian for inflaton self interaction in the $\zeta$-gauge:
\begin{align}
  \mathcal{L}_3 \supset - V''' \frac{\dot\phi^3}{H^3} \zeta^3 = \mathcal{O}(\epsilon^2)\times \dot\phi^2 \zeta^3 = H^2\mathcal{O}(\epsilon^3) \zeta^3~.
\end{align}
This belongs to the second line of \eqref{eq:l3}, which is subdominant. In other words, gravity is not the weakest force during inflation. The inflaton self interaction is weaker (although it is not a gauge interaction, in which case gravity is conjectured to be the weakest force \cite{ArkaniHamed:2006dz, Huang:2007mf}).

\subsection{From action to Hamiltonian}
As in classical mechanics, a Legendre transform 
$H=\int d^3x\left(\pi\dot\phi-\mathcal{L}\right)$ brings Lagrangian to Hamiltonian. Especially, potential energy $V(\phi)$ flips sign in the Legendre transform.

For the gravitational action, the situation is a bit non-trivial because there is derivative coupling. Fortunately, the third order interaction Hamiltonian is still just flipping the sign of the third order action:
\begin{align}\label{eq:h3}
  H_3 = -\int d^3x \mathcal{L}_3~.
\end{align}
We shall derive the above equation, together with a non-trivial relation for the 4th order relation between $H_4$ and $\mathcal{L}_4$ \cite{Huang:2006eha}. Readers who are not interested in this detail are suggested to skip the remainder of this subsection.

We consider a Lagrangian up to 4th order in perturbations:
\begin{align}
  \mathcal{L} = & f^{(0)}_{ab} \dot\alpha^a \dot\alpha^b + j^{(2)}
  \nonumber\\ &
  + g^{(0)}_{abc} \dot\alpha^a \dot\alpha^b \dot\alpha^c
  + g^{(1)}_{ab} \dot\alpha^a \dot\alpha^b + g^{(2)}_{a} \dot\alpha^a + j^{(3)}
  \nonumber\\ &
  + h^{(0)}_{abcd}\dot\alpha^a\dot\alpha^b\dot\alpha^c\dot\alpha^d
  + h^{(1)}_{abc} \dot\alpha^a \dot\alpha^b \dot\alpha^c
  + h^{(2)}_{ab} \dot\alpha^a \dot\alpha^b + h^{(3)}_{a} \dot\alpha^a + j^{(4)}~,
\end{align}
where $\alpha^a$ are the dynamical fields where $a$ runs over the number of those dynamical fields. The coefficients $f^{(n)},g^{(n)},h^{(n)},j^{(n)}$ are functions of $\alpha^a$ but do not contain $\dot\alpha^a$. The $n$ index in the parenthesis denotes the coefficient is $n$th order in $\alpha$.

First one define the conjugate momentum $\pi_a$:
\begin{align}
  \pi_a \equiv \frac{\partial\mathcal{L}}{\partial\dot\alpha^a} ~, \qquad \pi^a \equiv F^{ab} \pi_b~,
\end{align}
where $F^{ab}$ is the inverse of $ f^{(0)}_{ab} $ such that $ F^{ab}f^{(0)}_{bc} =\delta^a_c$.

Now $\dot\alpha^a$ can be solved in terms of $\pi_a$ order by order (up to 3rd order here) as
\begin{align}\label{eq:hl1}
  \dot\alpha^a_{(1)} = \frac{\pi^a}{2} ~,
  \qquad \dot\alpha^a_{(2)} = -\frac{1}{2} F^{ab}g^{(2)}_b - \frac{1}{2} F^{ac}g^{(1)}_{bc}\pi^b
                    - \frac{3}{8} F^{ac}g^{(0)}_{dbc}\pi^b\pi^d~.
\end{align}
\begin{align}\label{eq:hl2}
  \dot\alpha^i_{(3)} = & \frac{1}{2} F^{ak}F^{ij}g^{(1)}_{jk}g^{(2)}_a - \frac{1}{2} F^{ij}h^{(3)}_j
  + \frac{1}{2} F^{bk}F^{ij}g^{(1)}_{ak}g^{(1)}_{bj}\pi^a
  + \frac{3}{4} F^{bk}F^{ij}g^{(0)}_{ajk}g^{(2)}_{b}\pi^a
  + \frac{3}{4} F^{ck}F^{ij}g^{(0)}_{bjk}g^{(1)}_{ac}\pi^a\pi^b
\nonumber\\ &
  + \frac{3}{8} F^{ck}F^{ij}g^{(0)}_{abk}g^{(1)}_{cj}\pi^a\pi^b
  + \frac{9}{16} F^{dk}F^{ij}g^{(0)}_{ajk}g^{(0)}_{bcd}\pi^a\pi^b\pi^c
  - \frac{1}{2} F^{ij}h^{(2)}_{jk}\pi^k - \frac{3}{8} F^{ik}h^{(1)}_{ajk}\pi^a\pi^j
  - \frac{1}{4} F^{ij}h^{(0)}_{abjk}\pi^a\pi^b\pi^k ~.
\end{align}
Now the Hamiltonian density is
\begin{align}
  \mathcal{H} = \dot\alpha^a \pi_a - \mathcal{L}~,
\end{align}
where all the $\dot\alpha^a$ on the RHS are to be replaced by Eqs. \eqref{eq:hl1}, \eqref{eq:hl2}. Finally, define the interaction picture derivative as discussed in Sec.~\ref{sec:inter-hamilt}:
\begin{align}
  \dot\alpha_I^a \equiv \frac{\pi^a}{2} ~.
\end{align}
Inserting this definition into the Hamiltonian, the Hamiltonian density becomes
\begin{align} \label{eq:lag-to-ham}
  \mathcal{H} = & f^{(0)}_{ab} \dot\alpha_I^a\dot\alpha_I^b -j^{(2)}
  \nonumber\\ &
  - g^{(0)}_{abc} \dot\alpha_I^a \dot\alpha_I^b \dot\alpha_I^c
  - g^{(1)}_{ab} \dot\alpha_I^a \dot\alpha_I^b - g^{(2)}_{a} \dot\alpha_I^a - j^{(3)}
  \nonumber\\ &
  - h^{(0)}_{abcd}\dot\alpha_I^a\dot\alpha_I^b\dot\alpha_I^c\dot\alpha_I^d
  - h^{(1)}_{abc} \dot\alpha_I^a \dot\alpha_I^b \dot\alpha_I^c
  - h^{(2)}_{ab} \dot\alpha_I^a \dot\alpha_I^b - h^{(3)}_{a} \dot\alpha_I^a - j^{(4)}
  \nonumber\\ &
  +\frac{9}{4} F^{ab} g^{(0)}_{acd}g^{(0)}_{bef}
  \dot\alpha_I^c\dot\alpha_I^d\dot\alpha_I^e\dot\alpha_I^f
  + 3 F^{ab} g^{(0)}_{acd}g^{(1)}_{be} \dot\alpha_I^c\dot\alpha_I^d\dot\alpha_I^e
  + \frac{3}{2} F^{ab} g^{(0)}_{acd}g^{(2)}_{b} \dot\alpha_I^c\dot\alpha_I^d
  \nonumber \\ &
  + F^{ab} g^{(1)}_{ac}g^{(1)}_{bd} \dot\alpha_I^c\dot\alpha_I^d
  + F^{ab} g^{(1)}_{ac}g^{(2)}_{b} \dot\alpha_I^c
  + \frac{1}{4} F^{ab} g^{(2)}_{a}g^{(2)}_{b} ~.
\end{align}
We observe that at the third order, we indeed have $\mathcal{H}_3 = -\mathcal{L}_3$ even when there are derivative couplings. On the other hand, at the 4th order non-trivial terms arises in the last two lines.

\subsection{3-point correlation function (bispectrum)}

Now we have derived the Hamiltonian, and are in the position to use in-in formalism to calculate the correlation function. We only do it at the leading order of slow roll. 

\begin{figure}[htbp]
  \centering
  \includegraphics[width=0.45\textwidth]{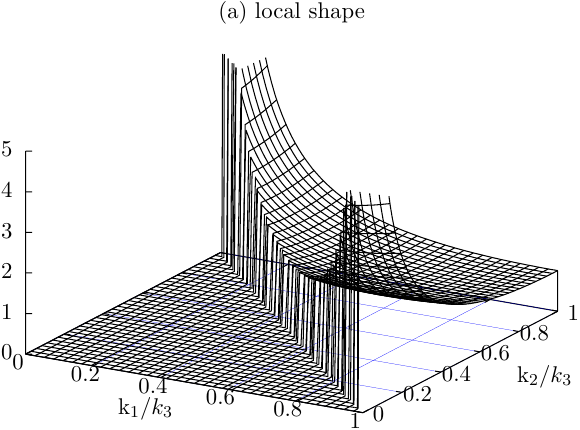}
  \\ \vspace{0.03\textheight}
  \includegraphics[width=0.45\textwidth]{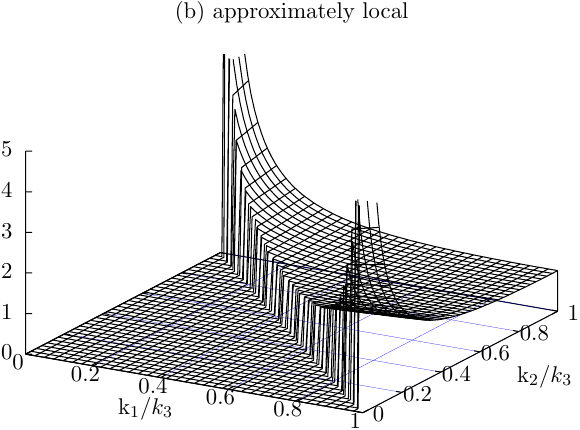}
  \hspace{0.03\textwidth}
  \includegraphics[width=0.45\textwidth]{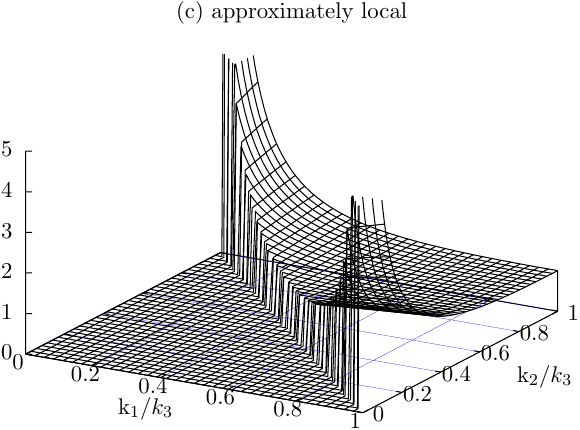}
  \caption{\label{fig:local-shapes} The panels (a), (b) and (c) are the shape functions $\frac{(k_1^3+k_2^3+k_3^3)}{3k_1k_2k_3} $, $\frac{(k_1k_2^2+k_1k_3^2+k_2k_1^2+k_2k_3^2+k_3k_1^2+k_3k_2^2)}
         {6k_1k_2k_3}$ and $\frac{(k_1^2k_2^2+k_1^2k_3^2+k_2^2k_3^2)}
         {k_1k_2k_3(k_1+k_2+k_3)}$ respectively. Note that only the configurations satisfying the triangle inequality $k_1+k_2>k_3$ makes sense in physics. The part with $k_1+k_2<k_3$ is set to zero by hand and should be ignored from physical consideration. The shape function in panel (a) is known as ``local shape'', because local processes could only generate this shape. The physical meaning of local shape is discussed extensively in Sec.~\ref{sec:local-non-gauss}. The other two panels look similarly to local shape. Especially all those 3 shapes have the same scaling when $k_1/k_3\rightarrow 0$, that $\mathcal{F}\propto k_3/k_1$ in this limit.}
\end{figure}

Inserting the mode function \eqref{eq:zeta-mode-function} together with the Hamiltonian (first line of \eqref{eq:l3} and Eq.~\eqref{eq:h3}), and perform the time integration. The result is \footnote{Here the slow roll approximation is assumed. Numerical code without slow roll approximation is available in \cite{Hazra:2012yn, Hazra:2012kq}.} \cite{Maldacena:2002vr}
\begin{align}
  \langle\zeta_{\mathbf{k}_1}\zeta_{\mathbf{k}_2}\zeta_{\mathbf{k}_3}\rangle = (2\pi)^7 \delta^3(\mathbf{k}_1+\mathbf{k}_2+\mathbf{k}_3)
  \frac{P_\zeta^2}{k_1^2 k_2^2 k_3^2} \mathcal{F}(k_1/k_3,k_2/k_3)~,
\end{align}
where there is a factor of $P_\zeta^2$ on the RHS. Thus the three point function is also called the bispectrum. The dimensionless shape function 
\begin{align} \label{eq:single-field-shape3}
  \mathcal{F}(k_1/k_3,k_2/k_3) \equiv 
  \frac{3(\eta-\epsilon)}{8} 
  \frac{(k_1^3+k_2^3+k_3^3)}{3k_1k_2k_3} 
  + \frac{3\epsilon}{4}
  \frac{(k_1k_2^2+k_1k_3^2+k_2k_1^2+k_2k_3^2+k_3k_1^2+k_3k_2^2)}
         {6k_1k_2k_3} 
  + \epsilon
  \frac{(k_1^2k_2^2+k_1^2k_3^2+k_2^2k_3^2)}
         {k_1k_2k_3(k_1+k_2+k_3)}~.
\end{align}
The shape functions are plotted in Fig~\ref{fig:local-shapes}. Note that when $k_1/k_3\rightarrow0$, $\mathcal{F}\propto k_3/k_1$. This $k_1/k_3\rightarrow 0$ limit (or limits $k_2/k_1\rightarrow0$ or $k_3/k_1\rightarrow0$) is called the squeezed limit. The squeezed limit is of particular importance in the study of non-Gaussianities. 

\subsection{Local non-Gaussianity and the local $f_{NL}$}
\label{sec:local-non-gauss}

We explain in this subsection the physical meaning of local non-Gaussianity \cite{Komatsu:2001rj}. The local non-Gaussianity is originally introduced as a position space expansion of non-Gaussian perturbations around Gaussian perturbations\footnote{The funny numerical coefficients come from that $f_{NL}$ is originally defined in terms of the gravitational potential $\Phi$.}\footnote{In Maldacena's paper, there was a minus sign in front of the $f_{NL}$ term. Here we use the WMAP convention. A summary of conventions can be found in \cite{Chen:2007gd, Wands:2010af}. }:
\begin{align} \label{eq:local-ansatz}
  \zeta(\mathbf{x}) = \zeta_g(\mathbf{x}) + \frac{3}{5} f_{NL} \zeta_g^2(\mathbf{x})
  + \frac{9}{25} g_{NL} \zeta_g^3(\mathbf{x})
  + \frac{27}{125} h_{NL} \zeta_g^4(\mathbf{x})
  + \frac{81}{625} i_{NL} \zeta_g^5(\mathbf{x}) 
  + \frac{243}{3125} j_{NL} \zeta_g^6(\mathbf{x}) 
  + \cdots~,
\end{align}
where $\zeta_g(\mathbf{x})$ is a perturbation variable satisfying Gaussian statistics. Note that this is an ansatz, instead of a general expansion. Because in general, the non-Gaussian estimators $f_{NL}, \cdots$ could depend on position, and terms like $\zeta_g^2(\mathbf{x})$ is not necessarily expanded at the same spatial point. As a result, the terms like $f_{NL} \zeta_g^2(\mathbf{x})$ has to be in general written as a convolution. 

The local ansatz has clear physical meaning. This ansatz assumes non-Gaussianities are generated independently at different spatial points. Translating this requirement into the context of inflation, typically this implies the non-Gaussianity is generated on super-Hubble scales.\footnote{Note the logic here: Local mechanisms (super-Hubble generation of non-Gaussianity) produces local shape non-Gaussianity. However, local shape non-Gaussianity may not necessarily be produced by local mechanisms. Nevertheless, counter examples are unknown to the best of my knowledge. On the other hand, super-Hubble conversion of non-Gaussianity does not necessarily produce local shape non-Gaussianity, as long as the non-Gaussianity is generated on sub-Hubble or Hubble scales \cite{Cai:2010rt, Gong:2009dh}.}

Especially, note that, for positive $f_{NL}$, the part of contribution from positive $\zeta_g$ gets enhanced, and the part of contribution from negative $\zeta_g$ gets suppressed. This provides an intuitive way to visualize the sign of $f_{NL}$.

\begin{figure}[htbp]
  \centering
  \includegraphics[width=0.7\textwidth]{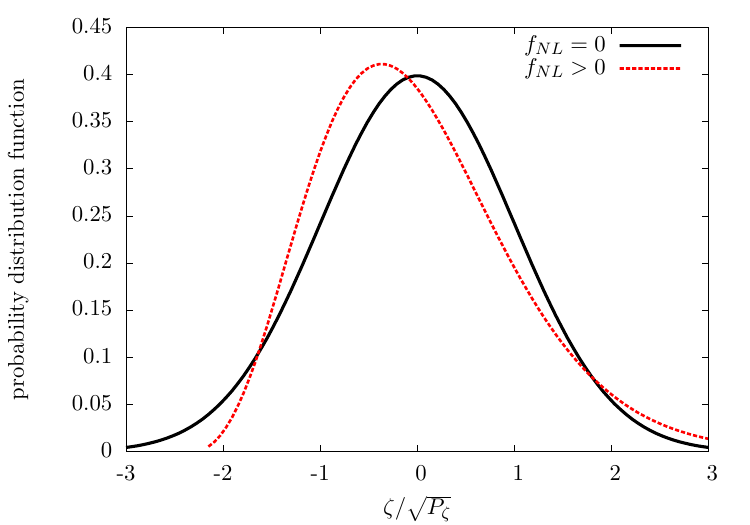}
  \caption{\label{fig:prob-fnl} Probability distribution of $\zeta$ for different $f_{NL}$ parameters. Positive $f_{NL}$ corresponds to the red curve in the figure. Negative $f_{NL}$ corresponds to $\zeta\rightarrow-\zeta$ of the red curve. This figure is only for illustration purpose. The real world $|f_{NL}|$ is much smaller than the one plotted here.}
\end{figure}

In position space, this non-Gaussianity can be characterized by the probability distribution function of $\zeta$. The probability distribution is plotted in Fig.~\ref{fig:prob-fnl}. Note that the tail at large and positive $\zeta$ indicates where structures are formed in the universe. Thus positive $f_{NL}$ means more galaxies are formed given a fixed power spectrum. For the CMB sky, positive $f_{NL}$ means there are more very cold spots than very hot spots on the CMB, and more modestly hot spots than modestly cold spots\footnote{On CMB cold spot means higher actual energy density. Because the CMB photon needs to climb up a bigger gravitational potential before reaching us.}.

The position space features of $f_{NL}$ is intuitive to understand. However, the inflationary perturbation manifests itself in momentum space. There are much more configurations of three point function to look at, than just looking at position space probability distributions point by point. Thus the momentum space correlation functions provides a much more powerful test of non-Gaussianity than the isolated position space test.

In momentum space, the corresponding expression is
\begin{align}
  \zeta_\mathbf{k} = \zeta^g_\mathbf{k} 
  + \frac{3}{5} f_{NL} \int \frac{d^3q}{(2\pi)^3} \zeta^g_\mathbf{q} \zeta^g_{\mathbf{k}-\mathbf{q}}~.
\end{align}
Inserting the expansion into the three point function, the three point function can be calculated as
\begin{align} \label{eq:fnl-local-def}
  \langle\zeta_{\mathbf{k}_1}\zeta_{\mathbf{k}_2}\zeta_{\mathbf{k}_3}\rangle = 
  (2\pi)^7 \delta^3(\mathbf{k}_1+\mathbf{k}_2+\mathbf{k}_3) 
  \frac{3f_{NL}}{10k_1^3k_2^3} P_\zeta(k_1) P_\zeta(k_2) 
  + (\mathbf{k}_2\Leftrightarrow\mathbf{k}_3) + (\mathbf{k}_1\Leftrightarrow\mathbf{k}_3)~,
\end{align}
This non-Gaussianity has the same shape as the first term of \eqref{eq:single-field-shape3}. The other two terms in Eq.~\eqref{eq:single-field-shape3} has different shape, but behave quite similarly. The observational bound of local shape $f_{NL}$ is \cite{Ade:2013ydc}
\begin{align}
  f_{NL}=2.7\pm 5.8~ \quad \mbox{(68\% CL statistical)}~.
\end{align}

The local ansatz \eqref{eq:local-ansatz} can be generalized to multiple species:
\begin{align} \label{eq:local-multi}
  \zeta(\mathbf{x}) = \sum_i N_i \delta\phi_i(\mathbf{x}) + \frac{1}{2} \sum_{ij} N_{ij} \delta\phi_i(\mathbf{x}) \delta\phi_j(\mathbf{x}) 
  + \frac{1}{6} \sum_{ijk} N_{ijk} \delta\phi_i(\mathbf{x}) \delta\phi_j(\mathbf{x}) \delta\phi_k(\mathbf{x}) + \cdots ~.
\end{align}
Note that in this generalization, the locality of non-Gaussianity is not modified. Only that multi-source of fluctuations become possible. Here the notation $N_{ij\cdots}$ and $\delta\phi_i$ is used, having in mind that one can interpret $N_{ij\cdots}$ as $\partial^n N/(\partial\phi_i\partial\phi_j\cdots)$. In this case the expansion \eqref{eq:local-multi} is related to the $\delta N$-formalism.

For uncorrelated Gaussian field fluctuations with identical power spectra, inserting the expansion into correlation functions gives, at tree level,
\begin{align}
  f_{NL} = \frac{5\sum_{ij}N_{ij}N_iN_j}{6 \left( \sum_i N_i^2 \right)^2} ~.
\end{align}
While for the 4-point function, we have
\begin{align}
  \langle\zeta_{\mathbf{k}_1}\zeta_{\mathbf{k}_2}\zeta_{\mathbf{k}_3}\zeta_{\mathbf{k}_4}\rangle_c =   (2\pi)^9 \delta^3(\mathbf{k}_1+\mathbf{k}_2+\mathbf{k}_3+\mathbf{k}_4) 
  \left[ 
    \frac{27}{100} g_{NL} \left( \frac{P_\zeta^3}{k_1^3k_2^3k_3^3} + 3 \mathrm{~perm}  \right)
    + \frac{1}{8} \tau_{NL} \left( \frac{P_\zeta^3}{k_1^3k_2^3k_{13}^3} + 11 \mathrm{~perm}  \right)
  \right] ~.
\end{align}
Because of the $P_\zeta^3$ factor, the four point function is also known as the trispectrum. Now even for the local shape, there are two estimators $g_{NL}$ and $\tau_{NL}$. In terms of the ansatz \eqref{eq:local-multi}, we have
\begin{align}
  g_{NL} = \frac{25\sum_{ijk}N_{ijk}N_iN_jN_k}{54 \left(\sum_iN_i^2 \right)^3} ~,
  \qquad
  \tau_{NL} = \frac{\sum_{ijk}N_{ij}N_{ik}N_jN_k}{\left(\sum_iN_i^2 \right)^3} ~.
\end{align}
Note that $g_{NL}$ depends on $N_{ijk}$, which does not appear in the expression of three point functions. On the other hand, $\tau_{NL}$ is determined by $N_{ij}$ and $N_i$, the same coefficients that determines $f_{NL}$. Thus it is interesting to see if there are relations between $\tau_{NL}$ and $f_{NL}$. Interestingly, such relation indeed exist. Using the Cauchy-Schwarz inequality
\begin{align}
  \sum_{ij} (A_iA_i)(B_jB_j) \geq \sum_{ij} (A_iB_i)(A_jB_j)~,
\end{align}
where $A_i$ and $B_j$ are real vectors, we have
\begin{align}
  \tau_{NL} \geq \frac{36}{25} f^2_{NL}~.
\end{align}
This relation is known as the Suyama-Yamaguchi relation \cite{Suyama:2007bg}. Especially, when there is only one source of fluctuation, the above inequality takes the equal sign. Similar relations also exist for higher order correlation functions \cite{Lin:2010ua} or loop corrections \cite{Tasinato:2012js}.

\subsection{Soft limits of correlation functions}

\subsubsection{The squeezed limit}

Here we are to consider the physical meaning of the squeezed limit. When one of the momenta is much smaller than the other two, say, $k_1/k_3\ll 1$, the $k_1$ mode is already super-Hubble and become classical and conserved when the modes $k_2 \simeq k_3$ exit the horizon \cite{Maldacena:2002vr}. This constant mode can be considered as just a shift of background.\footnote{On the other hand, if $\zeta$ is for some reason not conserved on super-Hubble scales, it cannot be considered as a shift of background. Because one can distinguish $\zeta$ and a redefinition of the background scale factor by the time evolution of $\zeta$. Examples of this type include multi-field inflation, and a growing mode of $\zeta$, as discussed in Secs.~\ref{sec:multi-field-infl} and \ref{sec:gener-attr-behav} respectively.}\footnote{There could also be initial correlations between the three modes on sub-Hubble scales, which are stretched to super-Hubble scales as they were, and happen to provide non-vanishing squeezed limit. An example of this type is discussed in Sec.~\ref{sec:gener-init-cond-1}.} Thus the three point function should reduce to the background shift of a two point function. 

To see this explicitly, one can expand $(\zeta_{\mathbf{k}_2}\zeta_{\mathbf{k}_3})$ with the presence of a background $\mathbf{k}_1$ mode \cite{Creminelli:2004yq}\footnote{Equivalently, define the dimensional spectrum $P_{\rm dim}(k)\equiv2\pi^2P_\zeta(k)/k^3$. A constant long mode $\lambda$ gives $P_{\rm dim}(k)|_\lambda=e^{-3\lambda}P_{\rm dim}(ke^{-\lambda})$, hence $\left.\partial_\lambda P_{\rm dim}(k)|_\lambda\right|_0=-(3+k\partial_k)P_{\rm dim}(k)=-(n_s-1)P_{\rm dim}(k)$.}:
\begin{align}\label{eq:soft-exp}
  \zeta_{\mathbf{k}_2}\zeta_{\mathbf{k}_3} = (\zeta_{\mathbf{k}_2}\zeta_{\mathbf{k}_3})_0
  + \zeta_{\mathbf{k}_1} \frac{\partial}{\partial\zeta_{\mathbf{k}_1}} (\zeta_{\mathbf{k}_2}\zeta_{\mathbf{k}_3}) + \cdots~.
\end{align}
Here $\partial/\partial\zeta_{\mathbf{k}_1}$ describes a background shift. Adding a constant long mode $\lambda$ is the inverse of the coordinate transformation \eqref{eq:global-shift}:
\begin{align}
  \zeta_{\mathbf{k}_1} \rightarrow \zeta_{\mathbf{k}_1}+\lambda(2\pi)^3\delta^3(\mathbf{k}_1)~,
  \qquad \mathbf{k}\rightarrow\mathbf{k}e^{-\lambda}~,
  \qquad (2\pi)^3\delta^3(\mathbf{k}_1)\frac{\partial}{\partial\zeta_{\mathbf{k}_1}}=\frac{\partial}{\partial\lambda}~.
\end{align}
Including the Fourier Jacobian, the short modes transform as $\zeta_{\mathbf k}\rightarrow e^{-3\lambda}\zeta_{e^{-\lambda}\mathbf k}$. Thus we have\footnote{The delta function on the denominator is not well defined. However, one can consider perturbations in a box and take the box size to infinity to cure this problem.}
\begin{align}
  \zeta_{\mathbf{k}_1}\frac{\partial}{\partial\zeta_{\mathbf{k}_1}}(\zeta_{\mathbf{k}_2}\zeta_{\mathbf{k}_3})
  =-\frac{\zeta_{\mathbf{k}_1}}{(2\pi)^3\delta^3(\mathbf{k}_1)}
  \left(6+\sum_{a=2}^3\mathbf{k}_a\cdot\frac{\partial}{\partial\mathbf{k}_a}\right)
  (\zeta_{\mathbf{k}_2}\zeta_{\mathbf{k}_3})~.
\end{align}
After taking the short-mode expectation value, the momentum delta function contributes $-3$ to this dilation operator, leaving the spectral tilt $n_s-1$.
Contract this term with $ \langle\zeta_{\mathbf{k}_1} \cdots \rangle$, we get
\begin{align}
  \lim_{k_1/k_3\rightarrow0} \langle\zeta_{\mathbf{k}_1}\zeta_{\mathbf{k}_2}\zeta_{\mathbf{k}_3}\rangle 
  = - (2\pi)^7 \delta^3(\mathbf{k}_1+\mathbf{k}_2+\mathbf{k}_3)
  \frac{P_\zeta^2}{4k_1^3k_3^3} (n_s-1) ~.
\end{align}
This consistency relation should be satisfied when $\zeta_{\mathbf{k}_1}$ is indeed a shift of background, especially for the standard model of single field inflation. To check that, we note the shape function takes the limit
\begin{align}
  \lim_{k_1/k_3\rightarrow0} \mathcal{F}(k_1/k_3, k_2/k_3) 
  = \frac{k_3}{4k_1} (2\epsilon+\eta)~,
\end{align}
Thus
\begin{align}
  \lim_{k_1/k_3\rightarrow0} \langle\zeta_{\mathbf{k}_1}\zeta_{\mathbf{k}_2}\zeta_{\mathbf{k}_3}\rangle 
  = (2\pi)^7 \delta^3(\mathbf{k}_1+\mathbf{k}_2+\mathbf{k}_3)
  \frac{P_\zeta^2}{4k_1^3k_3^3} (2\epsilon+\eta) ~.
\end{align}
Recall that the spectral index is $n_s-1 = -2\epsilon-\eta$. Thus the consistency relation is indeed satisfied for the three point function that we just reviewed.

Comparing with the squeezed limit of Eq.~\eqref{eq:fnl-local-def}
\begin{align}
  \lim_{k_1/k_3\rightarrow0}  \langle\zeta_{\mathbf{k}_1}\zeta_{\mathbf{k}_2}\zeta_{\mathbf{k}_3}\rangle = 
  (2\pi)^7 \delta^3(\mathbf{k}_1+\mathbf{k}_2+\mathbf{k}_3) 
  \frac{3f_{NL}}{5k_1^3k_2^3} P_\zeta(k_1) P_\zeta(k_2) ~,
\end{align}
we have for this minimal model
\begin{align}
  f_{NL} = - \frac{5}{12} (n_s-1)~.
\end{align}

Under the same assumptions, higher-point single-soft limits are determined by the common momentum-scale response of the full lower-point correlation function, including its power spectra and shape dependence.

Those consistency relations can also be checked by resumming the perturbation Hamiltonian with the presence of a classical mode of perturbation \cite{Li:2008gg}, or a result of unitarity \cite{Assassi:2012zq} or operator product expansion \cite{Assassi:2012et}. 

\subsubsection{Soft limit of internal momenta}
\label{sec:soft-limit-internal}
\begin{figure}[htbp]
  \centering
  \includegraphics[width=0.6\textwidth]{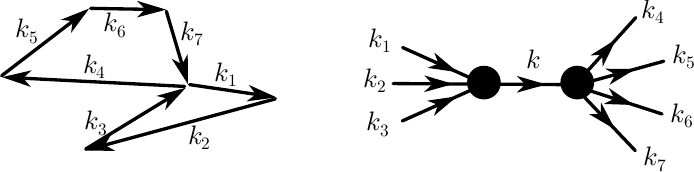}
  \caption{\label{fig:sy-ineq} Soft limit of correlation functions. In this example, we have $\mathbf{k}_1+\mathbf{k}_2+\mathbf{k}_3\rightarrow0$. The left panel illustrates momentum conservation, and the right panel illustrates the soft limit is dominated by a propagation of a soft momentum.}
\end{figure}

Additional soft limits arises for $n$-point function with $n>3$. For example, now we can have $\mathbf{k}_1+\mathbf{k}_2+\cdots + \mathbf{k}_m \rightarrow 0~, (2\leq m\leq n-2)$ without any external momentum being zero (note the ordering of momentum is not important here. Thus this covers any partition of summation goes to zero). The situation is illustrated in Fig.~\ref{fig:sy-ineq}.

Note that the exchanged scalar has a divergent behavior $1/k^3$ when $\mathbf{k}\equiv\mathbf{k}_1+\mathbf{k}_2+\cdots + \mathbf{k}_m \rightarrow 0$. Thus in this limit, this propagation should be the dominate diagram. 
The leading single-exchange contribution to the connected $n$-point correlation factorizes into an $(m+1)$-point correlation and an $(n-m+1)$-point correlation, each with an additional soft leg. With $\mathbf{q}=\sum_{a=1}^m\mathbf{k}_a$, this gives
\begin{align}
 \left\langle\prod_{a=1}^n\zeta_{\mathbf{k}_a}\right\rangle'_{c,\mathrm{exch}}
 \simeq
 \frac{\left\langle\prod_{a=1}^m\zeta_{\mathbf{k}_a}\,\zeta_{-\mathbf{q}}\right\rangle'_c
 \left\langle\zeta_{\mathbf{q}}\prod_{a=m+1}^n\zeta_{\mathbf{k}_a}\right\rangle'_c}
 {P_{\rm dim}(q)}~,
 \qquad q\ll k_a~.
\end{align}
Here a prime removes the overall momentum-conserving delta function, $c$ denotes the connected part, and $P_{\rm dim}(q)=2\pi^2P_\zeta(q)/q^3$ is the dimensional power spectrum. For the local shape with a single soft source, the four-point case gives\footnote{For more general shapes the soft response depends on the hard momenta and angles; the corresponding relations can also be expressed in position space \cite{Assassi:2012zq, Shiraishi:2013vja}.}
\begin{align}
  \tau_{NL} = \frac{36}{25} f_{NL}^2~.
\end{align}
Once we relax the single degree of freedom assumption, there can be more than one mode propagating as the soft mode. Also the propagating mode could contribute a different shift of the background. In this case
\begin{align}
  \tau_{NL} \geq \frac{36}{25} f_{NL}^2~.
\end{align}
This is again the Suyama-Yamaguchi relation.

\subsubsection{Usage of unitarity}

Here we derive the Suyama-Yamaguchi relation once more, by showing that the Suyama-Yamaguchi relation is actually a special example of optical theorem in cosmology \cite{Assassi:2012zq}. The main purpose of this subsection is to illustrate the unitarity method, and how to cut cosmological Feynman diagrams in general, despite that the same relation has already been derived twice in this review\footnote{I thank Miao Li, Chunshan Lin and Shinji Mukohyama for discussion for the contents in this section.}.

In the in-in formalism, an expectation value can be calculated as
\begin{align}
  \langle Q(t) \rangle = \langle F^\dagger Q_I(t) F \rangle~,
\end{align}
where $F$ is a unitary evolution operator on a real-time contour with Hermitian $H_I$, defined by
\begin{align}
  F \equiv F(t,t_0) = T \exp\left(-i \int_{t_0}^{t}H_I(t)dt\right)~,
\end{align}
and $Q_I(t)$ is the corresponding operator in the interaction picture. 

To be explicit, we have
\begin{align}
  \langle \zeta_{\mathbf{k}_{1}} \zeta_{\mathbf{k}_{2}} \zeta_{\mathbf{k}_{3}}\rangle
  = 
  \langle 
    F^\dagger 
    \zeta_{\mathbf{k}_{1}} \zeta_{\mathbf{k}_{2}} \zeta_{\mathbf{k}_{3}}
    F
  \rangle
\end{align}
and
\begin{align}
  \langle \zeta_{\mathbf{k}_{1}} \zeta_{\mathbf{k}_{2}} 
  \zeta_{\mathbf{q}_{1}}\zeta_{\mathbf{q}_{2}}\rangle
  = 
  \langle 
    F^\dagger 
    \zeta_{\mathbf{k}_{1}} \zeta_{\mathbf{k}_{2}}
    \zeta_{\mathbf{q}_{1}} \zeta_{\mathbf{q}_{2}}
    F
  \rangle~,
\end{align}

For the time being, we restrict our attention to states created by a single field operator, which describe the leading single-propagator contribution in the collapsed limit. In the free theory the sum over one-particle states is the projector
\begin{align}\label{eq:unitarity}
  \Pi_1 = \sum_i\int\frac{d^3p}{(2\pi)^3}
  |\phi^{(i)}_{\mathbf p}\rangle\langle\phi^{(i)}_{\mathbf p}|
  =\sum_i\int\frac{d^3p}{(2\pi)^3}
  \frac{\phi^{(i)}_{\mathbf p}|0\rangle\langle0|(\phi^{(i)}_{\mathbf p})^\dagger}
  {|u_p^{(i)}|^2}~,
\end{align}
where $u^{(i)}_p$ is the mode function of $\phi^{(i)}_{\mathbf{p}}$, defined as 
\begin{align}
  \phi^{(i)}_{\mathbf{p}} = u^{(i)}_p a^{(i)}_\mathbf{p} 
  + (u^{(i)}_p)^* (a^{(i)}_\mathbf{-p})^\dagger~.
\end{align}
A one particle state (well before horizon crossing) is defined as
\begin{align}
  |\phi^{(i)}_{\mathbf{p}}\rangle = (a^{(i)}_\mathbf{-p})^\dagger |0\rangle
  = \frac{\phi^{(i)}_{\mathbf p}}{(u_p^{(i)})^*}|0\rangle~.
\end{align}

These free one-particle states provide a useful basis well before the soft mode crosses the Hubble radius.

In the interaction picture, let $|\Psi\rangle=F|0\rangle$ be the normalized state and define $|v_A\rangle=\phi_A|\Psi\rangle$, where $A$ includes the field species and momentum. Their Gram matrix and the projector onto their span are
\begin{align}\label{eq:unitarity-prime}
 G_{AB}=\langle v_A|v_B\rangle~,
 \qquad \Pi_\phi=\sum_{A,B}|v_A\rangle(G^{-1})_{AB}\langle v_B|~.
\end{align}
The sums include momentum integrals, and the inverse is taken on the independent span of the states. The inner products are full two-point functions. In particular, evolving the free one-particle projector gives $F\Pi_1F^\dagger$, whereas $\Pi_\phi$ is constructed from $\phi_A F|0\rangle$; the two need not coincide. The vacuum contribution is removed when considering connected correlations.

Inserting this projector between the two pairs of hard fields isolates the single-soft-field channel. For a local response, choose a basis in which the soft two-point functions are diagonal and define the mixed squeezed amplitudes by
\begin{align}
 \langle\phi^{(i)}_{\mathbf q}\zeta_{\mathbf k}\zeta_{-\mathbf k-\mathbf q}\rangle'
 \simeq \frac{12}{5}f_{NL}^{(i)}(q,t)P_{\rm dim}^{(i)}(q,t)P_{\rm dim}(k,t)~,
 \qquad q\ll k~.
\end{align}
With this convention, the projected contribution gives
\begin{align}\label{eq:main}
  \tau_{NL}(k,t) 
  = \sum_{i} \frac{P^{(i)}(k, t)}{P(k,t)} \left(\frac{6}{5} f^{(i)}_{NL}(k, t)\right)^2~ 
  + (\text{states created by more field operators}) + \cdots,
\end{align}
Here $P^{(i)}$ and $P$ are the dimensionless soft spectra, and $\cdots$ denote contributions beyond the leading collapsed limit.

\section{Inflation Models}

So far, we have almost restricted our attention to the minimal model of inflation. On the other hand, there are a vast number of inflation models around. Those models brings various novel predictions. Here we introduce some of those models.

\subsection{General single field inflation}

In the case of single field inflation, there are already a lot of possibilities for generalization.

\subsubsection{Generalizing the Lagrangian}

The dynamics of inflation is determined by its Lagrangian. Thus towards generalization of the minimal inflation model, it is natural to consider how to generalize its Lagrangian. Here we restrict our attention to the scalar sector\footnote{One may use higher spin fields or modified gravity as inflaton as well. We shall not step into this huge literature in this review.}

An immediate question is if we can generalize the potential. The answer is almost negative, considering we have already considered the potential as a general function of $\phi$ in the minimal model.

However, this is not the end of the story. We have assumed $V(\phi)$ is single valued. One could consider multiple valued $V(\phi)$, for example $\phi$ moving on a monodromy potential. This opens up interesting possibilities such as large field inflation in string theory \cite{Silverstein:2008sg}. One could also consider features or small oscillations on the potential. The inflaton could either tunnel though the potential \cite{Freese:2004vs, Cline:2011fi}, or rolling but with slow roll conditions violated (see Sec.~\ref{sec:scale-depend-feat}).

Now let us go beyond modifying potential. The task is to find a general scalar Lagrangian, which is well behaved theoretically. Especially the Lagrangian should be ghost free. Otherwise either the theory is not consistent, or we are not expanding around the correct vacuum\footnote{In this case, a ghost condensate \cite{ArkaniHamed:2003uy, ArkaniHamed:2003uz} may drive us to the correct vacuum, as in the case of tachyon condensate.} or imposing the correct boundary conditions\footnote{We typically impose simplest boundary conditions such that the field approach its classical value and quantum fluctuation vanishes at infinity. However, if alternative boundary conditions are imposed, such as that imposed in $\mathcal{PT}$-symmetric quantum mechanics, the ghost may be eliminated \cite{Bender:1998ke, Bender:2007wu}.}. 

The ghost free requirement can be expressed in more explicit terms. Consider we have some background, and want to quantize small perturbations on this background:

First of all, we would like to restrict our attention to the theories with correct sign of the kinetic term: $S=\int d^4x (+\dot{\delta\phi}^2+\cdots)$ instead of $S=\int d^4x (-\dot{\delta\phi}^2+\cdots)$. Ghost is not welcome to physics \cite{Cline:2003gs}. To see this, note that the propagator is the inverse of the second order action. Thus a wrong sign kinetic term results in a wrong sign of the propagator
\footnote{Here we work in flat space. Because in the UV, cosmological background returns to flat space. And the ghost is already troublesome enough in the UV.}
\begin{align} \label{eq:ghost-prop}
  \frac{-i}{p^2-m^2} \quad\Rightarrow\quad 
  \frac{-i}{p^2-m^2+i\epsilon} \quad\mathrm{or}\quad \frac{-i}{p^2-m^2-i\epsilon} ~.
\end{align}
We yet have the freedom to choose the sign of $i\epsilon$ term for the ghost. However, if we choose the plus sign, the imaginary part of \eqref{eq:ghost-prop} has different sign from normal particle. Thus the optical theorem is violated and there is no probabilistic interpretation of the theory. On the other hand, if the minus sign is chosen, negative energy modes propagate forward in time. Thus the ghost particle carry negative energy. In this case the vacuum decays into ghost particles even if ghost and other particles only interact gravitationally. Even worse, the decay rate for vacuum to decay into two photon and two ghost particles is
\begin{align}
  \Gamma \sim \frac{\Lambda^8}{M_p^4} ~,
\end{align}
where $\Lambda$ is the UV cutoff of Lorentz violation. Thus in a Lorentz invariant theory, the decay rate is divergent. The divergence is physical: in each frame there is a non-vanishing decay rate and one has to integrate over the infinite volume of Lorentz group. Any small decay rate is forbidden.

Secondly, we require the equation of motion for perturbations to be second order in time derivatives. This is because, in a higher order derivative theory (for example, 4th order), the propagator can be written as
\begin{align}
  \frac{i}{(p^2-m_1^2)(p^2-m_2^2)} = 
  \frac{1}{m_1^2-m_2^2} \left[ \frac{i}{p^2-m_1^2} - \frac{i}{p^2-m_2^2}  \right]~.
\end{align}
Thus one of the two terms behaves as a ghost\footnote{It may be fine if the Lagrangian has infinite orders of time derivatives which resums to a ghost-free form \cite{Barnaby:2007yb, Barnaby:2008fk, Barnaby:2008vs}.}.\footnote{Note added in 2026: The absence of an Ostrogradsky mode can also follow from degeneracy of a higher-derivative scalar--tensor theory. Beyond-Horndeski and degenerate higher-order scalar--tensor (DHOST) theories extend the class discussed here; their constraints can remove the additional degree of freedom even when the equations contain higher derivatives. See \cite{Langlois:2015cwa}.}

We would like to emphasize that the above requirements are imposed on the perturbations. The background could still have higher orders derivatives as long as perturbations are well behaved.

The minimal inflation model is definitely ghost free. However, it is not the most general ghost free theory. Instead, the most general theory of a scalar, coupled to gravity and with second order equation of motion for perturbations, is \cite{Horndeski:1974wa, Deffayet:2011gz, Kobayashi:2011nu}
\begin{align} \label{eq:lgginf}
  \mathcal{L} = & P(\phi,X)-G_3(\phi,X) \Box \phi + G_4(\phi,X) R 
  + G_{4,X} \left[ (\Box \phi)^2 - (\nabla_\mu\nabla_\nu\phi)^2\right]
  \nonumber\\ &
  + G_5(\phi,X) G_{\mu\nu}\nabla^\mu\nabla^\nu\phi
  - \frac{1}{6} G_{5,X} \left[ (\Box \phi)^3 - 3\Box \phi(\nabla_\mu\nabla_\nu\phi)^2 
                       + 2(\nabla_\mu\nabla_\nu\phi)^3 \right]~.
\end{align}
where $P, G_3, G_4, G_5$ are arbitrary functions of $\phi$ and $X\equiv -(\partial\phi)^2/2$, and $G_{4,X}$ and $G_{5,X}$ denote derivatives with respect to $X$. For perturbations, the high order time derivatives are removed by integration by parts. One still need to be careful about the correct sign of the kinetic term, which cuts about half of the parameter space. The linear perturbations are considered in \cite{Kobayashi:2011nu} and the non-Gaussianities are calculated in \cite{Gao:2011qe, RenauxPetel:2011sb, Gao:2012ib}. 

Historically, the $P(\phi,X)$ term in the \eqref{eq:lgginf} is considered as K-inflation \cite{ArmendarizPicon:1999rj, Garriga:1999vw}. Later the $G_3$ term is added inspired by the Galilean symmetry \cite{Kobayashi:2010cm, Kobayashi:2011pc}. And finally the most general form \eqref{eq:lgginf} is rediscovered in the context of cosmology.

The full theory of \eqref{eq:lgginf} is complicated and we are not going into the details here. However, it is interesting to consider the simplest non-minimal case of the theory, where $G_4=M_p^2/2$, $G_3=G_5=0$, and the scalar Lagrangian is $P(\phi,X)$ \cite{ArmendarizPicon:1999rj, Garriga:1999vw, Seery:2005wm, Chen:2006nt}. In this case, by varying the action with respect to the metric, the energy density takes the form
\begin{align}
  \rho = 2X P_{,X} - P~.
\end{align}
And one can find the pressure is $p = P(\phi,X)$. The Friedmann equation and the continuity equation still hold:
\begin{align}
  3M_p^2H^2 = \rho~, \qquad \dot\rho + 3H(\rho+p) = 0~.
\end{align}
For the theory of perturbations, we work in the $\zeta$-gauge. The second order action can be derived as
\begin{align}
  S_2 = M_p^2\int dt d^3x \left[ a^3 \frac{\epsilon}{c_s^2} \dot\zeta^2 - a\epsilon(\partial\zeta)^2  \right]~,
\end{align}
where
\begin{align}
  c_s^2 = \frac{P_{,X}}{P_{,X}+2X P_{,XX}} ~.
\end{align}
Note that in the minimal inflation scenario, where $P(\phi,X)=X-V(\phi)$ and $c^2_s=1$. However here this parameter is generalized, and can be understood as the sound speed of the perturbations. This is because now the temporal and spatial directions have different weight in the action.

To solve the linear perturbation equations, one typically assumes slow variation of $c_s$: $|s|\ll 1$, where $s\equiv \dot c_s/(Hc_s)$. In this case, the $\zeta$ field can be quantized and solved as
\begin{align} \label{eq:zeta-mode-function-cs}
  \zeta_\mathbf{k} = u_k a_\mathbf{k} + u^*_k a^\dagger_{-\mathbf{k}}~,
  \qquad u_k = \frac{H}{M_p\sqrt{4\epsilon c_sk^3}} (1+ikc_s\tau)e^{-ikc_s\tau}~.
\end{align}
Note again that now the oscillation factor in the time direction is $e^{-ikc_s\tau}$. On the other hand, the spatial Fourier transform still has the same factor $e^{i \mathbf{k} \cdot \mathbf{x}}$. Thus a propagating wave described by this Lagrangian has a modified speed.

The power spectrum can be calculated as
\begin{align}
  P_\zeta = \frac{H^2}{8\pi^2 M_p^2 c_s \epsilon}~.
\end{align}
The only difference from the minimal inflation model is the factor of $c_s$. When $c_s^2 < 1$, this is an enhancement factor. The quantum fluctuations are larger because the perturbations are frozen at a smaller horizon.

\begin{figure}[htbp]
  \centering
  \includegraphics[width=0.45\textwidth]{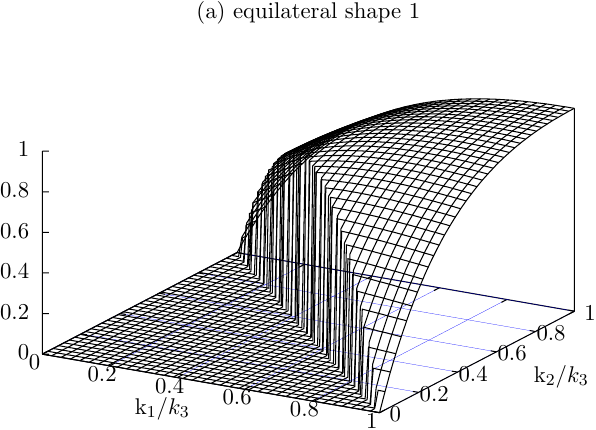}
  \hspace{0.03\textwidth}
  \includegraphics[width=0.45\textwidth]{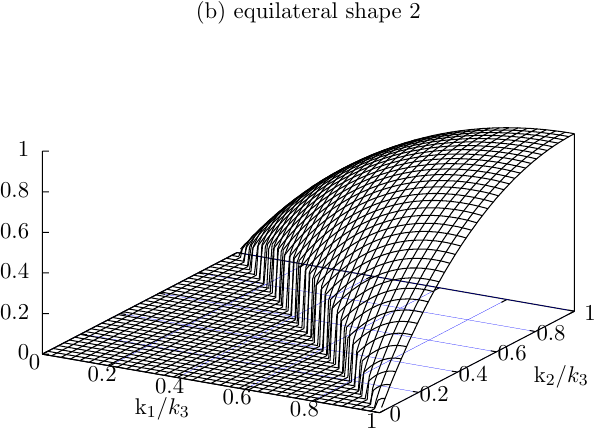}
  \caption{\label{fig:equil-shapes} The left and right panels are the first and second terms of Eq.~\eqref{eq:two-equi}. They are equilateral shapes.}
\end{figure}

\begin{figure}[htbp]
  \centering
  \includegraphics[width=0.45\textwidth]{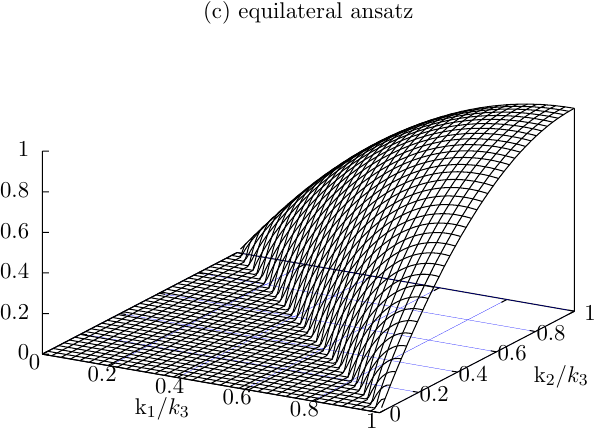}
  \hspace{0.03\textwidth}
  \includegraphics[width=0.45\textwidth]{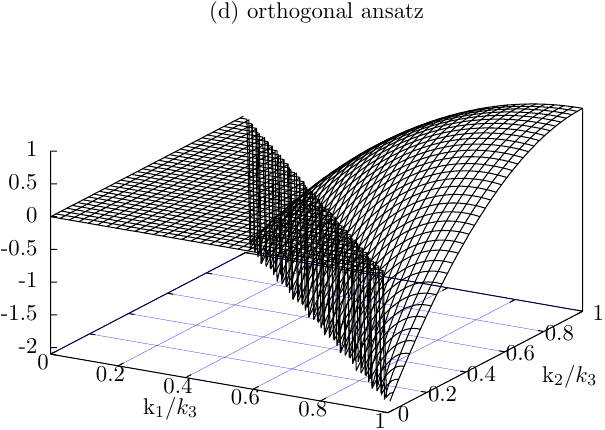}
  \caption{\label{fig:equil-ansatz} Left panel: the equilateral template $10\mathcal{F}_e/9$. Right panel: the orthogonal template $10\mathcal{F}_o/9$.}
\end{figure}

For non-Gaussianities, the leading order terms in the slow roll approximation is \cite{Chen:2006nt}
\begin{align} \label{eq:l3k1}
  S_3 = \int dt d^3x ~ a^3 \left\{ 
    -\left[ \Sigma\left( 1-\frac{1}{c_s^2}\right) + 2\lambda \right] \frac{\dot\zeta^3}{H^3}
    - \frac{3M_p^2\epsilon}{c_s^4} (1-c_s^2) \zeta\dot\zeta^2
    + \frac{M_p^2\epsilon}{a^2c_s^2} (1-c_s^2)\zeta(\partial\zeta)^2
  \right\} ~,
\end{align}
where 
\begin{align}
  \Sigma \equiv  X P_{,X} + 2X^2 P_{,XX} = \frac{M_p^2H^2\epsilon}{c_s^2} ~,
  \qquad 
  \lambda \equiv X^2 P_{,XX} + \frac{2}{3} X^3P_{,XXX}~.
\end{align}
Using some integration by parts and making use of the equation of motion, \eqref{eq:l3k1} can further be simplified into\footnote{This same form of Lagrangian can also be obtained by starting from the $\delta\phi$-gauge and do a gauge transformation to the $\zeta$-gauge. To get those leading order terms, the gravitational interaction in the $\delta\phi$-gauge can be ignored. And the gauge transformation only needs to be expanded into the leading order $\delta\phi = - \dot\phi \zeta / H$. Thus the calculation is much simplified in the $\delta\phi$-gauge plus gauge transformation way.}
\begin{align} \label{eq:l3k2}
  S_3 = \int dt d^3x ~ a^3 \left\{ 
    - 2\frac{\lambda}{H^3} \dot\zeta^3
    + \frac{\Sigma}{a^2H^3} (1-c_s^2)\dot\zeta (\partial\zeta)^2
  \right\} ~,
\end{align}
Using the in-in formalism, the bispectrum can be calculated as
\begin{align}
  \langle\zeta_{\mathbf{k}_1}\zeta_{\mathbf{k}_2}\zeta_{\mathbf{k}_3}\rangle = (2\pi)^7 \delta^3(\mathbf{k}_1+\mathbf{k}_2+\mathbf{k}_3)
  \frac{P_\zeta^2}{k_1^2 k_2^2 k_3^2} \mathcal{F}(k_1/k_3,k_2/k_3)~,
\end{align}
where
\begin{align}
  \mathcal{F} = & \left( \frac{1}{c_s^2}-1-\frac{2\lambda}{\Sigma} \right) 
         \frac{3k_1k_2k_3}{2(k_1+k_2+k_3)^3} 
         \nonumber\\ &
         + \left( \frac{1}{c_s^2}-1\right) 
         \left[
           -\frac{k_1^2k_2^2+k_1^2k_3^2+k_2^2k_3^2}
             {k_1k_2k_3(k_1+k_2+k_3)} 
           +\frac{k_1^2k_2^3+k_1^2k_3^3+k_2^2k_3^3
             +k_2^2k_1^3+k_3^2k_1^3+k_3^2k_2^3}{2k_1k_2k_3(k_1+k_2+k_3)^2} 
           +\frac{k_1^3+k_2^3+k_3^3}{8k_1k_2k_3}
         \right]~.
\end{align}
In the squeezed limit $k_1/k_2\rightarrow 0$, it is clear that the first term vanishes proportional to $k_1/k_2$. The second term appears divergent. However, both the divergence and the constant term near $k_1/k_2\rightarrow 0$ actually cancel. To see this, consider the limit $k_1\rightarrow 0$ with $k_3\rightarrow k_2+\alpha k_1$. In this limit, 
\begin{align}
  \mathcal{F} = 
  \left( \frac{1}{c_s^2}-1-\frac{2\lambda}{\Sigma} \right) \frac{3k_1}{16k_2} 
  + \left( \frac{1}{c_s^2}-1\right)  \frac{(5\alpha^2-11)k_1}{16k_2} 
  + \mathcal{O}(k_1^2)~.
\end{align}
Thus $\mathcal{F} \propto  k_1 / k_2$ near $k_1/k_2\rightarrow 0$, where the linear term depends on how the squeezed limit is taken. The shape functions are plotted in Fig.~\ref{fig:equil-shapes}. The left and right panels are respectively the normalized shape functions
\begin{align} \label{eq:two-equi}
  \frac{27k_1k_2k_3}{(k_1+k_2+k_3)^3}  ~, \quad  
  -\frac{24}{7} \left[
           -\frac{k_1^2k_2^2+k_1^2k_3^2+k_2^2k_3^2}
             {k_1k_2k_3(k_1+k_2+k_3)} 
           +\frac{k_1^2k_2^3+k_1^2k_3^3+k_2^2k_3^3
             +k_2^2k_1^3+k_3^2k_1^3+k_3^2k_2^3}{2k_1k_2k_3(k_1+k_2+k_3)^2} 
           +\frac{k_1^3+k_2^3+k_3^3}{8k_1k_2k_3}
         \right]~.
\end{align}
Those shapes are known as equilateral shapes. Note that those shape functions are not separable. In other words, the shape functions cannot be written in the form $\sum_i c_i f^{(1)}_i(k_1)f^{(2)}_i(k_2)f^{(3)}_i(k_3)$. Technically it is much harder to do data analysis of non-separable shapes than separable ones. For this purpose, some shape templates are proposed to mimic the equilateral shapes \cite{Creminelli:2005hu}: The equilateral template 
\begin{align}
  \mathcal{F}_e = \frac{9}{10} \left( -2- \frac{k_1^2}{k_2k_3} -\frac{k_2^2}{k_1k_3} -\frac{k_3^2}{k_2k_1} 
    + k_1/k_2 + k_1/k_3 + k_2/k_1 + k_2/k_3 + k_3/k_1 + k_3/k_2
  \right)~,
\end{align}
and the orthogonal template
\begin{align}
  \mathcal{F}_o = \frac{9}{10} \left( -8- \frac{3k_1^2}{k_2k_3} -\frac{3k_2^2}{k_1k_3} -\frac{3k_3^2}{k_2k_1} 
    + 3k_1/k_2 + 3k_1/k_3 + 3k_2/k_1 + 3k_2/k_3 + 3k_3/k_1 + 3k_3/k_2
  \right)~.
\end{align}
Those templates are plotted in Fig.~\ref{fig:equil-ansatz}. The coefficients multiplying those shapes are constrained as \cite{Ade:2013ydc}
\begin{align}
  f_{NL}^\mathrm{equil} = -42\pm 75~, \quad f_{NL}^\mathrm{ortho}=-25\pm39~ \quad \mbox{(68\% CL statistical)}~.
\end{align}

The trispectrum of this $P(X,\phi)$ theory can also be calculated using the in-in formalism \cite{Chen:2009bc, Arroja:2009pd}. It is useful to note that non-trivial transformation \eqref{eq:lag-to-ham} between the Lagrangian to the Hamiltonian has to be considered. Also the scalar propagation diagram (like the right panel of Fig.~\ref{fig:sy-ineq}) needs to be included.

This $P(X,\phi)$ class of models can also be generalized to multiple field case, where multi-field effects can appear even if the field space trajectory is a geodesic \cite{Langlois:2008mn}. In the multiple field case, local and equilateral non-Gaussianities may get combined \cite{RenauxPetel:2010uw}.

\subsubsection{Effective field theory of perturbations}

In this subsection, we take a different point of view for extending the cosmic perturbation theory. Unlike the previous section, here we assume an inflationary background, and see what is the general perturbation theory living on that background. This approach is known as the effective field theory (EFT) approach for inflationary perturbations \cite{Cheung:2007st} (see also, \cite{Burgess:2003zw, Weinberg:2008hq, Baumann:2011su}).

As in the general discussion of EFT, any operator respecting the symmetry of the theory could appear in the Lagrangian. The relative importance of those operators are determined by the dimension of the operator. Operators with higher dimensions are suppressed more by the UV cutoff scale of the EFT. 

Now we are to write down all possible terms for perturbations. Considering that there are many possibilities, it is helpful to choose a gauge to reduce the number of terms. The gauge to choose here is the $\zeta$-gauge, where $\delta\phi=0$. In this gauge, the most general Lagrangian for single field inflation, with the foliation-preserving diff-invariance $x^i\rightarrow x^i+\xi^i(t,\mathbf{x})$ is
\begin{align}\label{eq:eft-uni}
  S = & \int d^4x \sqrt{-g} \left[ 
    \frac{1}{2} M_p^2 R - c(t) g^{00} - \Lambda(t) + \frac{1}{2} M_2^4(t) (g^{00}+1)^2 
    + \frac{1}{6} M_3^4(t) (g^{00}+1)^3 \right.
    \nonumber\\ & \left.
    - \frac{\bar M_1^3(t)}{2} (g^{00}+1)\delta K^\mu_\mu
    - \frac{\bar M_2^2(t)}{2} (\delta K^\mu_\mu)^2
    - \frac{\bar M_3^2(t)}{2} \delta K^\mu_\nu\delta K^\nu_\mu + \cdots
  \right]~,
\end{align}
where $\delta K_{\mu\nu}\equiv K_{\mu\nu}-Hh_{\mu\nu}$. The dots stand for higher order terms (including higher order in perturbation variable $\zeta$ and higher order in derivatives), that can be written down straightforwardly if needed. Curvature operators are related by integration by parts and the Gauss--Codazzi relation; independent higher-derivative curvature operators are included in the ellipsis. Expanding around the FRW background, one finds
\begin{align}
  c(t) = -M_p^2\dot H~, \qquad \Lambda(t) = M_p^2(3H^2 + \dot H)~.
\end{align}
The form of \eqref{eq:eft-uni} is not yet very useful. This is because we have to solve all the complicated gravitational interactions, which is possible but messy. Fortunately, there are a hierarchy of energy scales in the theory, among which the Planck mass is the highest. This implies the existence of a range of energy where gravitational interactions decouple. 

To find this range of energy, together with the simplified action with decoupled gravity, we should go away from the $\zeta$-gauge. As we discussed in previous sections, there are a lot of cancellations in the $\zeta$-gauge, which makes decoupling not transparent at all.

The $\zeta$-gauge is chosen by picking up a specific time coordinate. Thus the $\zeta$-gauge can be undone by promoting the $\zeta$-gauge time to a scalar field $T(t,\mathbf{x})$, such that in the $\zeta$-gauge, $T_\zeta(t_\zeta,\mathbf{x})=t_\zeta$ (where the subscript $\zeta$ denotes quantity in the $\zeta$-gauge). Now consider a general gauge
\begin{align} \label{eq:eft-time-shift}
  t = t_\zeta - \pi(t,\mathbf{x})~.
\end{align}
$T(t,\mathbf{x})$ transforms as a scalar field thus
\begin{align}
  T(t,\mathbf{x}) = T_\zeta(t_\zeta,\mathbf{x}) = T_\zeta(t+\pi(t,\mathbf{x}),\mathbf{x}) = t + \pi ~.
\end{align}
Now we are ready to make the $\zeta$-gauge quantities into tensors from $T(t,\mathbf{x})$:
\begin{align}
  \delta_\mu^0 = \partial_\mu t \qquad\Rightarrow\qquad \partial_\mu T =\partial_\mu (t+\pi) = \delta_\mu^0 + \partial_\mu \pi~,
\end{align}
\begin{align}
  g^{0\mu} = \delta^0_\nu g^{\mu\nu} \qquad\Rightarrow\qquad g^{0\mu}+ g^{\mu\nu}\partial_\nu\pi~,
\end{align}
\begin{align}
  g^{00} = \delta^0_\mu\delta^0_\nu g^{\mu\nu} \qquad\Rightarrow\qquad 
  g^{00} + 2g^{0\mu}\partial_\mu\pi + g^{\mu\nu}\partial_\mu\pi \partial_\nu\pi ~,
\end{align}
\begin{align}
  n_\mu = - \frac{\delta_\mu^0}{\sqrt{-g^{00}}} \qquad\Rightarrow\qquad
  -\frac{ \delta_\mu^0 + \partial_\mu \pi}{\sqrt{-g^{00}-2g^{0\mu}\partial_\mu\pi -g^{\mu\nu}\partial_\mu\pi \partial_\nu\pi}} ~,
\end{align}
where the left column are unitary gauge quantities and the right column are quantity without choosing a gauge. By doing the above replacements to the action \eqref{eq:eft-uni}, together with the replacement \eqref{eq:eft-time-shift}, we return from the $\zeta$-gauge to a general coordinate system without choosing a gauge:
\begin{align} \label{eq:eft-df-mix}
  S = & \int d^4x \sqrt{-g} \Big\{
    \frac{1}{2} M_p^2R - M_p^2\left[ 3H^2(t+\pi)+\dot H(t+\pi)\right]
    \nonumber\\ & 
    + M_p^2 \dot H(t+\pi) \left[ 
      (1+\dot\pi)^2g^{00} + 2(1+\dot\pi)\partial_i\pi g^{0i}+ g^{ij}\partial_i\pi\partial_j\pi 
    \right] 
    \nonumber\\ &
    + \frac{M_2^4(t+\pi)}{2} \left[  
      (1+\dot\pi)^2g^{00} + 2(1+\dot\pi)\partial_i\pi g^{0i}+ g^{ij}\partial_i\pi\partial_j\pi +1
    \right]^2
    \nonumber\\ &
    + \frac{M_3^4(t+\pi)}{6} \left[  
      (1+\dot\pi)^2g^{00} + 2(1+\dot\pi)\partial_i\pi g^{0i}+ g^{ij}\partial_i\pi\partial_j\pi +1
    \right]^3
    + \cdots
  \Big\}~,
\end{align}
where we neglect more terms in $\cdots$ for simplicity.

After re-obtaining the gauge invariant action, we are free to choose the $\delta\phi$-gauge (or any other gauge close to the $\delta\phi$-gauge, like the Newtonian gauge). The advantage of $\delta\phi$-gauge, or any gauge which is close to that, is that \textit{near the horizon crossing}, gravity is weakly coupled and only provide slow roll suppressed corrections. An example of this statement can be found in Eq.~\eqref{eq:S2-fg}. More generally, we can compare the terms with and without dynamical gravity in the new action, and conclude: The decoupling energy range for matter perturbations from gravity is\footnote{This estimate assumes $c_s\simeq1$. For a nontrivial dispersion relation, the energy and physical-momentum conditions for decoupling must be distinguished.}
\begin{align}
  \frac{k}{a}  \gg  \omega_\mathrm{mix} = \mathrm{max}\left( \frac{|\ddot\phi|}{H M_p} , \frac{|\dot\phi|}{M_p}, \frac{M_2^2}{M_p}, \cdots \right) ~,
\end{align}
where the dots denotes other scales (not shown in \eqref{eq:eft-df-mix}) that could appear, which are also suppressed by $M_p$. Note that $\dot\phi/M_p \sim \sqrt\epsilon H$, and $\ddot\phi/(HM_p)=(\eta/2-\epsilon)\dot\phi/M_p$. This result is intuitive to understand: when all scales appearing in the theory are smaller than $M_p$, gravity should be weak.

Below the mixing scale, gravity cannot be decoupled. The leading decoupled action also omits gravitational and slow-variation corrections of order $\epsilon$ and $\eta$, which must be retained when calculating the slow-roll bispectrum, including its squeezed limit \cite{Cheung:2007st, Maldacena:2002vr}.

Nevertheless, if we are interested in physics of $k/a\gg \omega_\mathrm{mix}$, or non-Gaussianity with $f_{NL}\gg \mathcal{O}(\epsilon,\eta)$, it is safe to use the action with gravity (including the constraints in the gravity sector) decoupled:
\begin{align} \label{eq:eft-df}
  S = \int d^4x \sqrt{-g} \left\{
    \frac{1}{2} M_p^2R - M_p^2\dot H\left( \dot\pi^2 - \frac{(\partial_i\pi)^2}{a^2}\right)
    + 2M_2^4 \left( \dot\pi^2+\dot\pi^3 - \dot\pi \frac{(\partial_i\pi)^2}{a^2}  \right)
    - \frac{4}{3} M_3^4 \dot\pi^3 + \cdots
  \right\}~.
\end{align}
Eq.~\eqref{eq:eft-df} recovers the action \eqref{eq:l3k1}, without referring to a background model explicitly. Moreover, in the $\cdots$ terms we neglected here (but can be recovered straightforwardly), a much wider class of models are covered, for example, ghost inflation \cite{ArkaniHamed:2003uz, Huang:2010ab, Izumi:2010wm}.

By construction, EFT requires in principle a UV completion. The ultimate UV completion of inflation is not yet known. But it is interesting to see in toy models how UV degrees of freedoms are integrated out to get the EFT in cosmology. For example, EFT operators arise when a very massive scalar is integrated out \cite{Tolley:2009fg, Cremonini:2010ua}.

Although we have put the EFT approach here as a generalization of single field inflation, the usage of EFT is not limited to single field (or field theory at all). Instead, the EFT approach can be used widely in cosmology, including applications in multi-field inflation \cite{Senatore:2010wk}, quasi-single field inflation \cite{Baumann:2011nk, Noumi:2012vr} and large scale structures \cite{Carrasco:2012cv}. 

\subsubsection{Generalizing the initial condition}
\label{sec:gener-init-cond-1}

In the previous subsections we have discussed how the operators of inflation could be generalized, on the full and perturbation level respectively. However, physical observables not only depend on operators but also depend on states. We have argued that it is natural to choose the initially lowest energy state as the vacuum state. But yet there can be generalizations.

The simplest generalization follows from Eq.~\eqref{eq:vk-with-c2}. Written in terms of the mode function of comoving curvature perturbation $\zeta$, we have \cite{Chen:2006nt}
\begin{align} 
  u_k = \frac{H}{M_p\sqrt{4\epsilon k^3}}\left[C_k^{(+)} (1+ik\tau)e^{-ik\tau} + C_k^{(-)} (1-ik\tau)e^{ik\tau}\right] ~.
\end{align}
One can turn on the  $C_k^{(-)}$ component, with $|C_k^{(+)}|^2-|C_k^{(-)}|^2=1$. In this case, we have for the power spectrum
\begin{align}
  P_\zeta = \frac{H^2}{8\pi^2\epsilon M_p^2} \left| C_k^{(+)} + C_k^{(-)} \right|^2 ~.
\end{align}
In case $C_k^{(\pm)}$ have some phase-sensitive $k$-dependence, there could be small oscillations on top of the nearly scale invariant power spectrum.

\begin{figure}[htbp]
  \centering
  \includegraphics[width=0.45\textwidth]{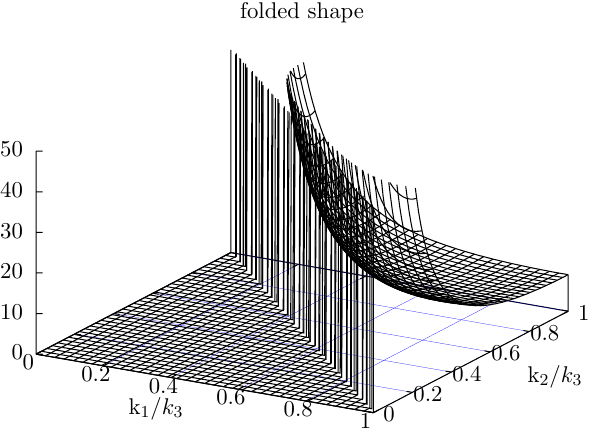}
  \caption{\label{fig:folded} A plot of the shape function \eqref{eq:fold-shape}.}
\end{figure}

Things become more interesting at the level of non-Gaussianity. We restrict our attention to a small non-BD component $|C_k^{(-)}|\ll |C_k^{(+)}|$. Including a factor of $C_k^{(-)}/C_k^{(+)}$ effectively flips the sign of $k$ (sign of amplitude, not the vector direction of $\mathbf{k}$). Recall that we had a shape function
\begin{align}
  \frac{k_1^2k_2^2+k_1^2k_3^2+k_2^2k_3^2}
         {k_1k_2k_3(k_1+k_2+k_3)}
\end{align}
in Eq.~\eqref{eq:single-field-shape3} of the minimal inflation model. With sign flip of the $k_i$'s\footnote{Note that only the $k$ in the $(1+ik\tau)e^{-ik\tau}$ part of the mode function is flipped. The $\frac{H}{\sqrt{4\epsilon k^3}}$ part of the mode function does not change.}, we get
\begin{align}\label{eq:fold-shape}
  \mathrm{Re}[C_k^{(-)}]
  \frac{k_1^2k_2^2+k_1^2k_3^2+k_2^2k_3^2} {k_1k_2k_3} 
  \left[
    \frac{1}{-k_1+k_2+k_3} + \frac{1}{k_1-k_2+k_3} + \frac{1}{k_1+k_2-k_3}
  \right]~.
\end{align}
This shape function blows up at the folded triangle limit $k_2+k_3\rightarrow k_1$, or its permutations. This folded shape is plotted in Fig.~\ref{fig:folded}. In the squeezed limit $k_1/k_2\rightarrow0$, $k_3 = k_2+\alpha k_1$ with $|\alpha|<1$, this shape function behaves as
\begin{align}
    \mathrm{Re}[C_k^{(-)}] \frac{k_2^2}{k_1^2} \frac{2}{1-\alpha^2} ~,
\end{align}
which blows up dramatically -- More quickly than the local shape, and depends on how the limit is taken.

However, we need to be careful in interpreting the divergence at the folded limit. This is because this divergence originates in the UV. On the one hand, it is tricky to define the UV limit of a non-vacuum state. On the other hand, the $i\epsilon$ prescription of selecting the interacting vacuum no longer works for an excited state. One may put a cutoff by hand to cure those problems phenomenologically \cite{Chen:2006nt}. However, the prediction is sensitive to the detailed cutoff scheme and thus sensitive to the UV physics \cite{Chen:2013aj}. An explicit UV completion of physics above the cutoff can be found in \cite{Chen:2010bka}.

\subsubsection{Generalizing the attractor behavior}
\label{sec:gener-attr-behav}

Inflation is attractive as an attractor solution. And we have implicitly used this nature in various places. In case that during a period of time the attractor behavior is violated, the inflationary perturbations could behave differently.

For example, typically we consider an inflaton field rolling down its potential. However, what happens when the inflaton use its kinetic energy to climb up a potential?

Note that the kinetic energy of the inflaton itself has equation of state $p_\mathrm{kin}=\rho_\mathrm{kin}$. Thus even there is no potential, the kinetic energy dilutes as $\rho_\mathrm{kin}\propto  a^{-6}$. With a potential, the kinetic energy could only dilute faster. Thus we get $\eta\leq-6+2\epsilon\simeq-6$ for $\epsilon\ll1$ during the climbing up stage of the potential.

Recall the equation of motion for the comoving curvature perturbation:
\begin{align}
  \ddot\zeta + H(3+\eta)\dot\zeta + \frac{k^2}{a^2} \zeta = 0~.
\end{align}
When $\eta<-3$, $\zeta$ effectively live in a contracting universe (though the actual universe is expanding). As a result, the conventionally decaying mode of $\zeta$ becomes a growing mode and dominates over the constant mode. One can show that when $\eta\simeq -6$, the power spectrum is still scale invariant. However the non-Gaussianity becomes $f_{NL}\sim 5/2$, which is not only order one but also breaks the consistency relation of single field inflation \cite{Namjoo:2012aa, Martin:2012pe}. One may further enhance this local non-Gaussianity by considering a small sound speed \cite{Chen:2013aj}. Qualitatively, the behavior of the perturbations are similar to the case of bouncing cosmology \cite{Cai:2009fn}. 

\subsection{Multi-field inflation}
\label{sec:multi-field-infl}

Here we generalize inflation towards another direction: including more light fields.

\subsubsection{Adiabatic and entropy perturbations}
\label{sec:curv-isoc-pert}

The status of perfect fluid can be described by its energy density $\rho(t,\mathbf{x})$, pressure $p(t,\mathbf{x})$ and velocity $v^i(t,\mathbf{x})$. On super-Hubble scales $v^i(t,\mathbf{x})$ decays exponentially thus the status of the perfect fluid is determined by $\rho(t,\mathbf{x})$ and $p(t,\mathbf{x})$.

For a barotropic fluid, $p$ is determined by $\rho$ as $p=p(\rho)$. In this case, we denote the fluctuation of the fluid as \textit{adiabatic perturbation}. If it is not the case, we can parameterize $p$ as
\begin{align}
  p = p(\rho, s_i)~.
\end{align}

Recall that when there is no entropy perturbation, the comoving curvature perturbation is conserved, following Eq.~\eqref{eq:local-energy-conservation}, on uniform energy density slice $\rho=\rho(t)$:
\begin{align}
  H(t)  + \dot\zeta(t,\mathbf{x}) = -\frac{1}{3} \frac{\dot\rho(t)}{\rho(t)+p(\rho, s_i)} ~.
\end{align}
Thus we can define the non-adiabatic part of pressure according to the derivation from this equation. 

Consider a general situation without choosing a gauge. We note:
\begin{align}
  p \equiv  p_0(t) + \delta p(t,\mathbf{x}), \qquad
  \delta p(t,\mathbf{x}) = \frac{\dot p_0}{\dot\rho_0} \delta\rho 
  + \frac{\partial p}{\partial s_i} \left( \delta s_i - \frac{\dot s_i}{\dot\rho} \delta\rho \right) + \cdots ~,
\end{align}
where $\cdots$ denotes higher order terms in perturbations.

It is the second term which make $\zeta$ change on super-Hubble scales. Thus we can define the non-adiabatic part of $p(\rho, s_i)$ as \cite{Wands:2000dp}
\begin{align}
  \delta p_\mathrm{nad} \equiv \frac{\partial p}{\partial s_i} \left( \delta s_i - \frac{\dot s_i}{\dot\rho} \delta\rho \right) + \cdots
  = \delta p -\frac{\dot p_0}{\dot\rho_0} \delta\rho ~.
\end{align}
This non-adiabatic part, or entropy part of pressure has clear physical meaning: It is the difference of equation of state, between perturbing the fluid and evolving the fluid in time. When there is such a difference, perturbation in the universe cannot be considered as a pure time shift between different local universes.

\begin{figure}[htbp]
  \centering
  \includegraphics[width=0.6\textwidth]{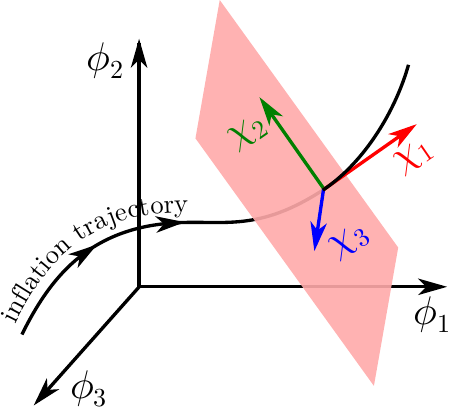}
  \caption{\label{fig:multi-orth} Redefining the field directions for multi-field inflation. The $\chi_1$ direction is the inflationary direction and the perturbation on $\chi_1$ direction is the adiabatic perturbation. The other directions are on the hypersurface perpendicular to $\chi_1$. Fluctuations on those directions are entropy perturbations.}
\end{figure}

Now we identify the adiabatic and entropy perturbations for inflation. In case there are multiple rolling\footnote{Originally very massive fields are not rolling, and remain not rolling after this rotate of basis.} fields $(\phi_1, \phi_2, \cdots, \phi_n)$. We can choose an instantaneous basis in field space such that the background velocity has only one nonzero component \cite{Gordon:2000hv, GrootNibbelink:2001qt}. This can be done by considering basis vector 
\begin{align}
  e_1^i \equiv \frac{\dot\phi^{(0)}_i}{\sqrt{\sum_j(\dot\phi^{(0)}_j)^2}} ~,
\end{align}
where superscript $(0)$ denotes the background level of the field, without perturbations. And the other $(n-1)$ basis vector are defined as the orthogonal directions. The background speed and the perturbation projections are defined as
\begin{align}
  \dot\chi_1\equiv\sqrt{\sum_i(\dot\phi_i^{(0)})^2}~,\qquad \delta\chi_a\equiv e_a^i\delta\phi_i~.
\end{align}
The adiabatic direction is $e_1^i$. The orthogonal directions satisfy $e_a^i\dot\phi_i^{(0)}=0$ for $a>1$, and their perturbations $\delta\chi_a$ are entropy perturbations. Their potential gradients need not vanish: the background field equations give $e_a^i V_{,i}=-\dot\chi_1 e_a^i\dot e_1^i$. This decomposition is illustrated in Fig.~\ref{fig:multi-orth}.

When the inflation trajectory is straight in field space, the basis vectors $e_a^i$ are time independent and the rotation is trivial. Otherwise the basis vectors $e_a^i$ are time dependent. 

For models with standard kinetic term, and at linear order in perturbations, $\chi_1$ is coupled to $\chi_2, \cdots, \chi_n$ direction only through a rotation of basis. In the case of straight trajectory those directions decouple. For a straight trajectory, the transverse background velocities and potential gradients vanish. 

On super-Hubble scales, a time dependence of $\zeta$ can be generated by this turning of trajectory:
\begin{align}
  \dot\zeta = - \frac{H}{\rho+p}  \delta p_\mathrm{nad} = - \frac{2H}{\dot\chi_1} \dot e_1^i \delta\phi_i~.
\end{align}
Thus the adiabatic perturbation is sourced by the entropy perturbation along the $\dot e_1^i$ direction. Note that the $\dot e_1^i$ direction is perpendicular to $e_1^i$. It is sometimes convenient \cite{GrootNibbelink:2000vx,  Langlois:2008mn, Tye:2008ef} to take this direction as $e_2^i$:
\begin{align}
  e_2^i = \frac{\dot e_1^i}{\sqrt{\sum_j (\dot e_1^j)^2}} ~.
\end{align}
In this case the $e_2^i$ direction is the source of time dependent $\zeta$ on super-Hubble scales.

Note that those new add-ons of curvature perturbation come from different source from the quantum fluctuation of the original inflaton direction. Thus the power spectrum $P_\zeta$ tends to increase instead of decrease, with the presence of turning of trajectory. It is in general difficult to suppress the adiabatic perturbation, once it is generated \cite{Sloth:2005yx, Bartolo:2005jg, Li:2008jn}. 

Before going into detailed models, a few comments are in order here:
\begin{itemize}
\item The adiabatic/entropy perturbations are also called curvature/isocurvature perturbations.
\item Note that we have defined two versions for adiabatic and entropy perturbations: one for fluid and the other for field from rotating basis. At first order on super-Hubble scales, field-space entropy perturbations can source the non-adiabatic pressure through the turning of the background trajectory. At second order perturbation theory, the classification of adiabatic/isocurvature perturbation becomes not unique. One can consistently use either of them. Also on sub-Hubble scales, even a single field has non-vanishing $\delta p_\mathrm{nad}$.
\item Sometimes (e.g. the curvaton scenario), the fields which are rolling much more slowly than the inflaton (approximately aligned with the instantaneous entropy directions) are also considered isocurvature directions and generating entropy perturbations. 
\item We have shown that at linear order, entropy perturbation can convert to the comoving curvature perturbation only if the inflationary trajectory turns. However, this is under the assumption of standard kinetic term for both fields. Otherwise, kinetic coupling can couple between adiabatic and entropy perturbations without any turning of trajectory.
\item At non-linear order, there can be coupling of adiabatic and entropy fluctuations (when defined as in the field theory way as Fig.~\ref{fig:multi-orth}) even without turning of trajectory. This is because the adiabatic direction is defined at the background level. At the perturbation level, the real field trajectory is perturbed slightly \cite{Langlois:2006vv}.
\end{itemize}

In the next a few sections, we are to discuss a few models where the entropy perturbation converts to adiabatic perturbation. A summary of those models in light of the Planck data can be found in \cite{Suyama:2013nva}.

\subsubsection{Curvaton mechanism}

\begin{figure}[htbp]
  \centering
  \includegraphics[width=0.95\textwidth]{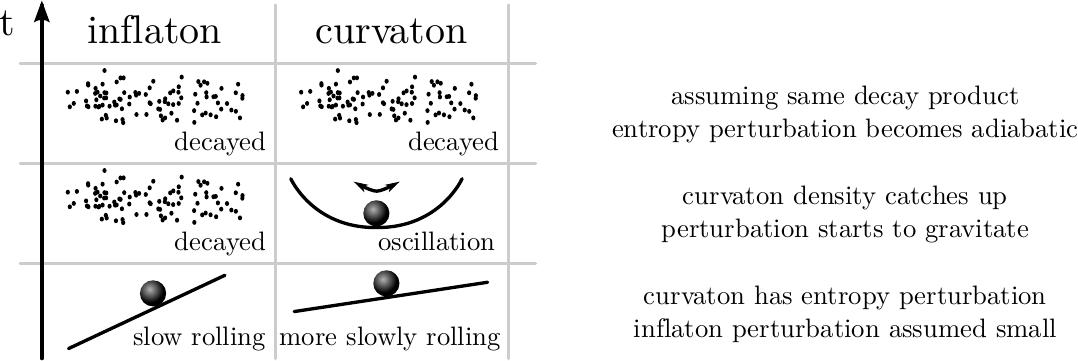}
  \caption{\label{fig:curvaton} The curvaton mechanism.}
\end{figure}

In the curvaton mechanism \cite{Mollerach:1989hu, Linde:1996gt, Enqvist:2001zp, Lyth:2001nq, Moroi:2001ct}, it is assumed that there is a curvaton field $\sigma$, which rolls much more slowly than the inflaton during inflation. After inflation, the inflaton decays first. Later the curvaton energy density catches up. When the curvaton energy density becomes non-negligible, the entropy fluctuation converts into adiabatic perturbation. This is illustrated in Fig.~\ref{fig:curvaton}.

For simplicity here we review the simplest scenario of curvaton: the inflaton decays to radiation. And the curvaton oscillate around the minima of its potential $V(\sigma)=\frac{1}{2} m^2\sigma^2$ before curvaton decay, which happens later than the inflaton decay. We use the method developed in \cite{Sasaki:2006kq}.

Recall the discussion in Sec.~\ref{sec:comov-curv-pert}, there are separately conserved quantities for non-interacting field species. The conserved quantity associated to the curvaton field is
\begin{align}\label{eq:zeta-sigma-curvaton}
  \zeta_\sigma^\mathrm{(gi)} = \delta N 
  + \frac{1}{3} \int_{\bar \rho_\sigma(t)}^{\rho_\sigma(t,\mathbf{x})} \frac{d\rho_\sigma}{\rho_\sigma} ~,
\end{align}
where we have used $p_\sigma=0$ for a fast oscillating field. Here and later in this subsection, we shall omit the superscript (gi) on $\zeta_\sigma$ for simplicity. 

This conserved quantity provides connection between initial condition and observations. Thus we have to: (a) relate the conserved quantity with the horizon crossing value of the curvaton field $\sigma_*$; and (b) relate the conserved quantity with the curvature perturbation $\zeta$. 

To relate $\zeta_\sigma$ with $\sigma_*$, we note that on spatial flat slice ($\delta N=0$), Eq.~\eqref{eq:zeta-sigma-curvaton} can be integrated as
\begin{align}\label{eq:curv-zeta}
  \rho_\sigma = e^{3\zeta_\sigma} \bar \rho_\sigma~.
\end{align}
The curvaton energy density is
\begin{align}\label{eq:curv-rho}
  \rho_\sigma = \frac{1}{2} m^2\sigma^2~,
\end{align}
where $\sigma$ is the amplitude of curvaton oscillation. 

However, the quantum initial condition for $\sigma$ perturbations are determined at Hubble-crossing of the corresponding perturbation modes. Thus we have to relate $\sigma$ at the oscillation stage to the value at Hubble crossing\footnote{Note that, here we are aiming at a full non-linear analysis of perturbations. Thus rigorously speaking the perturbation is not a single $k$ mode and $k=aH$ is not well defined. However, observationally we are only interested in a small range of $k$. Considering the $\sigma$ perturbation is nearly scale invariant, the horizon-crossing value do not differ very much.}: $\sigma_* \equiv \sigma|_{k=aH}$. At Hubble crossing, the curvaton is a spectator field, thus
\begin{align}\label{eq:curv-sigma}
  \sigma_* = \bar \sigma_* + \delta\sigma_*~,
\end{align}
where $\delta\sigma_*$ is Gaussian and has power spectrum
\begin{align}
  P_* = \frac{H_*^2}{4\pi^2} ~.
\end{align}

The relation between $\sigma$ and $\sigma_*$ is model dependent. In general, we can write this dependence as
\begin{align}\label{eq:curv-g}
  \sigma = g(\sigma_*).
\end{align}

Combining \eqref{eq:curv-zeta}, \eqref{eq:curv-rho}, \eqref{eq:curv-sigma} and \eqref{eq:curv-g}, we have\footnote{Note here $\bar \rho_\sigma$ is no longer the average energy density because there is also a contribution from $\frac{1}{2} m^2\langle(\delta\sigma)^2\rangle$. However, it is still consistent and convenient to define the background as $\delta\sigma=0$ instead of $\delta\rho_\sigma$=0~.}
\begin{align} \label{eq:curv-sigma-nl}
  e^{3\zeta_\sigma} = \frac{1}{\bar g^2} 
  \left[
    \bar g + \sum_{n=1}^{\infty} \frac{g^{(n)}}{n!} \left( 
        \delta\sigma_*
    \right)^n
  \right]^2
\end{align}
where 
$\bar g\equiv g(\bar\sigma_*)$ and 
$g^{(n)}\equiv \partial^n g(\sigma)/\partial\sigma^n |_{\sigma=\bar \sigma_*}$. Eq.~\eqref{eq:curv-sigma-nl} is a nonlinear relation between the $\sigma$-comoving curvature perturbation and the Gaussian fluctuation $\delta\sigma_*$. 

We need yet a relation between $\zeta$ and $\zeta_\sigma$. To derive this, we
make use of \eqref{eq:zdn-gi} and its counter part for radiation, on uniform total energy density slice. As a result,
\begin{align}
  \zeta_r = \zeta + \frac{1}{4} \ln \left( \frac{\rho_r}{\bar \rho_r}  \right)~, 
  \quad\rightarrow\quad \rho_r=\bar \rho_r e^{4(\zeta_r-\zeta)}~,
\end{align}
\begin{align}
  \zeta_\sigma = \zeta + \frac{1}{3} \ln \left( \frac{\rho_\sigma}{\bar \rho_\sigma}  \right)~,
  \quad\rightarrow\quad \rho_\sigma=\bar \rho_\sigma  e^{3(\zeta_\sigma-\zeta)}~,
\end{align}
Thus we have, on uniform total energy density slice\footnote{Here we assume curvaton decays on uniform total energy density slice. Otherwise, there is additional slice matching \cite{Cai:2010rt}.},
\begin{align}
  \bar \rho_r + \bar \rho_\sigma = \rho_r + \rho_\sigma 
  = \bar \rho_r e^{4(\zeta_r-\zeta)} + \bar \rho_\sigma  e^{3(\zeta_\sigma-\zeta)}~.
\end{align}
In case the final adiabatic perturbation is dominated by the curvaton perturbation, we have $\zeta_r=0$. In this case, the above equation reduces to
\begin{align}
  (1-\Omega) e^{-4\zeta} + \Omega e^{3(\zeta_\sigma-\zeta)} = 1~,
\end{align}
where
\begin{align} \label{eq:curv-zeta-nl}
  \Omega \equiv \frac{\bar \rho_\sigma}{\bar \rho_\sigma+\bar \rho_r} ~.
\end{align}

Thus finally one can solve $\zeta$ from Eqs. \eqref{eq:curv-sigma-nl} and \eqref{eq:curv-zeta-nl} order by order in terms of the Gaussian perturbation $\delta\sigma_*$. As a result, 

\begin{align}
  P_\zeta = r^2 \left( \frac{g' H_*}{3\pi g}  \right)^2 ~,
\end{align}
\begin{align}
  f_{NL} = \frac{5}{4r} \left( 1+\frac{gg''}{g'{}^2}  \right) -\frac{5}{3} -\frac{5r}{6} ~,
\end{align}
\begin{align}
  g_{NL} = \frac{25}{54} \left[ 
    \frac{9}{4r^2} \left( \frac{g^2g'''}{g'{}^3} + \frac{3gg''}{g'{}^2} \right)
    - \frac{9}{r} \left( 1+\frac{gg''}{g'{}^2}  \right)
    + \frac{1}{2} \left( 1-\frac{9 g g''}{g'{}^2}  \right) + 10r + 3r^2 \right]~,
\end{align}
where we have omitted the bar on $\bar g$, and 
\begin{align}
  r \equiv \frac{3\Omega}{4-\Omega} = \frac{3 \bar \rho_\sigma}{3\bar \rho_\sigma + 4\bar \rho_r} ~,
\end{align}
with $\bar \rho_\sigma$ and $\bar \rho_r$ evaluated at curvaton decay time.

Note that in the $r\ll 1$ limit, $f_{NL}\gg 1$. It is intuitive to understand why there is large non-Gaussianity in the curvaton scenario: when $r\ll 1$, the curvaton is subdominant. Thus it has to fluctuate harder to have the correct $P_\zeta$. The harder curvaton fluctuates, the more non-Gaussian it is. Also, the inflationary slow roll parameter $\epsilon$ doesn't enter this later time calculation thus $f_{NL}$ is not $\epsilon$-suppressed. This is because the curvaton potential near the curvaton oscillation is not slow roll suppressed.

The other trispectrum estimator is trivial, considering the single-source nature of curvaton mechanism:
\begin{align}
  \tau_{NL} = \frac{36}{25} f_{NL}^2~.
\end{align} 

In the simplest case, if the curvaton has only quadratic potential, there is no nonlinear evolution on super-Hubble scales. Thus in this case $g''=0$ and $g'''=0$. The above formulas simplify considerably in this limit. Otherwise the dynamics can be solved perturbatively or numerically in $g$ \cite{Enqvist:2005pg, Huang:2008bg, Kawasaki:2011pd}. 

\subsubsection{Non-trivial reheating surface in field space}
\label{sec:non-triv-rehe}
Another way to convert entropy perturbation to adiabatic perturbation is to make use of a non-trivial reheating surface. In this section we consider a non-dynamical reheating surface but dynamical trajectory due to entropy perturbations. A dynamical reheating surface is discussed in the next subsection.

\begin{figure}[htbp]
  \centering
  \includegraphics[height=0.24\textheight]{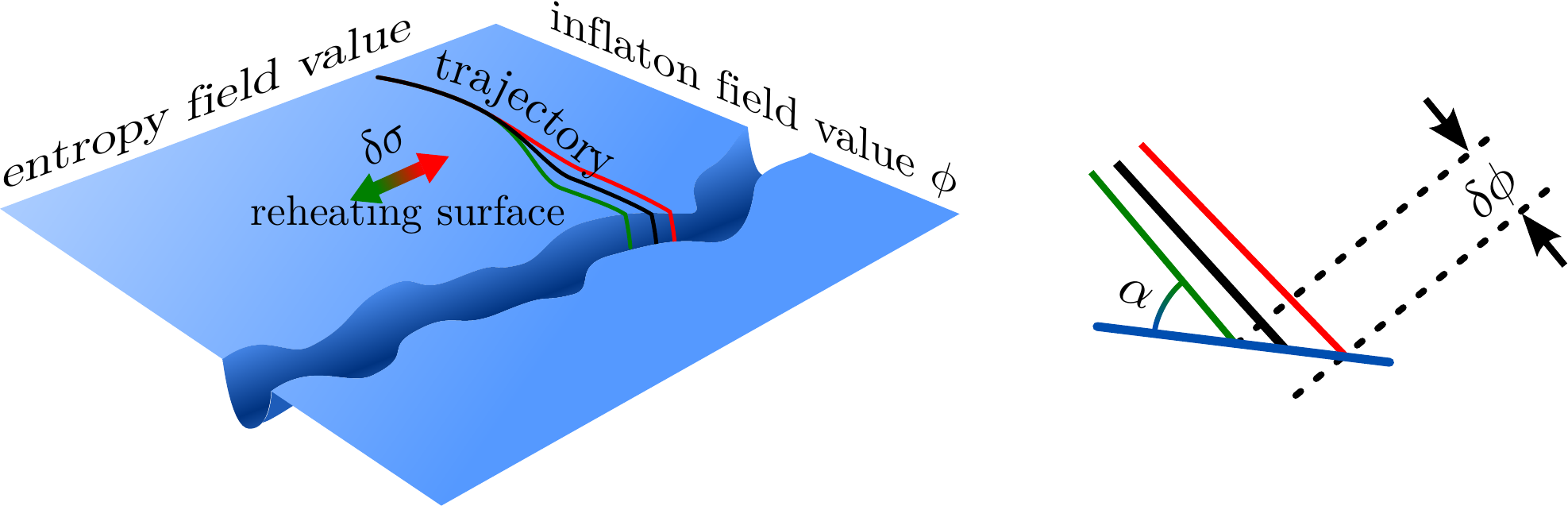}
  \caption{\label{fig:reh-suf} Non-Gaussianity from a non-trivial reheating surface in field space. When the reheating surface is not perpendicular to the inflationary trajectory, entropy fluctuation converts into e-folding number difference, and thus adiabatic perturbation.}
\end{figure}

Conventionally, we assume inflation terminates when the inflation direction $\phi$ (or $\chi_1$ after a change of basis) reach a certain (range of) value, where potential is no longer flat or the inflaton field suddenly decays to something else. 

However, the above simple picture is not necessarily the full story. In general, the reheating surface could be a hypersurface in field space, which is not perpendicular to the inflationary trajectory. The reheating surface may be even curved instead of flat. Such a generalized reheating surface can convert entropy perturbation into curvature perturbation, with a possibility for large non-Gaussianity. This situation is illustrated in Fig.~\ref{fig:reh-suf}. An explicit example of non-trivial reheating surface is known as multi-brid inflation \cite{Sasaki:2008uc}, and more general analysis is performed in \cite{Huang:2009vk}.

As shown in the right panel of Fig.~\ref{fig:reh-suf}, the amount of isocurvature perturbation (in terms of field space distance), converted to curvature perturbation, is characterized by the angle $\alpha$, between the reheating surface and the inflation trajectory:
\begin{align}
  \delta\phi = \delta\sigma \cot \alpha~.
\end{align}
Note that $\zeta = -H \delta\phi/\dot\phi$. Thus in case the $\delta\sigma$ fluctuation is conserved from Hubble crossing to reheating, then the adiabatic power spectrum induced by the entropy perturbation is
\begin{align}
  \Delta P_\zeta = \cot^2\alpha \frac{H^2}{8\pi^2M_p^2\epsilon} ~.
\end{align}
For standard inflaton sector, the total power spectrum is
\begin{align}
  P_\zeta = \left( 1+\cot^2\alpha\right) \frac{H^2}{8\pi^2M_p^2\epsilon} ~.
\end{align}
More generally, the $\sigma$ field could change its value from Hubble crossing to reheating. In this case the technique in the previous section can be employed to calculate the density perturbation. 

The non-Gaussianity of this scenario follows from how the reheating surface is curved \cite{Huang:2009vk}. When the inflaton trajectory hits a convex reheating surface, a positive $f_{NL}$ is created. 

\begin{figure}[htbp]
  \centering
  \includegraphics[width=0.9\textwidth]{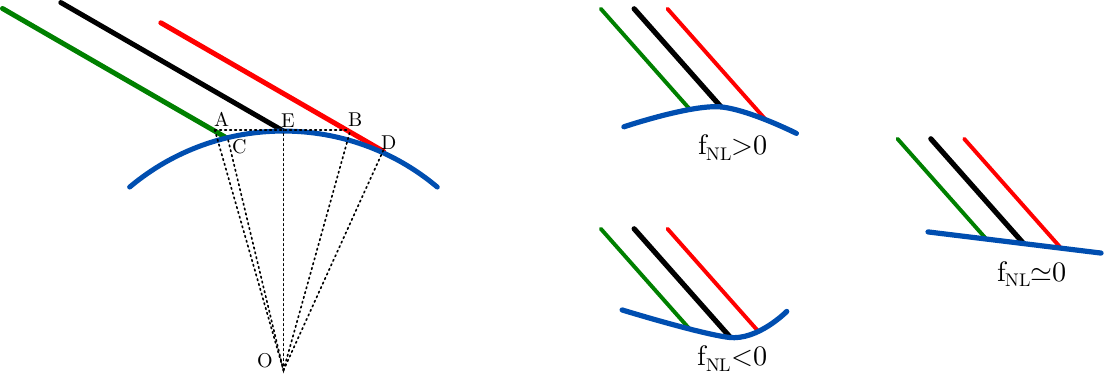}
  \caption{\label{fig:reh-suf-ng} The curvature of reheating surface (or any other hypersurface on which entropy perturbation is converted into adiabatic perturbation) and the sign of non-Gaussianity. }
\end{figure}

This fact can be observed geometrically. For small isocurvature perturbations, the curved reheating surface can be approximated as a circle. Consider the left panel of Fig.~\ref{fig:reh-suf-ng}. Here the green and red solid lines denote different inflation trajectories, with perturbation $-\delta\sigma$ and $\delta\sigma$ respectively. Note that we have $|BD|>|AC|$.\footnote{Proof: in the figure the line intervals satisfy $|AO|=|BO|$ and $|CO|=|DO|$, where $O$ is the center of the circle. From the construction we have angles $\angle BOD > \angle AOC$ (this can be proved by rotating triangle $\triangle AOC$ by angle $\angle AOB$). Thus interval $|BD|>|AC|$}

Recall the observation of section \ref{sec:local-non-gauss}. $\zeta = \zeta_g + \frac{3}{5}f_{NL}\zeta_g^2 + \cdots$. Thus positive $f_{NL}$ corresponds to an enhancement of positive $\zeta_g$. Note that the red line (positive $\delta\sigma$) is positive $\zeta_g$ (because of greater $\delta N$), and $|BD|>|AC|$ means the second order effects enhance the $\delta N$ of the red line compared with the green. Thus when the trajectory hits a convex reheating surface, a positive $f_{NL}$ is produced.

Note that if the inflation trajectory is perpendicular to the reheating surface, there is vanishing three-point function because of the symmetry of the configuration. However, once the trajectory hits the curved reheating surface with an angle, the sign of $f_{NL}$ only depends on how the reheating surface is curved.

Similarly, hitting a concave surface produces a negative $f_{NL}$. A reheating surface without any curvature converts the isocurvature perturbation to curvature perturbation linearly and thus there is no non-Gaussianity. This situation is illustrated in the right panel of Fig.~\ref{fig:reh-suf-ng}.

The relation between curved reheating surface and non-Gaussianity can also be generalized. Note that the curved reheating surface is nothing but a hypersurface to convert entropy perturbation to adiabatic perturbation. If during inflation, or after reheating, such conversion happens, the blue surface in Fig.~\ref{fig:reh-suf-ng} can be considered as the conversion surface from entropy to adiabatic perturbations.

\begin{figure}[htbp]
  \centering
  \includegraphics[width=0.8\textwidth]{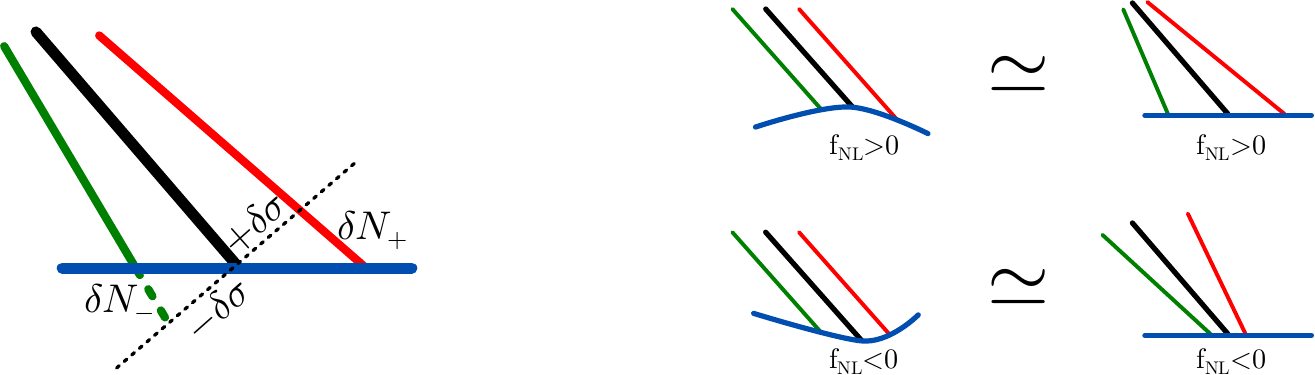}
  \caption{\label{fig:reh-suf-gen} A bundle of diverging trajectories corresponds to positive $f_{NL}$. This is because the positive part of $\zeta_g$ converts more and the negative part converts less. Similarly, a bundle of converging trajectories corresponds to negative $f_{NL}$. }
\end{figure}

Further, it is still possible that the reheating surface (or other hypersurface with entropy to adiabatic conversion) is not curved, however the field trajectories themselves are diverging or converging when they hit the surface. This situation is illustrated in Fig.~\ref{fig:reh-suf-gen}.

Finally, it is interesting to generalize the above argument from a ``hypersurface'' of entropy to adiabatic conversion, into some continuous conversion mechanism (see, e.g. \cite{Seery:2012vj}). Or consider originally non-Gaussian $\delta\sigma$ fluctuations (thus they are no longer $\zeta_g$). 

\subsubsection{Modulated reheating}
\label{sec:modulated-reheating}
\begin{figure}[htbp]
  \centering
  \includegraphics[height=0.24\textheight]{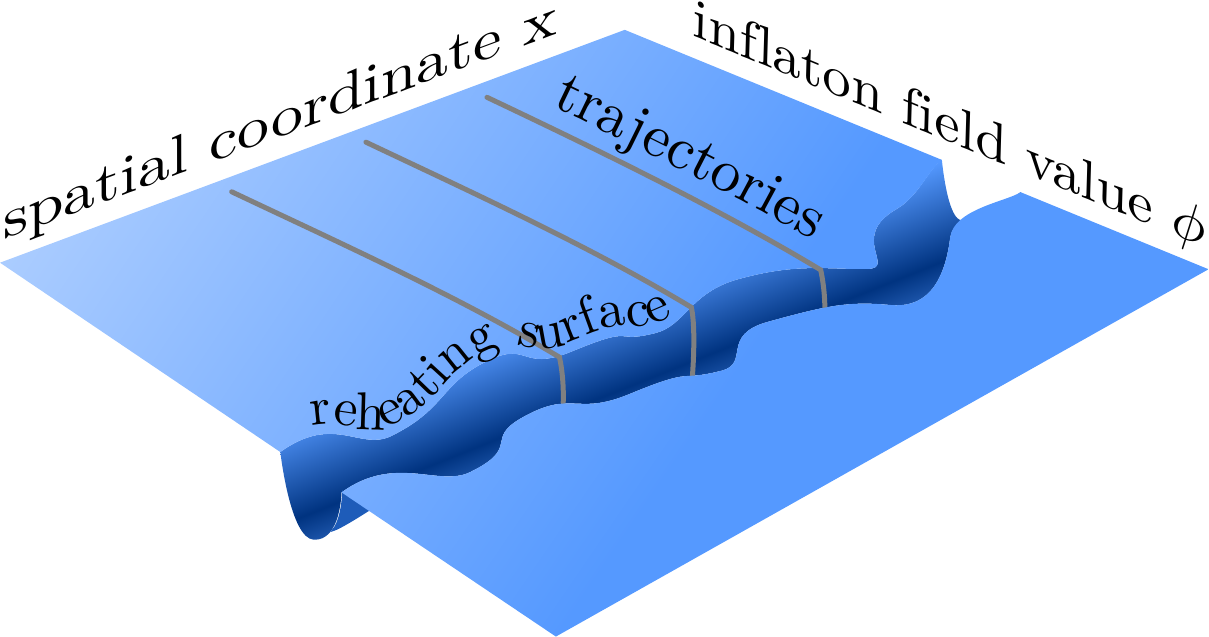}
  \caption{\label{fig:mod-reh} Modulated reheating. Note that despite the similarly looking of this figure with Fig.~\ref{fig:reh-suf}, here the ``/'' direction axis denotes spatial coordinate, instead of field value. The other axis has the same meaning (inflaton field value).}
\end{figure}

In this subsection we again play with the reheating surface of multi-field inflation. However, instead of the way to hit reheating surface, here we make the shape of the reheating surface dynamically depend on the entropy fluctuations. This scenario is known as the modulated reheating \cite{Dvali:2003em,Kofman:2003nx, Suyama:2007bg}. This situation is illustrated in Fig.~\ref{fig:mod-reh}. 

Consider that the decay rate of inflaton depends on the entropy field:
\begin{align}
  \Gamma_\phi = \Gamma(\sigma)~.
\end{align}

During reheating, the inflaton energy density $\rho_\phi$ decays into radiation $\rho_r$. This process can be described by
\begin{align}
  \frac{d\rho_\phi}{dN} + 3\rho_\phi = -\frac{\Gamma}{H} \rho_\phi~,
\end{align}
\begin{align}
  \frac{d\rho_r}{dN} + 4\rho_r = \frac{\Gamma}{H} \rho_\phi~,
\end{align}
\begin{align}
  3M_p^2H^2 = \rho_\phi + \rho_r ~.
\end{align}
The above equations are to be solved with initial condition, at a time $t_0$ before the inflaton decay, as $\rho_\phi(t_0)=3M_p^2H_0^2$ and $\rho_r(t_0)=0$. The e-folding number during this process can be calculated as
\begin{align} \label{eq:mod-reh-N}
  N = \frac{1}{2} \ln \frac{H_0}{H_f} + Q \left( \frac{\Gamma}{H_0}  \right)~,
\end{align}
where $Q$ satisfies
\begin{align}
  \exp \left[ 4Q \left( \frac{\Gamma}{H_0}  \right) \right] = \int_0^\infty dN' \frac{e^{4N'}\Gamma}{H(N')} \frac{\rho_\phi(N')}{\rho_\phi(t_0)} ~.
\end{align}
Expanding Eq.~\eqref{eq:mod-reh-N}, one obtains 
\begin{align}
  \Delta P_\zeta = \left( \left.\frac{dQ(x)}{dx}\right|_{x=\Gamma/H_0} \frac{\Gamma'}{H_0} \frac{H_*}{2\pi} \right)^2 ~.
\end{align}
We again assumed extremely slow-rolling of $\sigma$. Otherwise the dynamics of $\sigma$ should be tracked similarly to the curvaton scenario. The non-Gaussianities can be calculated by expanding Eq.~\eqref{eq:mod-reh-N} to higher orders \cite{Suyama:2007bg}.

For the sign of non-Gaussianities, the intuition in the last section (Fig.~\ref{fig:mod-reh}) still applies. Now we work in position space instead of field space. The distribution of $\sigma$ field is Gaussian. However, the conversion $\Gamma(\sigma)$ makes the reheating surface non-Gaussianly curved. The sign of non-Gaussianity depends on both derivatives of $\Gamma(\sigma)$ and the nonlinear response of the expansion history. In the slow decay limit, $N\simeq-(1/6)\ln\Gamma+\mathrm{const}$ gives $f_{NL}=5[1-\Gamma\Gamma''/(\Gamma')^2]$.

\subsubsection{Modulated expansion history}

In this section we discuss a mechanism similar to the modulated reheating, where the expansion history, instead of the decay rate, is modulated by an isocurvature direction \cite{Dvali:2003ar}\footnote{The discrepancy in \cite{Vernizzi:2003vs} was resolved by correcting the initial conditions in Sec.~5 and Eq.~(33) of \cite{Vernizzi:2005mass}, recovering $\zeta=\zeta_*+\frac13\delta m/m$ in the limit of complete mass domination. Here we follow \cite{Dvali:2003ar}, reformulated using the $\delta N$ formalism.}.

For example, after inflation ends, the inflaton field decays to thermalized particles $\chi$, with mass $m$. The reheating temperature is $T_I \gg m$. The $\chi$ particles are relatively long lived, such that it decays after the temperature of the universe drops to $T_F\ll m$. Thus from reheating to $\chi$-decay, the universe is initially radiation dominated and later matter dominated. After the $\chi$-decay to lighter particles the universe becomes hot again. 

To track the local expansion history of this period, here we use a sudden transition approximation, neglecting the transition period between radiation domination and matter domination. This should be a reasonable approximation, because after $m$ varies at different local universes, those different local universes undergoes the same transition (at different times). Thus the error cancels when we calculate the e-folding number difference $\delta N$.

We use $T_{E}$ and $H_{E}$ to denote the temperature and Hubble parameter at matter radiation equality time, for the $\chi$-particle dominated universe. The expansion from the initial time to the equality time is
\begin{align}
  N_{IE} = \log (a_E/a_I) = \frac{1}{2} \log (H_I/H_E)~,
\end{align}
and the expansion from the equality time to the final time is
\begin{align}
  N_{EF} = \log (a_F/a_E) = \frac{2}{3} \log(H_E/H_F)~.
\end{align}
The total e-folds is a summation of the two stages:
\begin{align}
  N_{IF} = \frac{1}{2} \log H_I + \frac{1}{6} \log H_E
  - \frac{2}{3} \log H_F ~.
\end{align}

To relate the expansion to the $\delta N$-formalism, we have to choose the initial slice to be the flat slice and the final slice to be the uniform energy density slice. Assuming the original density fluctuation of the $\chi$ particles (originates from inflaton fluctuation) is negligible. In this case the uniform $H_I$ slice is indeed the initial flat slice. On the other hand, the final uniform $H_F$ slice is the uniform energy density slice because $H_F\propto\sqrt{\rho_F}$ from the Friedmann equation. Thus when we vary $N_{IF}$, $H_I$ and $H_F$ should be fixed. The only variation comes from $H_E$. The mass, energy density and Hubble parameter can be related as\footnote{There is an additional factor of $7/8$ in the energy density for fermions, which do not affect the final result.}
\begin{align}
  \rho_E\simeq \frac{\pi^2}{30} g T_E^4 \propto m^4 ~,\qquad H_E \propto \sqrt{\rho_E} \propto m^2~.
\end{align}
Here we still approximately used the energy density of radiation dominated epoch. The difference results in an error of the sudden transition approximation. 

Thus the comoving curvature perturbation can be calculated as
\begin{align}
  \zeta = \delta N_{IF} = \frac{1}{6} \frac{\delta H_E}{H_E} = \frac{1}{3} \frac{\delta m}{m} ~.
\end{align}

Now we need to calculate $\delta m$ in the $\delta N$-formalism. As a simplest case: we may directly have $m=m(\sigma)$, where $\sigma$ is an isocurvature field which gets fluctuations during inflation. In this case, $\delta m = m'(\sigma)\delta\sigma$. If $\sigma$ is rolling extremely slowly during inflation, such that the Hubble-exit value is the same as the value at equality time, the induced power spectrum for curvature perturbation is
\begin{align}
  P_\zeta = \frac{1}{9} \left( \frac{m'}{m}  \right)^2  P_{\delta\sigma} 
  = \frac{H_*^2}{36\pi^2} \left( \frac{m'}{m}  \right)^2~,
\end{align}
where $H_*$ is the horizon crossing value of Hubble parameter. Similarly, the non-Gaussian estimators can be calculated.

On the other hand, if $\sigma$ is rolling not so slowly, but follow linear evolution (i.e. the potential is just a mass term), $\delta\sigma/\sigma$ is a constant on super-Hubble scales. In this case, when $m\propto\sigma^{\alpha}$, we have $\delta m/m = \alpha \delta\sigma/\sigma$ and
\begin{align}
  P_\zeta = \frac{\alpha^2 H_*^2}{36\pi^2\sigma_*^2} ~,
\end{align}

Finally, if the $\sigma$ potential is dominated by terms other than mass during some periods of time evolution, one can define a function $g$ such that $\sigma = g(\sigma_*)$, and calculate $g$ perturbatively, as is done in the curvaton mechanism.

As the last step, $\chi$ has to decay, preferably to radiation components and re-reheat the universe. Note that the decay rate of $\chi$ may also depend on $m$. There are at least two possibilities:

\textit{No modulation in re-reheating:} If the mass retains its modulation while the decay rate $\Gamma_\chi$ is independent of $\sigma$, decay occurs on a uniform-Hubble slice and introduces no additional modulation.

\textit{Modulated re-reheating:} In this case, the discussion for the extra curvature perturbation transfer is in parallel to the modulated reheating.

As alternative realizations, one may either modulate the expansion rate at other stages of the universe. For example, one can modulate the dilution rate of isocurvature direction, to generate entropy perturbation. As another example, during the oscillation of the inflaton, before reheating, the shape of the potential may get modulated. For example, near the end of inflation, the mass term of inflaton may get an anomalous dimension\footnote{We assume other terms drives inflation, and this potential only applies near the end of inflation. Otherwise severe fine tuning is needed to make $\sigma$ massless.}:
\begin{align} \label{eq:d-coupling}
  V 
  =  \frac{1}{2} m^2\phi^2 \exp \left[ \frac{\sigma}{\Lambda} \log \left| \frac{\phi}{m}  \right| \right]~,
\end{align}
As a result, the dilution of inflaton no longer has equation of state $p/\rho=0$, but instead
\begin{align}
  \frac{p}{\rho}  = \frac{\sigma}{\sigma+4\Lambda} \simeq \frac{\sigma}{4\Lambda} ~.
\end{align}
Thus the expansion history is modified as a function of the $\sigma$ value in local universes. Assuming $\sigma\ll\Lambda$,
\begin{align}
  \zeta \simeq - w\Delta N \simeq -\frac{\sigma}{4\Lambda} \Delta N~,
\end{align}
where $\Delta N$ is the number of e-folds of inflaton oscillation before it decays. The inflationary power spectrum follows from squaring this equation.

\subsubsection{A landscape of inflatons and isocurvatons?}

In the previous subsections, we have been making the multi-field inflationary potential complicated. Inspired by the string landscape \cite{Bousso:2000xa, Giddings:2001yu, Kachru:2003aw, Susskind:2003kw}, it is argued that inflation may take place on a rather complicated potential \cite{Tye:2008ef, Agarwal:2011wm, Frazer:2011br}. Yet there is a question: how complicated the inflationary potential could be?

To see the physics more closely, we decomposed the question into a few more concrete ones:

\begin{figure}[htbp]
  \centering
  \includegraphics[width=0.6\textwidth]{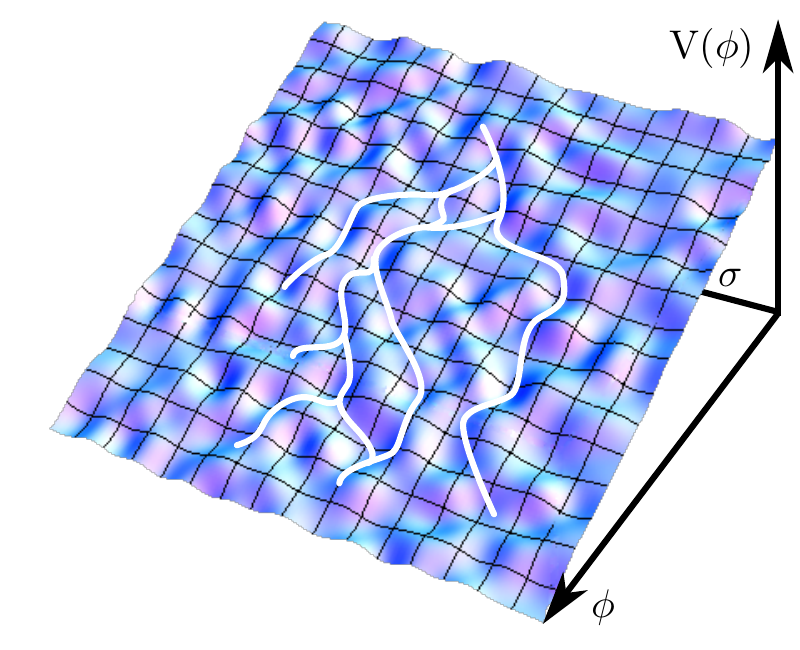}
  \caption{\label{fig:multi-stream} Once inflation takes place on a complicated multi-field potential, the inflationary trajectory (white lines in the figure) may bifurcate. The bifurcation, if happened during observable state, and not under control, brings disasters.}
\end{figure}

\begin{itemize}
\item How complicated could a potential be on one field direction? Note that slow roll of inflationary direction already tightly constraint the potential along the inflation direction. We postpone the discussion of possible break down of slow roll to Subsection \ref{sec:scale-depend-feat}.

\item How many fields could there be during inflation? The constraint \cite{Linde:1993xx, Huang:2007zt} is quite weak. Nevertheless a weak constraint is stronger than no constraint \cite{Adshead:2008gk}. 

Note that during inflation, each massless (actually $m<H$ should be enough) field has a fluctuation $\sigma_i\sim H$ per Hubble volume. Thus the total energy density of $n$ massless fields is
\begin{align}
  \rho_n \sim \sum_i^n (\partial\sigma_i)^2 \sim n H^4~.
\end{align}
This must be much smaller than the background energy density $\rho$. Thus we have
\begin{align} \label{eq:fn-bound}
  n \ll \frac{M_p^2}{H^2} ~.
\end{align}
This constraint can also be argued as that, near Hubble crossing, the quantum perturbation of each massless field becomes classical and carry of order one bit entropy (about its phase). However, de Sitter space has an entropy bound of order $M_p^2/H^2$. Thus the field number is bounded by \eqref{eq:fn-bound}.

Note that this number is greater than $10^{10}$. Thus most models are unconstrained, unless one wants the number of fields to be greater than $1/P_\zeta$ or $1/P_h$.

\item What extras could a multi-field complicated potential bring compared with a single field one?

As we already discussed, in multi-field inflation, much more things could happen compared to the case of single field inflation. Here regarding to a complicated potential, there is at least one more: The classical trajectory of inflation may bifurcate during inflation \cite{Li:2009sp}. Inflation with such a bifurcation is called multi-stream inflation, as illustrated in Fig.~\ref{fig:multi-stream}.

In case the bifurcation of inflationary trajectory is under control (with symmetry or fine-tuning), those bifurcations can be responsible to non-Gaussianities \cite{Li:2009sp} or position space features such as asymmetries and/or cold spots on the CMB \cite{Afshordi:2010wn}. Or alternatively the bifurcation could happen before the observable stage of inflation \cite{Li:2009me, Wang:2010rs}. However, if the bifurcation is due to a random potential, the e-folding number difference between different trajectories are not under control. In this case the CMB sky becomes extremely anisotropic and non-Gaussian. This kind of models should be ruled out.

It is shown that in a double-field random potential, bifurcation could happen with some not-so-fine tuning of potential parameters \cite{Duplessis:2012nb}. When the random potential has more dimensions, the probability for bifurcation should increase thus there may be a tighter constraint.

There is also a danger of rapid tunneling in a random potential, as discussed in \cite{HenryTye:2006tg, Tye:2009rb}.

\end{itemize}

\subsection{Quasi-single field inflation}

\subsubsection{Energy scales during inflation}

We have discussed single field inflation and multi-field inflation. Definitely we cannot cover even a small portion of the huge literature. Nevertheless at first look the classification of single/multiple field covers all possibilities.

However, it is actually not the case. Note that inflation is a theory with multiple energy scales built in. The inflationary background defines an energy scale $E_B \sim \sqrt{|\eta_V|} H$, whereas the perturbations defines another energy scale $E_P=H$. There is a hierarchy between those two energy scales. 

The hierarchy of energy scales allows a class of inflation models, in which at the homogeneous and isotropic background level, the models look like single field inflation. However, when we consider cosmic perturbations, which live at a higher energy scale, the physics becomes different from single field. This class of models is called quasi-single field inflation \cite{Chen:2009we}\cite{Chen:2009zp}\cite{Baumann:2011nk}.

\begin{figure}[htbp]
  \centering
  \includegraphics[width=0.8\textwidth]{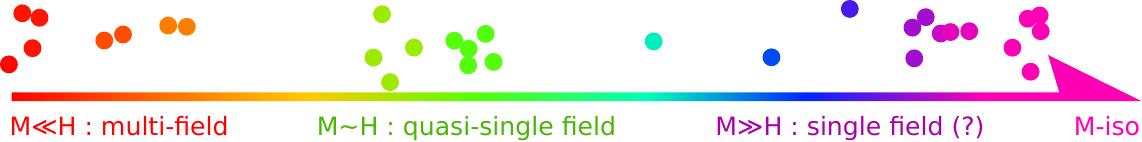}
  \caption{\label{fig:qsfi-scale} Energy scales during inflation. }
\end{figure}

For example, consider a scalar field $\sigma$ with mass $m\sim H$ during inflation (Fig.~\ref{fig:qsfi-scale}). At the background level, this field stays at its minimum of the potential, because the mass is higher than the background energy scale and not slow rolling. Thus at the background level we have single field inflation. 

However, at the energy scale of perturbations, $m\sim H$ is not too heavy. It takes quite a few e-folds for the $\sigma$ perturbation to roll back to the origin. As a result, there is room for the $\sigma$ perturbation, which appears as entropy perturbation, to convert to curvature perturbation.

Note that the potential of the $\sigma$ field is not restricted by slow roll condition. Thus the potential need not to be very flat. As a result, the self interactions of $\sigma$ can be strong. When the entropy perturbations in the $\sigma$ direction converts to curvature perturbation, the non-Gaussianity of $\sigma$, from the self interactions, are also converted. The generation and conversion of non-Gaussianity is near Hubble crossing. This is different from the local non-Gaussianity (which is generated on super-Hubble scales) or equilateral non-Gaussianity (which is generated on sub-Hubble scales). As a result, a family of new shapes of non-Gaussianity arises, controlled by the mass $m$ of $\sigma$.

\subsubsection{Mass uplifting of flat directions}

As we discussed in Sec.~\ref{sec:open-questions}, there is an $\eta_V$-problem of inflation: Originally massless fields (unless protected by symmetry) get uplifted to $m\sim H$ during inflation. Note that this is true not only for the inflaton field, but also for other originally massless fields. 

If we have some originally massless fields in field theory. Now inflationary back-reaction typically uplift them to $m\sim H$. Some of them may have $m\ll H$ by pure chance\footnote{The chance may not be small if there are many $m\sim H$ fields.}  or some fine tuning. They can behave as the inflaton, because we want inflation. The other fields has $m\sim H$. A string theory realization of quasi-single field inflation is considered in \cite{McAllister:2012am}.

Among those $m\sim H$ fields, some may decouple from inflaton. The fluctuation of those fields are diluted after a few e-folds of super-Hubble evolution and thus not observable. However, others may couple to inflaton, via operators \cite{Baumann:2011nk}
\begin{align}
  \dot\phi^2 \sigma, \quad (\partial_\mu \phi)^2 \sigma ~,  \quad \sigma \partial_\mu \phi \partial^\mu \sigma~, \quad \cdots~.
\end{align}

\subsubsection{A simple model}
\label{sec:simple-model}
\begin{figure}[htbp]
  \centering
  \includegraphics[width=0.5\textwidth]{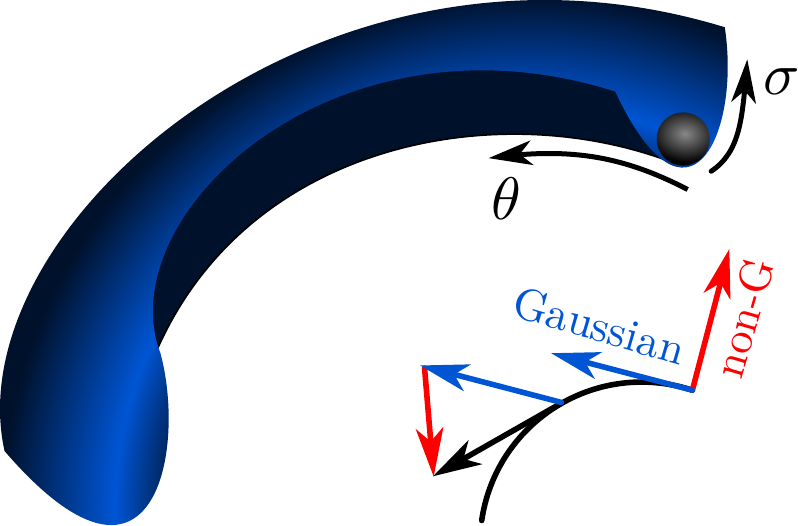}
  \caption{\label{fig:qsfi} A simple example of quasi-single field inflation. The isocurvature direction is not slow roll restricted. Thus the entropy fluctuation can be non-Gaussian. During the turning of trajectory, the non-Gaussian entropy fluctuation is converted to the adiabatic perturbation, along the inflation direction.}
\end{figure}

It is helpful to consider a simple model to see explicitly how the above argument works \cite{Chen:2009we}\cite{Chen:2009zp}. To do that, we consider a model in which the inflation trajectory is along an arc. The angular direction is the inflaton and the radius direction is the massive field $\sigma$. We illustrate the case in Fig.~\ref{fig:qsfi}.

For simplicity, we further assume the potential for the two scalars can be written in the factorized form when written in the polar coordinates. In this case, we can write down the theory in polar coordinates as
\begin{align} \label{eq:qsfi-action}
  S_m = \int d^4 x \sqrt{-g} \left[  
    -\frac{1}{2} \left(  \mathcal{\widetilde R} + \sigma \right)^2 g^{\mu\nu} \partial_\mu\theta \partial_\nu\theta - \frac{1}{2} g^{\mu\nu}\partial_\mu\sigma\partial_\nu\sigma - V_\mathrm{sr}(\theta) - V(\sigma)
  \right]~,
\end{align}
where $V_\mathrm{sr}(\theta)$ is the slow roll potential in the inflation direction; $V(\sigma)$ is the potential in the $V''\sim m^2\sim H^2$ direction, without slow roll restriction; and $ \mathcal{\widetilde R}$ is the radius of the turning potential. Gravity minimally couples to this matter sector. The background equations of motion are
\begin{align}
  3M_p^2H^2 = \frac{1}{2} \mathcal{R}^2 \dot\theta_0^2 + V + V_\mathrm{sr}~, \qquad -2M_p^2\dot H = \mathcal{R}^2\dot\theta_0^2~,
\end{align}
where $\mathcal{R}$ is the radius of turning trajectory, which has a shift by a constant compared with $\mathcal{\widetilde R}$ due to the centrifugal force: $\sigma_0\equiv R - \widetilde R$, such that $V'(\sigma_0) = R \dot\theta_0^2$.

As we have shown in the general discussion of adiabatic and entropy perturbations, the turning of trajectory couples the adiabatic and entropy directions. This can be verified in the present model by expanding \eqref{eq:qsfi-action} in perturbations. The interaction is
\begin{align}
  \Delta\mathcal{L}_2  = 2a^3  \dot\theta_0 \times ( \mathcal{R}\dot{\delta\theta} ) \times \delta\sigma ~.
\end{align}
Near Hubble-crossing, the canonically normalized field $\mathcal R\delta\theta$ has $\mathcal R\dot{\delta\theta}\sim H^2$, and for $m\sim H$ the entropy fluctuation has amplitude $\delta\sigma\sim H$. For $m\gg H$, the low-frequency virtual propagator scales as $m^{-2}$; integrating out $\delta\sigma$ gives $\Delta\mathcal L_\mathrm{eff}\sim a^3(\dot\theta/m)^2(\mathcal R\dot{\delta\theta})^2$ \cite{Chen:2012ge, Pi:2012gf}.

The corresponding dimensionless mixing strength can be estimated as
\begin{align}\label{eq:qsfi-g2}
  g_2 \sim \begin{cases}
    \dot\theta/H, & \text{if } m\sim H \\
    \dot\theta/m, & \text{if } m\gg H \\
    \text {cutoff~needed} & \text{if } m\ll H
  \end{cases} ~.
\end{align}

\begin{figure}[htbp]
  \centering
  \includegraphics[width=0.9\textwidth]{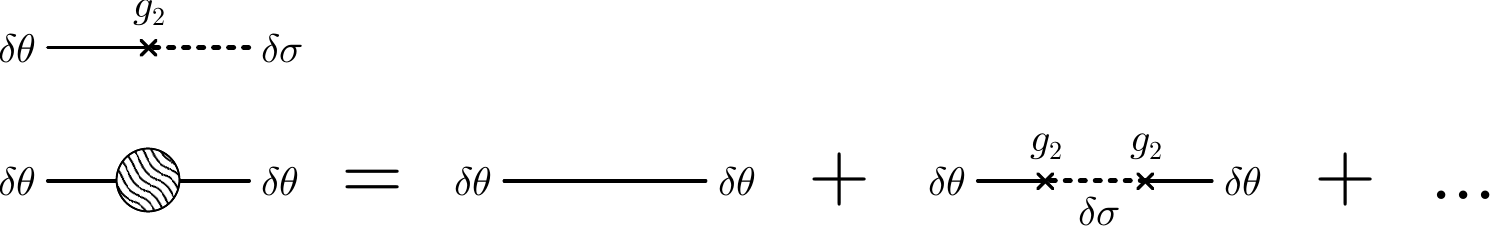}
  \caption{\label{fig:qsfi-int} The transfer vertex in quasi-single field inflation, and the correction to the power spectrum, due to the transfer vertex.}
\end{figure}

Because of the transfer vertex, the power spectrum gets a relative correction proportional to $(g_2)^2$. This situation is illustrated in Fig.~\ref{fig:qsfi-int}.

As a result, the power spectrum goes as
\begin{align}
  P_\zeta = \frac{H^4}{4\pi^2 R^2 \dot\theta_0^2} \left( 1+ \delta \right)~,
\end{align}
where $\delta$ is the relative correction due to the transfer vertex. The analytical expression of $\delta$ is given as $\delta=8\dot\theta^2(\mathcal{C}_1+\mathcal{C}_2)/H^2$ in \cite{Chen:2012ge}. The $g_2$ estimate in \eqref{eq:qsfi-g2} should apply for both $m\sim H$ and $m \gg H$. Thus in those cases, up to $\mathcal{O}(1)$ factors,
\begin{align}
  \delta \sim g_2^2~.
\end{align}
On the other hand, when $m\ll H$, there is a divergence that worth noticing. It can be shown that in this limit, if one take the bounds in the in-in integral as $-\infty<\tau<0$, $\delta$ ends up with a divergence in the small $m$ limit: $\delta\sim36(\dot\theta/H)^2H^4/m^4$. However this divergence is not real because inflation will eventually end. If we instead take $m=0$ and cut the in-in integral as $-\infty<\tau<\tau_c$, we have $\delta\sim (\dot\theta/H)^2\ln^2|k\tau_c|$. Thus in the $m\ll H$ limit, we have
\begin{align}
  \delta \sim \left(\frac{\dot\theta}{H}\right)^2 \min \left( \frac{H^4}{m^4} , N^2 \right), \qquad N\equiv-\ln|k\tau_c|\simeq60~,
\end{align}
where we have assumed $60$ e-folds of observable inflation, so that $N^2\simeq3600$. This perturbative estimate requires $\delta\ll1$. It turns out that when $m\simeq 0.3 H$ (which is actually not much smaller than H), we will already have to consider the e-folding number limit of inflation. 

The IR growth of $m\ll H$ case is straightforward to understand: for a massless field, when there is a fluctuation, the fluctuation will continues interacting with the adiabatic perturbation all the time until the interaction is turned off (for example, reheating). In flat space QFT, the cross sections and decay rates are defined as a finite value per unit time. Thus there is no wonder the interaction rate diverges with a diverge duration of time.

\subsubsection{Amplitude of non-Gaussianities}

The amplitude of non-Gaussianity can be estimated similarly to the power spectrum for quasi-single field inflation. The bispectrum can be estimated by the coupling
\begin{align}
  \Delta \mathcal{L}_3 = V''' \delta\sigma^3~.
\end{align}
Note that $V'''$ has mass dimension. Thus it is natural to have $V'''\sim H$ during inflation, similar to the $m\sim H$ argument. And again similarly to the $\eta_V$-problem, the $V'''$ term may have UV sensitivity driving $V''' \gg H$ straightly to the UV-sensitivity scale. One can estimate
\begin{align}\label{eq:qsfi-g3}
  g_3 \equiv \frac{\Delta\mathcal{L}_3}{\mathcal{L}_2} = \begin{cases}
    V'''/H, & \text{if } m\sim H \\
    V'''H^2/m^3, & \text{if } m\gg H \\
    \text {cutoff~needed} & \text{if } m\ll H
  \end{cases} ~.
\end{align}
Here we shall focus on the weakly coupled case $g_3\ll 1$, leaving the more general case to Sec.~\ref{sec:general-quasi-single}. 

We should note that this interaction $g_3$ is huge, compared with the inflaton counterpart, which is offered by a slow roll potential $V_\mathrm{sr}$. As shown in Eq.~\eqref{eq:Vppp-sr}, the slow roll potential is suppressed by
\begin{align}
  V_\mathrm{sr}''' \sim H \mathcal{O}(\epsilon^2)\sqrt{P_\zeta}~.
\end{align}
And we know this $V_\mathrm{sr}'''$ generates subdominant non-Gaussianity for single field inflation with amplitude $f_{NL}\sim \mathcal{O}(\epsilon^2)$. Thus we immediately know that the $f_{NL}^\mathrm{ent}$ in the entropy perturbations should be $f_{NL}^\mathrm{ent} \sim g_3/\sqrt{P_\zeta}$, which could be huge considering $1/\sqrt{P_\zeta}\sim 10^4$.

Here we obtain the $1/\sqrt{P_\zeta}$ factor by comparing with the single field potential. The same estimation could either be obtained as follows: The three point function is $\langle\zeta\zeta\zeta\rangle\sim P_\zeta^{3/2}\times$(coupling constant). On the other hand, the bispectrum estimator $f_{NL}$ is multiplied with $P_\zeta^2$ in its definition. To compare them we need to divide the result by $\sqrt{P_\zeta}$.

\begin{figure}[htbp]
  \centering
  \includegraphics[width=0.63\textwidth]{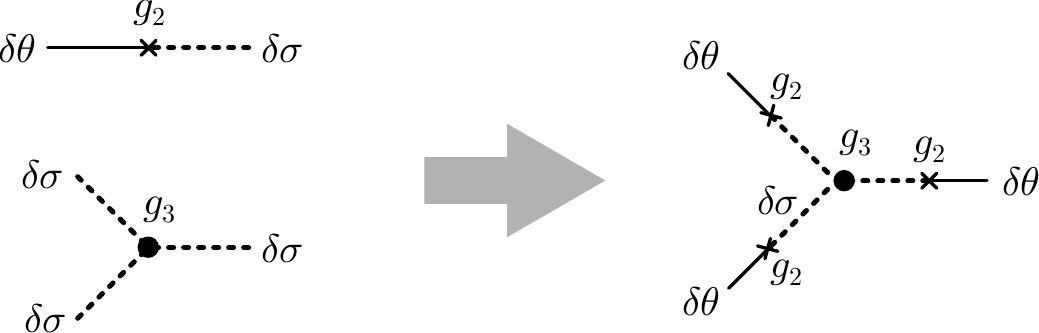}
  \caption{\label{fig:qsfi-int3} The interaction vertex and the three point correlation function.}
\end{figure}

Now we have got the amplitude of non-Gaussianity in the isocurvature direction. However, this entropy perturbation needs to be converted to adiabatic perturbation. We have shown the conversion is mediated by $g_2$ coupling. As a result, the amplitude of non-Gaussianity for quasi-single field inflation can be estimated as
\begin{align}
  f_{NL}\sim \frac{g_3 (g_2)^3 }{\sqrt{P_\zeta}} ~.
\end{align}
This amplitude is illustrated in Fig.~\ref{fig:qsfi-int3}. 

\begin{figure}[htbp]
  \centering
  \includegraphics[width=0.99\textwidth]{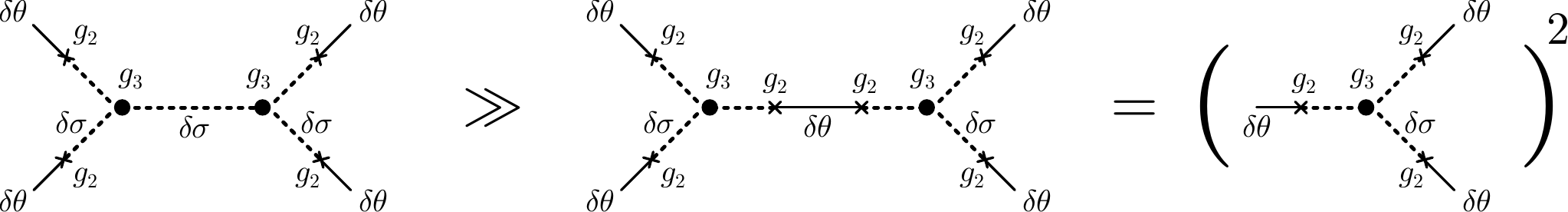}
  \caption{\label{fig:qsfi-sy} The Suyama-Yamaguchi relation. Here we have $\tau_{NL}\gg (36/25)f_{NL}^2$, as long as the transfer vertex is perturbative. In case $g_2\gg 1$, our perturbative calculation cannot be trusted. However, the Suyama-Yamaguchi relation should still hold. The implication is that, resummation should suppress the large coupling.}
\end{figure}

Similarly, the analogue of $g_{NL}$ and $\tau_{NL}$ can also be estimated. As we have shown in Sec.~\ref{sec:soft-limit-internal}, $\tau_{NL}$ is dominated by the diagram with an internal propagator, and $g_{NL}$ is dominated by the other diagram. Thus we have
\begin{align}
  g_{NL}\sim g_4 (g_2)^4/P_\zeta~, \qquad \tau_{NL} \sim (g_3)^2(g_2)^4 / P_\zeta~.
\end{align}
When $g_2$ is small (which is the case we can trust our calculation), we have
\begin{align}
  \tau_{NL} \gg f_{NL}^2~.
\end{align}
Thus the Suyama-Yamaguchi inequality is satisfied. The graphical understanding is given in Fig.~\ref{fig:qsfi-sy}. One can also estimate the higher order correlation functions \cite{Lin:2010ua}. The $\tau_{NL}$ analogue of the n-point function scales as
\begin{align}
  f^{(n)}_{NL} \sim \left( g_3/\sqrt{P_\zeta}\right)^{n-2} (g_2)^{n}~,
\end{align}
where we have factorized the product into two groups, one scales with power $n-2$ and another scales with $n$. Those two groups follows different interaction structures:
\begin{itemize}
\item The power $n-2$ group shows how the non-Gaussianity is generated. This is because for three point interaction (which generates $f_{NL}, \tau_{NL}, \cdots$, and $g_{NL}$ belongs to yet another case), we need $n-2$ vertices to have $n$-point function.
\item The power $n$ group shows how the non-Gaussianity is converted from entropy direction to curvature direction. This is because for $n$-point function, we need at least $n$ transfer vertices to convert all external legs to adiabatic perturbation.
\end{itemize}
This scaling behavior is discussed in more details in \cite{Barnaby:2011pe}, known as the feeding mechanism.

\begin{figure}[htbp]
  \centering
  \includegraphics[width=0.7\textwidth]{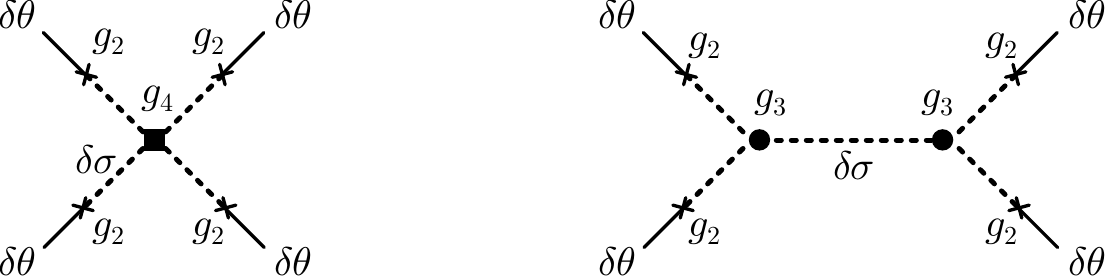}
  \caption{\label{fig:qsfi-int4} The quasi-local analog of $g_{NL}$ (left panel) and $\tau_{NL}$ (right panel).}
\end{figure}

\subsubsection{Shape of non-Gaussianities}

Now comes the shape of non-Gaussianity. To get some intuition, we may consider two limits: 
\begin{itemize}
\item When $m\ll H$, quasi-single field inflation approaches multi-field inflation and the shape should approach local shape. This is because when there is an entropy fluctuation, in the massless limit the fluctuation do not roll back. Thus the interaction continues on super-Hubble scales. From causality, the non-Gaussianity should be dominated by the local shape.
\item When $m\gg H$, the kinetic term of $\sigma$ becomes hard to excite. Thus the heavy field just stays at the effective minimal of the potential, determined by the centrifugal force $V'(\sigma)=R\dot\theta^2$. To perturb this equation, additional factors of $\dot{\delta\theta}^2$ comes into the action of $\delta\theta$, which modifies the sound speed $c_s$ of $\delta\theta$ perturbations. As a result, the non-Gaussianity should be very non-local.
\end{itemize}
In general, we expect the shape should be in between the above cases. In the squeezed limit $k_1\ll k_2\simeq k_3$, the $k$-dependence of the three-point function can be calculated explicitly \cite{Chen:2009zp}. Here we will not go into the details. Instead, as observed in \cite{Baumann:2011nk}, the squeezed limit of quasi-single field inflation can be estimated as follows: The long wave length mode crosses the horizon at $|k_1\tau_1|\sim 1$. Later, the short wave length mode crosses the horizon at $|k_2\tau_2|\sim 1$. For $0<m<3H/2$, between times $\tau_1$ and $\tau_2$, the dominant super-Hubble mode of the massive field $\sigma$ decays as:
\begin{align} \label{eq:qsfi-dec}
  \sigma_{k_1}(\tau_2) \sim \sigma_{k_1}(\tau_1) \left( \frac{\tau_2}{\tau_1}  \right)^{3/2-\nu}
  \sim \sigma_{k_1}(\tau_1) \left( \frac{k_1}{k_2}  \right)^{3/2-\nu}~,
\end{align}
where $\nu\equiv \sqrt{9/4-m^2/H^2}$.

In the squeezed limit, $k_1\ll k_2$. At $\tau_2$, the $k_1$ mode can still be considered as a shift of background. But the shift has a decay factor \eqref{eq:qsfi-dec}. As a result,
\begin{align}
  \lim_{k_1/k_2\rightarrow0}  \langle\zeta_{\mathbf{k}_1}\zeta_{\mathbf{k}_2}\zeta_{\mathbf{k}_3}\rangle
  \propto \frac{1}{k_1^3k_2^3} \left( \frac{k_1}{k_2}  \right)^{3/2-\nu}
  \propto \frac{1}{k_1^{3/2+\nu}} ~.
\end{align}
This analysis can also be made more rigorous \cite{Assassi:2012zq}. Also, in the special case $\nu=0$, a logarithm factor appears in the squeezed limit \cite{Chen:2009zp}.

\begin{figure}[htbp]
  \centering
  \includegraphics[width=0.45\textwidth]{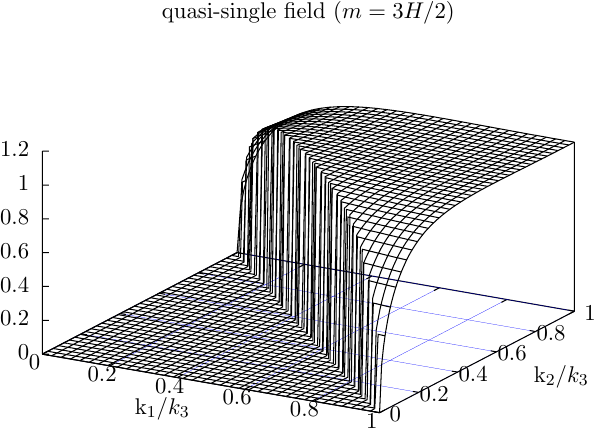}
  \hspace{0.03\textwidth}
  \includegraphics[width=0.45\textwidth]{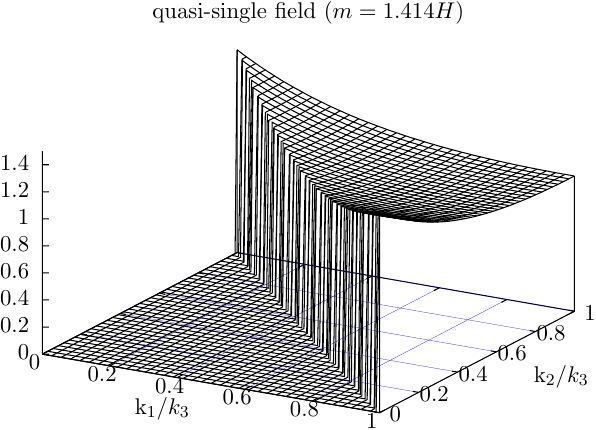}
  \\ \vspace{0.07\textheight}
  \includegraphics[width=0.45\textwidth]{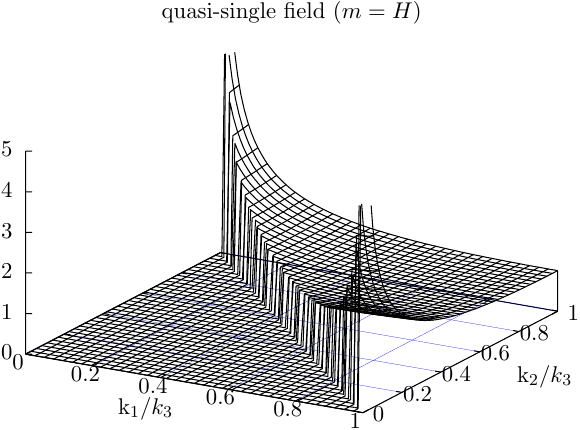}
  \hspace{0.03\textwidth}
  \includegraphics[width=0.45\textwidth]{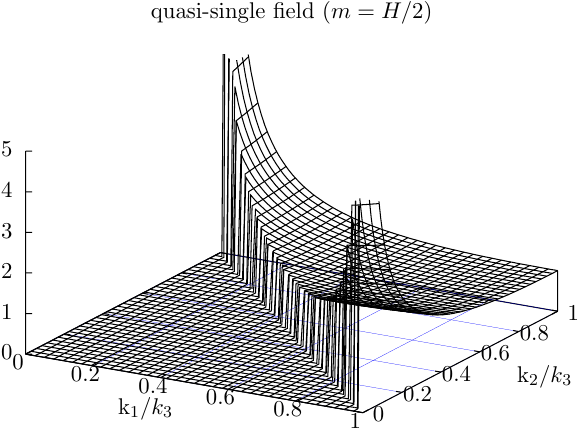}
  \caption{\label{fig:qsfi-shapes} The shapes of non-Gaussianity for quasi single field inflation, for $m=3H/2$ (i.e. $\nu=0$), $m=1.414H$ (i.e. $\nu=0.5$), $m=H$ (i.e. $\nu=1.12$) and $m=H/2$ (i.e. $\nu=1.414$) respectively. The shape ansatz \eqref{eq:shape-qsfi} is used. Note that the change of shapes are very sensitive to the $\sigma$ mass in Hubble unit.}
\end{figure}

The exact shape of three-point function is quite complicated and plotted in \cite{Chen:2009zp}. Phenomenologically, the dimensionless shape function can be approximated by 
\begin{align} \label{eq:shape-qsfi}
  3^{\frac{5}{2}-3\nu} (k_1^2+k_2^2+k_3^2)  (k_1k_2k_3)^{\frac{1}{2} - \nu} (k_1+k_2+k_3)^{-\frac{7}{2}+3\nu}
  ~.
\end{align}
This shape function is simple, but unfortunately not factorized. One can nevertheless write a factorized shape near the squeezed limit \cite{Sefusatti:2012ye}. The shape function \eqref{eq:shape-qsfi} is plotted in Fig.~\ref{fig:qsfi-shapes}.

It is interesting to notice that, the shape function for $m=H/2$ already looks very similar to the local shape. Indeed, a Fisher matrix analysis shows that the $m\leq H/2$ shape and the local shape are not distinguishable in the next 10 years' experiments \cite{Sefusatti:2012ye} (see also \cite{Norena:2012yi}). In case the local shape non-Gaussianity fits future data, it is challenging to distinguish if the non-Gaussianity comes from a massless field, or from a $m\leq H/2$ field and the corresponding quasi-local shapes. On the other hand, shapes from greater mass (such as $m= H$) will be distinguishable from local shape.

\begin{figure}[htbp]
  \centering
  \includegraphics[width=0.95\textwidth]{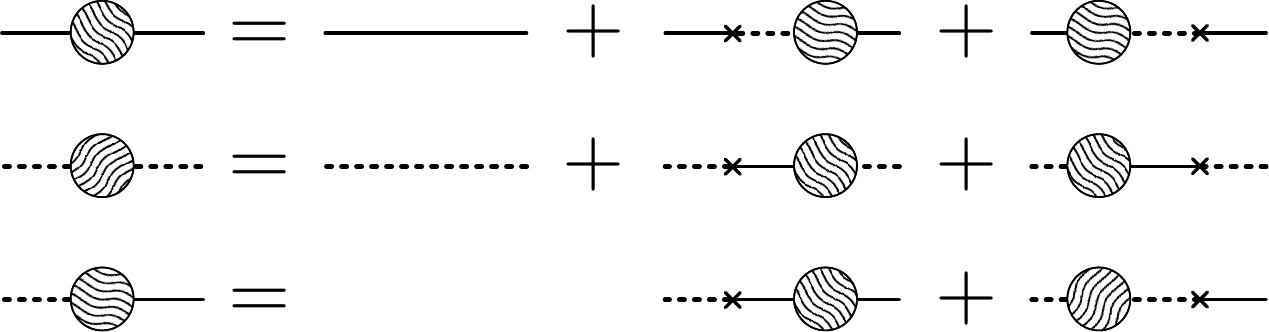}
  \caption{\label{fig:qsfi-resum} Resumming the transfer vertex. Note that the above diagrams represents a set of integral equations. When $V'''$ and/or other interactions are turned on, the resummation becomes more complicated but yet possible.}
\end{figure}

\subsubsection{Scaling dimensions for general operators}
\label{sec:general-quasi-single}

In the above discussion of quasi-single field inflation, we have restricted our attention to the weakly coupled case. In the strongly coupled case, perturbative methods of directly calculation breaks down. However, one can still try to resum the diagrams, seeking a weakly coupled effective field theory. For example, the transfer vertex can be in principle resummed using the Schwinger-Dyson equation, as illustrated in Fig.~\ref{fig:qsfi-resum}.

Alternatively, one can generalize the weakly coupled picture without doing the resummation. Especially for a conformal field theory (CFT) the interaction structure is highly constrained by symmetry \cite{Green:2013rd}. In CFT, mass and interaction operators are characterized by their scaling dimensions. In this case, the squeezed limit of correlation functions are as functions of the scaling dimensions of the operators. 

The scaling dimension of the mass operator is an analogue of the example of Sec.~\ref{sec:simple-model}. However, a non-trivial scaling dimension of interaction operators open up new possibilities. For example, one can imagine a sector of isocurvature perturbations which is strongly coupled on sub-Hubble scales, but weakly coupled on super-Hubble scales. The transition is controlled by the RG-flow and the anomalous dimension of the interaction operator. In this case the anomalous dimension dynamically generates a scale, analogous to $\Lambda_{QCD}$. One can compare this scale to the Hubble scale, and generate a one parameter family of non-Gaussianities. New possibilities such as the nearly orthogonal shape arise compared to the simple model that we reviewed in detail.

\subsection{Scale dependent features}
\label{sec:scale-depend-feat}

So far the inflation models we have reviewed are nearly scale invariant. Such models are natural because the observed power spectrum is nearly scale invariant. However, it is still possible that there are small ripples in the inflationary power spectrum, that we have not yet observed. Or the observational effects hide in higher order correlation functions or gravitational waves, which we haven't yet observed. Here we review a few models of this kind.

\subsubsection{Features in the potential}

Consider single field inflation, where the potential is approximately flat but there are small features on top of that.

\begin{figure}[htbp]
  \centering
  \includegraphics[width=0.5\textwidth]{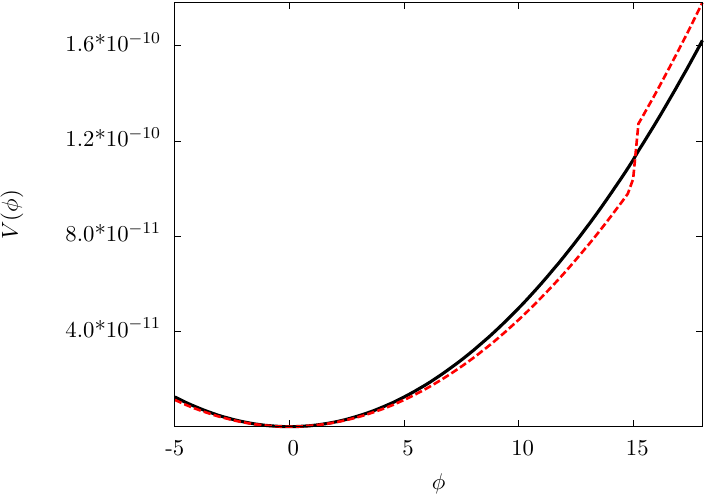}
  \caption{\label{fig:feature-pot} A feature in the potential. The solid line denote the actual feature considered in \cite{Chen:2006xjb}, with $c=0.0018$ and $d=0.022$. The feature is actually too small to be observable on the plot. The dot line is just an illustration of the feature, choosing by hand a huge feature ($c=0.1$), which is clearly ruled out by experiments.}
\end{figure}

For example, one can model a feature as \cite{Chen:2006xjb}
\begin{align}
  V(\phi) = \frac{1}{2} m^2\phi^2 \left[ 1+c\tanh \left( \frac{\phi-\phi_s}{d}  \right) \right]~.
\end{align}
The potential is illustrated in Fig.~\ref{fig:feature-pot}.

Alternatively, one can choose a feature with a non-continuous slope \cite{Martin:2011sn}:
\begin{align}
  V(\phi) = \begin{cases}
    V_0+A_+(\phi-\phi_0), & \text{if } \phi>\phi_0 \\
    V_0+A_-(\phi-\phi_0), & \text{if } \phi<\phi_0
  \end{cases}~.
\end{align}

A sharp feature can in general enhance slow roll parameters. The enhanced slow roll parameters result in a feature in the power spectrum, as well as large non-Gaussianity. However, it is useful to estimate which slow roll parameter is enhanced most. Considering the sharpness of the feature, higher time derivatives get enhanced most. 

In the third order action of standard single field inflation \eqref{eq:l3}, the highest derivative is $\dot\eta$. Thus the interaction is dominated by
\begin{align}
  \mathcal{H}_3 \supset -\frac{M_p^2a^3\epsilon}{2} \dot\eta \zeta^2\dot\zeta~.
\end{align}
The shape of the bispectrum can be calculated numerically. The feature time $\tau_*$ controls the oscillation $\sin[(k_1+k_2+k_3)\tau_*]$, while its smoothing timescale $\Delta\tau$ controls the high-frequency envelope \cite{Chen:2008wn}. For a sharp feature, $\Delta\tau\ll|\tau_*|$, so modes with $|(k_1+k_2+k_3)\tau_*|\gg1$ can also be affected; the envelope is suppressed for $(k_1+k_2+k_3)\Delta\tau\gg1$.

\subsubsection{Sharp turning of trajectory}

\begin{figure}[htbp]
  \centering
  \includegraphics[width=0.7\textwidth]{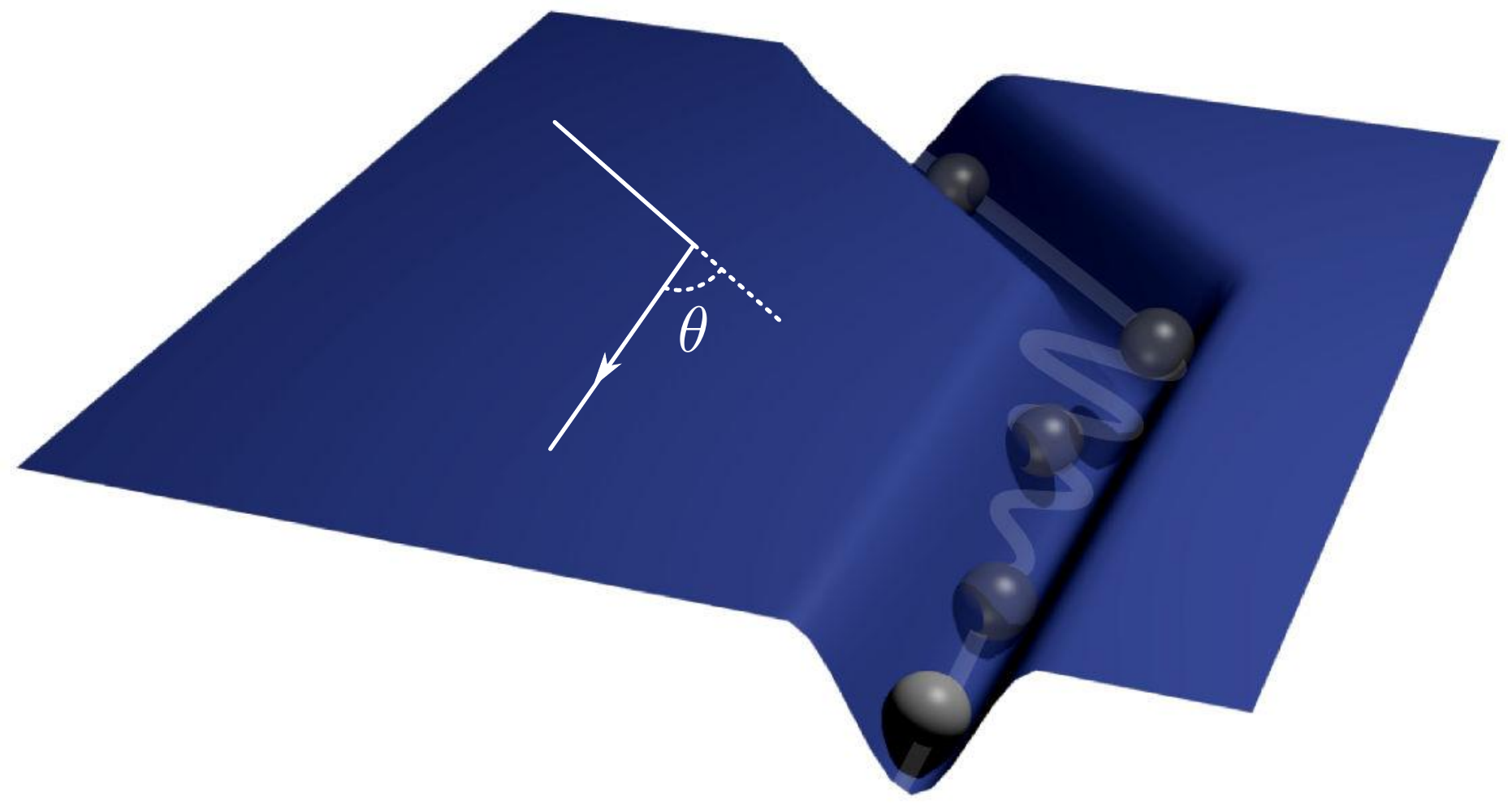}
  \caption{\label{fig:sharp} A sharp turning of trajectory.}
\end{figure}

Here we review a sharp turning of inflationary trajectory \cite{Achucarro:2010da, Achucarro:2010jv, Chen:2011zf, Shiu:2011qw, Cespedes:2012hu, Achucarro:2012sm, Avgoustidis:2012yc, Gao:2012uq, Achucarro:2012fd}. For example, there is one inflation direction, and another very massive direction. The turning of trajectory is between those two directions, as illustrated in Fig.~\ref{fig:sharp}.

During a sharp turning, a combination of effects may happen:
\begin{itemize}
\item \textit{Background oscillation:} When the trajectory turns very sharply, the kinetic energy of the inflaton gets projected on to the massive direction. The amount of transfer energy is $\frac{1}{2}\dot\phi^2\sin^2\theta$, where $\theta$ is the turning angle. This amount of energy drives the massive direction $\sigma$ to move as high as $\sigma \sim (\dot\phi/m)\sin\theta$. In case this offset in the $\sigma$ direction is not cancelled by a linear potential, the $\sigma$ field get released from this height and start to oscillate.

The background oscillation is not just a shift of background. Instead more dramatic things can happen. Note that the perturbations has varying physical wave length $k/a$. Once the $k/a$ for a perturbation mode gets close to the oscillation frequency of the background, parametric resonance can happen. This situation is similar to the case of preheating. As a result, both the linear perturbation and non-Gaussianity are enhanced.

It is interesting to note that, both the background oscillation itself, and the induced effects such as the resonant enhancement of perturbations, are periodic (with a decaying envelope) with respect to physical time. On the other hand, the comoving wave number $k$ is defined with respect to the conformal time. Thus in case one observes such a periodic feature in the sky, the function dependence between the physical time and conformal time is obtained. This function dependence differs significantly between inflation and alternative to inflation models \cite{Chen:2011zf, Chen:2011tu}.

In case the turning of trajectory is not sharp enough, or a linear potential comes to cancel the centrifugal force during the turning of trajectory, the background oscillation may not be present.

\item \textit{Sharp feature:} During the turning of trajectory, the slow roll parameters got boosted thus there is an similar effect as discussed in Sec.~\ref{sec:scale-depend-feat}.

\item \textit{Vacuum projection:} The wave function of a quantum harmonic oscillator depends on the frequency of the oscillator. The same is true for scalar fields. The wave function of a massive direction is different from the wave function of a massless direction. However, after a sharp turning of trajectory, the two wave functions gets mixed. As a result, the perturbations of the fields are no longer in the vacuum state.

In terms of $k$-modes, the perturbation modes with $H < k/a < m$ are affected. The mixing, which can be calculated using the in-in formalism, kicks the perturbations even if there is no background oscillation. 

\end{itemize}

\subsubsection{Particle production}

Despite the wording ``slow roll'', the inflaton field actually runs very fast, with $\dot\phi \gg H^2$. Thus it is useful to check if the fast motion of the inflaton pollutes the inflationary environment. For example, when other fields have coupling to the inflaton, there may be particle productions induced by the motion of the inflaton \cite{Chung:1999ve, Romano:2008rr, Barnaby:2010ke, Battefeld:2011yj}.

For example, consider an otherwise massless field $\chi$, with the interaction Lagrangian
\begin{align}
  \mathcal{L}_\mathrm{int} = - \frac{g^2}{2} (\phi-\phi_0)^2 \chi^2 
  \equiv  -\frac{1}{2} m^2_\mathrm{eff}(\phi) \chi^2  ~.
\end{align}
The motion of the inflaton provides an effective mass for $\chi$. The equation of motion for $\chi$ is
\begin{align}
  \ddot\chi + 3H \dot\chi - \frac{\partial^2}{a^2} \chi + m^2_\mathrm{eff}(\phi) \chi = 0~.
\end{align}
When the change of background is slow, the $\chi$ field has a WKB solution. As a result $\chi$ almost stays unchanged without particle production. However, when $\phi$ moves across $\phi_0$, the situation changes dramatically. Not only that $\chi$ gets effectively massless, but also the WKB condition breaks badly:
\begin{align}
  \frac{|\partial_t(m_\mathrm{eff})|} {m^2_\mathrm{eff}} = \frac{|\dot\phi|}{g(\phi-\phi_0)^2} \gg 1 ~.
\end{align}
Thus particles of $\chi$ get produced. To see the effect more closely, one can parameterize the behavior of $\phi$ near $\phi_0$ as $\phi\sim \phi_0 + vt$. Solving the EoM of $\chi$, the occupation number can be calculated as
\begin{align}
  n_k = \exp \left( -\frac{\pi k^2}{a_*^2g|v|}  \right)~.
\end{align}
Here $k$ is comoving momentum and $a_*$ is the scale factor at particle production. After the moment $\phi=\phi_0$, the produced particles quickly become massive. The mass increases while the number density is diluted. The back-reaction from the produced particles to the adiabatic perturbation can be calculated. The power spectrum gets a correction. For the bispectrum, for small $k$ the shape of non-Gaussianity is similar to local shape. For larger $k$ there are oscillations in the shape function \cite{Barnaby:2010ke}.

It is also interesting to put this scenario onto a Landscape type potential. When the number of independent particle-production loci increases, it is more likely, and sometimes exponentially likely, for the inflaton to encounter particle production events (with a small mass of $\chi$) \cite{Battefeld:2011yj}.












\section*{Acknowledgement}
I would like to thank my collaborators Niayesh Afshordi, Robert Brandenberger, Yi-Fu Cai, Bin Chen, Xingang Chen, James Cline, Francis Duplessis, Xian Gao, Jinn-Ouk Gong, A.Emir Gumrukcuoglu, Min-xin Huang, Bin Hu, Qing-Guo Huang, Miao Li, Chunshan Lin, Alexander Maloney, Guy Moore, Shinji Mukohyama, A. Riotto, Gary Shiu, An\v{z}e Slosar, Tower Wang and Wei Xue for collaborations and discussions on inflation and cosmic perturbations. I thank Valentin Assassi, Daniel Baumann, Xingang Chen, and David Seery for discussions during writing this review. I thank Xingang Chen for helpful comments on a draft of this review, and thank Evan Mcdonough for pointing out typos in a previous version. This work is supported by fundings from the Kavli Institute for the Physics and Mathematics of the Universe (Kavli IPMU), and from the Japan Society for the Promotion of Science (JSPS).

\appendix

\section{Computer Assisted Analytical Calculation}

Although the physics of inflationary perturbations is interesting, the calculations are sometimes tediously long. Fortunately, nowadays computers are in the position to assist us for this purpose.

\subsection{Algebra systems}

There are quite a few choices of algebra systems in the market. Among those, there is proprietary software such as Mathematica and Maple, and free software such as Sage and Maxima. Here we comment the pros and cons for our particular purpose (there are of course other choices as well):

\begin{itemize}
\item \textit{Mathematica:} The functional programming paradigm is typically convenient for math calculations. Also there is a powerful mechanism of pattern matching, related to its functional programming nature. There are also GR packages such as xAct available on Mathematica.
\item \textit{Maple:} There is a very powerful package on general relativity: GRTensor, on the Maple platform.
\item \textit{Sage:} Sage is a collection of free math software. Those lower level software use different languages and conventions. And sage unifies them with an interface using the Python programming language. Python is strong in object-oriented programming, in contrast with the functional programming approach taken by Mathematica.
\item \textit{Maxima:} Currently Sage use Maxima as back-end. In the analytical aspects, Maxima itself is feature complete, if one would not need the conveniences provided by Python.
\end{itemize}

The examples of calculating cosmic perturbations using Mathematica, Maple and Sage are given at 
\begin{verbatim}
https://dl.dropbox.com/u/49916494/Yi-Wang-pert-sample-code.tar.gz
\end{verbatim}
Not every part of the code is thoroughly checked thus please double check before usage.\footnote{Note added in 2026: The original Dropbox download link is no longer maintained. More developed implementations of the tensor-algebra, general-relativity and integration-by-parts methods discussed here are available in MathGR (Mathematica) \cite{Wang:2013MathGR}, \texttt{https://github.com/tririver/MathGR}, and MathGR-Py (Python/SymPy, with an MCP interface), \texttt{https://github.com/tririver/mathgr-py}.} From now on we take Mathematica as an example for discussion.

\subsection{Summed indices}

One annoying fact is that computers are usually not clever enough to combine quantities in desired forms. Especially they typically do not collect summed indices. Algebraic systems as Mathematica will leave you something like $ \partial_1\partial_1 f + \partial_2\partial_2 f + \partial_3\partial_3 f $ instead of $\partial_i\partial_if$. There are at least two tricks one can do for this:
\begin{itemize}
\item \textit{Replace the dummy indices:} Let \texttt{pd[f,x]} represent a partial derivative without evaluating it, and let \texttt{p2[f]} represent the spatial Laplacian. We can define a replacement rule
  \begin{verbatim}
    expression /. HoldPattern[pd[pd[f_,x1],x1]] :>
      p2[f] - pd[pd[f,x2],x2] - pd[pd[f,x3],x3]
  \end{verbatim}
  Note the underscore here. This underscore matches and replaces $\partial_1\partial_1$ acting on all variables, not only f itself. We will also meet a lot of other combinations that we need to design rules to match and replace. 
\item \textit{Keep them abstract from the beginning:} For most of the cases, we don't need to actually calculate the summation of spatial indices, we just need to write them in a compact form. For this purpose, we can directly write $\partial_i\partial_if$ in the code. It is (typically) elegant and not hard to do this. We have to note:
  \begin{itemize}
  \item The default derivative operator tries to calculate $\partial_if$ explicitly. This is not the desired behavior when we decide to keep $\partial_i\partial_if$ as abstract. To solve this problem, we can define derivative operators ourselves. The definition should include linearity and the Leibniz's rule for multiplication, without defining how to calculate $\partial_if$ explicitly. It is also convenient to use the ``Notation'' package to overload the $\partial$ operator.
  \item Dummy indices may conflict. For example, how does Mathematica write $(\partial^2 f)^2$ as $\partial_i\partial_if\partial_j\partial_jf$ instead of as $\partial_i\partial_if\partial_i\partial_if$? For this purpose, the Unique[] function is useful. We can write
    \begin{verbatim}
      With[{i = Unique[], j = Unique[]},
        pd[pd[f,i],i] pd[pd[f,j],j]
      ]
    \end{verbatim}
    Using the same \texttt{pd} derivative head, Mathematica assigns independent dummy indices to the two factors. It is also not hard to write a function to simplify dummies.
  \end{itemize}
\end{itemize}

\subsection{Integration by parts}

In the code calculating the action of perturbations, one frequently need to integrate by parts. The situations can be divided into 2nd order or higher order actions, and spatial or temporal directions. For 2nd order spatial direction, one can go to momentum space to get rid of integration by parts. However, for higher order, or temporal direction, going to momentum space does not help that much. 

One general way to do integration by parts is to use the Mathematica function ReplaceList[]. This function tries all possible replace rules (which we write ourselves). We just need to setup some measure to compare the result of each replacement and find a good one. We can repeatedly call ReplaceList[] (or multiple ReplaceList[] in a row) until the result is simplest. 


\end{document}